\documentstyle[12pt,twoside,axodraw]{report}
\newif\ifall\alltrue
\def\private#1{}            
\def\ILM{Instanton Liquid Model}
\def\eqd{\stackrel{\bullet}{=}}
\def\ff{\Longrightarrow}
\def\gdw{\Longleftrightarrow}

\def\gtapprox{\buildrel{\lower.7ex\hbox{$>$}}\over
                       {\lower.7ex\hbox{$\sim$}}}
\def\nq{\hspace{-1em}}
\def\look{\(\uparrow\)}
\def\ignore#1{}
\def\deltabar{{\delta\!\!\!^{-}}}
\def\hbar{h\!\!\!\!^{-}\,}
\def\dbar{d\!\!^{-}\!}
\def\eps{\varepsilon}
\def\beq{\begin{equation}}
\def\eeq{\end{equation}}
\def\beqn{\begin{displaymath}}
\def\eeqn{\end{displaymath}}
\def\bqa{\begin{eqnarray}}
\def\eqa{\end{eqnarray}}
\def\bqan{\begin{eqnarray*}}
\def\eqan{\end{eqnarray*}}

\begin{document}

\begin{titlepage}
\begin{center}       \vspace*{3cm}
  {\Huge     INSTANTONS IN QCD                       } \\[5mm]
  {\LARGE    Theory and Application of the           } \\[5mm]
  {\LARGE    Instanton Liquid Model                  } \\[70mm]
  {\large\it Ph.D. Thesis of the Faculty for Physics } \\[2mm]
  {\large\it at the Ludwig Maximilians University Munich } \\[30mm]
  {\large    Presented by                            } \\[2mm]
  {\large    \bf Marcus Hutter                       } \\[2mm]
  {\large    Munich, 17$^{th}$ of October, 1995      } \\[20mm]
\end{center}
\end{titlepage}


\begin{titlepage}

\begin{center}\large
Inofficial translation from the German Ph.D. thesis \\[6ex]\sl
INSTANTONEN IN DER QCD \\
Theorie und Anwendungen des \\
Instanton-Fl\"ussigkeit-Modells
\end{center}

\vspace*{14cm}
1. Advisor: Prof. H. Fritzsch \\
2. Advisor: Prof. S. Theisen \\
Defense: 9 February 1996
\end{titlepage}

\centerline{\bf Abstract}

The current theory of strong interactions, the quantum
chromodynamics (QCD), is a non-abelian gauge theory, based on the
gauge group SU(3). Despite its formal similarity to QED there are
significant differences. It was shown in 1973 already that the
coupling constant $g$ increases for large distances \cite{Gross}.
This gave hope for the possibility to explain quark and gluon
confinement. Soon after, in 1975, non-trivial solutions of the
Euclidian Yang Mills equations were found, nowadays called BPST
instantons \cite{BPST}, which significantly influence the low
energy structure of QCD. Many exact results are known for the
1-instanton vacuum \cite{Raj,Bro}, whereby an interesting
phenomenological result of it is the explicit breaking of the
axial $U(1)$ symmetry \cite{tHo}. On the other hand, a one
instanton approximation, similar to a tree approximation in
perturbation theory, cannot describe boundstates or spontaneous
symmetry breaking. The next step was the analysis of exact
\cite{Act} and approximate \cite{CDG} multi-instanton solutions.
There are two useful visualizations for these solutions. In
one of these, instantons are interpreted as tunneling processes between
different vacua. In the other interpretation, a solution describes
an ensemble of extended (pseudo) particles in 4 dimensions.

In conventional perturbation theory one computes fluctuations
around the trivial zero solution. The correct quantization process
is to consider all classical solutions of the field equations and
their fluctuations. In the path integral representation of QCD the
partition function is, hence, dominated by an ensemble of extended
particles (instantons) in 4 dimensions at temperature $g^2$. In
the simplest case the partition function describes a diluted ideal
gas of independent instantons. Unfortunately, this assumption
leads to an infinite instanton density caused by large instantons,
which obviously contradicts the assumption of a diluted gas. This
problem is known as the infrared problem.  The problem is avoided
by assuming a repulsive interaction \cite{Ilg} which prevents the
collapse. This is the model of a 4 dimensional liquid. Under
certain circumstances the interaction can be replaced by an
effective density. The \ILM\ in a narrow sense describes the QCD
vacuum as a sum of independent instantons with radius
$\rho=(600\mbox{MeV})^{-1}$ and effective density
$n=(200\mbox{MeV})^4$. The correctness of this model is still
being intensively investigated. So far the model is essentially
justified by its phenomenological success.

Numerical simulations of the \ILM\ allowed to determine a number of
hadronic quantities, especially meson masses, baryon masses,
hadron wave functions, and condensates \cite{Shu,ShV}.

For computing the quark propagators and the meson correlators
there are also analytical methods. The most important predictions
are probably the breaking of the chiral symmetry (SBCS) in the
axial triplet channel \cite{Dya} and the absence of Goldstone
bosons in the axial singlet channel.

The largest part of this thesis is devoted to extending the
analytical methods and to evaluating the results in (semi)analytical
form.

The meson correlators (also called polarization functions) will be
computed in the \ILM\ in zeromode and $1/N_c$ approximation,
whereby dynamic quark loops will be taken into account. A spectral
fit allows the computation of the masses of the $\sigma$, $\rho$,
$\omega$, $a_1$ and $f_1$ mesons in the chiral limit. A separate
consideration also allows computation of the $\eta'$ mass. The
results coincide on a 10\% level with the experimental values.
Furthermore, determining the axial form factors of the proton,
which are related to the proton spin (problem), will be attempted.
A gauge invariant gluon mass for small momentum will also be computed.

The thesis ends with several predictions which do not rely on the
\ILM. In the 1-instanton vacuum a gauge invariant quark propagator
will be computed and compared to the regular and singular
propagator. Rules for the choice of a suitable gauge, especially
between regular or singular, will be developed. A finite
relation between the quark condensate and the QCD scale $\Lambda$
will be derived, whereby neither an infrared cutoff, nor a specific
instanton model will be used.

{\parskip=0ex
 \tableofcontents
 \vspace{1cm}
 The chapters marked with * are more technical in nature.
 \listoftables
}



\chapter{Introduction}\label{Ch1}

The fundamental theories which describe the observed interactions
are all gauge theories.
They describe gravitation, electroweak and strong
interaction. In the quantized version the forces between
particles (fermions) are mediated by gauge bosons.
QCD describes the strong interaction between quarks and gluons.
It is a SU(3) gauge theory with the Lagrangian\footnote{
Throughout the entire work the Euclidian formulation of QCD is used
 \cite{SVZ}. }
\beq
  {\cal L} = {1\over 4g^2}G_{\mu\nu}^aG^{\mu\nu}_a +
    \sum_{i=1}^{N_f} \bar\psi_i(iD\!\!\!/+im_i)\psi.
\eeq
Taking into account only the light quarks $u$, $d$ and
sometimes $s$ and setting their mass to zero, one arrives at a
theory with only one parameter, the gauge coupling constant $g$,
which actually is no parameter due to dimensional transmutation.
Despite the fact that QCD looks so simple (e.g. compared to
electroweak interaction) it is very hard to solve this theory due
to nonperturbative effects.

\section{Methods to Solve QCD}\label{sec11}

To clarify the role of instantons in QCD let me first give a list
of the most important methods used to tackle QCD, starting with
very general methods applicable to any quantum field theory (QFT)
and ending with more specific approaches:

\begin{itemize}
 \item {\it Axiomatic field theory: }
   Wightman/Oswalder\&Schrader have stated a set of
   Minkowskian/Euclidian axioms for
   vacuum correlators a general QFT should respect (analyticity,
   regularity, Lorentz invariance, locality, \ldots).
   It is clear that the theorems derived from these axioms
   have to be very general because no Lagrangian is used \cite{Gli}.
 \item{\it Haag-Ruelle/LSZ Theory: }
   S-Matrix elements are related to vacuum correlators. 
   The S-Matrix
   is the central object for particle phenomenology, containing such
   important information as cross sections, form factors,
   structure functions, \ldots.
   Vacuum correlators are more suitable for theoretical studies
   because they avoid the need of explicitly
   constructing the Hilbert space, which is an extremely
   complicated task beyond perturbation theory.
 \item{\it Quantization: }
   One might think that quantization should not appear in a list
   of methods for solving QFT because it is the method to
   obtain and define a QFT. On the other hand, besides canonical
   quantization there are other ways of quantizing a theory.
   The most popular is the path integral quantization \cite{Fey}.
   In textbooks for particle physics
   it is usually only used as an abbreviation to derive
   theorems more quickly. In general (beyond perturbation theory), different
   quantization methods lead to different physical and mathematical
   insights and different methods to solve the theory. Variants of
   the path integral quantization
   are the random walk quantization used in lattice theories, and
   stochastic quantization.
 \item{\it (Broken) Symmetry: }
   Every degree of freedom like spin, flavor and color is the
   possible origin of (approximate) symmetries
   like $SU(2N_f)$ or subgroups and $SU(3)$ gauge invariance.
   Conserved currents and Ward identities \cite{War}
   can be obtained. In the case of light quarks,
   one further has approximate chiral symmetry leading to
   PCAC, axial Ward identities, current algebra theorems,
   soft pion physics,\ldots. 
   Furthermore, QCD with massless quarks
   possesses an anomalously broken scale invariance, which is the
   origin of the huge field of renormalization group techniques \cite{CSZ}.
 \item{\it Perturbation theory: }
   Due to asymptotic freedom the coupling constant $g$ decreases for
   high energies and perturbation theory in $g$ is applicable. 
   QCD can thus be solved for processes which involve only momenta of,
   say, more than 1 GeV. The small distance behaviour of vacuum correlators
   is, thus, calculable ($x\leq 0.2$ fm).
 \item{\it Operator Product Expansion (OPE): }
   An improvement of perturbation theory is to separate the small
   distance physics from large distance effects. The former is contained
   in the so-called Wilson coefficients calculated perturbatively.
   The latter nonperturbative effects are contained in a few vacuum
   or hadron expectation
   values of {\sl local} operators which have to be determined from
   phenomenology or uncertain assumptions \cite{SVZ}.
   Vacuum correlators can be obtained
   up to distances of $x\leq 0.3\ldots 0.5$ fm.
 \item{\it QCD Sum rules: }
   QCD sum rules are widely used to determine hadron masses and couplings.
   The general method is to assume the existence of certain hadrons and
   take a resonance+continuum ansatz in the Minkowskian region
   for some correlator. The Euclidian correlator is calculated
   from theory (OPE, lattice, instantons). Via dispersion relations
   one may match both in some Euclidian window by fitting the hadron
   parameters which leads to a prediction for them. 
 \item{\it Effective Theories: }
   One may construct effective Lagrangians containing mesons
   and/or baryons more or less motivated by QCD or history or
   other physical branches.
   Their parameters are determined from experiment or QCD as far
   as possible.
\end{itemize}

A variety of phenomena have been explained qualitatively and
calculated quantitatively with the methods listed above, but the large
distance behaviour of QCD is still unsolved. There are
at least two problems belonging to this domain: chiral symmetry
breaking and confinement. Up to now chiral symmetry breaking was
assumed and the consequences such as Golstone bosons
were discussed within this assumption.
In OPE one takes the nonzero values of the quark and other condensates
from experiment but has no possibility to predict them from theory.
The quark condensate is the order parameter of chiral symmetry and a nonzero
values indicates spontaneous breaking of this symmetry (SBCS).
Confinement has also to be assumed.

Approaches to solve these problems are:

\begin{itemize}
 \item{\it Lattice QCD: }
   In principle the method is very simple. The continuum
   is replaced by a fine lattice covering a large
   but finite volume. The path integral is thus replaced by
   a finite number of integrals evaluated numerically. All vacuum
   correlators can be obtained for arbitrary Euclidian distances.
   Confinement and SBCS have been shown and other hadronic parameters
   are obtained. In practice, lattice
   calculations are much less straightforward than this sketchy
   description might suggest \cite{Cre}.
 \item{\it Instantons: }
   As in lattice QCD one evaluates the Euclidian path integrals
   but now in semiclassical approximation. In addition to the
   global minimum of the QCD action $A_\mu^a=0$ used by perturbation theory
   there are many other local minima called instantons which have to be
   taken into account. Inclusion of light quarks leads to an effective
   $2N_f$ quark vertex responsible for SBCS. 
   Although confinement cannot be explained,
   a lot of hadron parameters can nevertheless
   be calculated \cite{Shu,ShV} suggesting
   that confinement is not essential for the properties of hadrons.
\end{itemize}

Often only a combination of the results from various approaches
allows contact to phenomenology.

\section{Contents}\label{sec12}

In this work I want to give a quantitative and systematic
study of the implications of instantons to hadron properties.
For an elementary introduction into the classical and semi-classical
theory of solitons and instantons I recommend \cite{Raj}.

{\it Chapter \ref{Ch2}} is an introduction to the semiclassical
evaluation of nontrivial integrals. After developing the method in
the finite dimensional case the partition function of QCD is
considered and the instanton liquid model is introduced.

In {\it Chapter \ref{Ch3}} the propagator of a light quark is
calculated. The approximations which have to be made are stated
and discussed. It is shown that for one quark flavor the $1/N_c$
expansion is exact. The constituent quark masses and the quark
condensates are calculated for $u$,$d$ and $s$ quarks.

The same approximations are used in {\it Chapter \ref{Ch4}} to
calculate the 4 point functions. Special attention is payed to the
singlet correlator where a chain of quark loops contributes and is
{\it not} suppressed in the large-$N_c$ limit. Within one and the
same approximation we get Goldstone bosons in the pseudoscalar
triplet correlator but no massless singlet boson.

In {\it Chapter \ref{Ch5}} meson correlators are discussed and
plotted. Employing a spectral ansatz it is possible to extract
various meson masses and couplings. They are compared to the
values obtained from extensive numerical studies of the instanton
liquid \cite{ShV} and to the experimental values.

In {\it Chapter \ref{Ch6}} the axial anomaly is examined. The most
interesting quantities in the axial singlet channel are the mass
of the $\eta'$ meson and the spin of the proton. A combination of
the axial anomaly and the scale anomaly with ideas from instanton
physics $m_{\eta'}$ can be determined with an accuracy of 10\%.
The focus is on the discussion of the proton spin and its
computation in the \ILM.

In {\it Chapter \ref{Ch7}} the ghost and gluon propagator in the
\ILM\ will be computed. For small momentum the ghost mass is 340 MeV and
the gluon mass is 480 MeV. The masses are independent of the gauge
parameter  $\xi$.

{\it Chapter \ref{Ch8}} attempts to predict several quantities
without relying on \ILM. In regular gauge it is possible to derive
a finite relation between the quark condensate and the QCD scale
$\Lambda$. Since the results depends critically on the choice of
the right gauge, the chapter starts with a detailed discussion on
the choice of a gauge and with a calculation of a gauge
independent propagator.

A summary of the results, new in this thesis, can be found in {\it
Chapter \ref{Ch9}}, followed by open questions, which deserve
further investigation. The {\it Appendices} contain a list of used
notation and explicit expressions for one instanton and its Dirac
zeromode in singular, regular, and axial gauge. Furthermore,
methods to compute Fourier transforms and convolutions of Lorentz
covariant functions are described.

All {\it Figures} can be found at the end of this work before the
{\it Reference} section (sorted with respect to topics).

The various chapters can be read independently of each other.
Sections marked with * are more technical in nature.



\chapter{Theory of the Instanton Liquid}\label{Ch2}

This chapter gives an introduction into the semiclassical
evaluation of the QCD partition function. For this, the
fluctuations around the classical solutions of the Euclidian
equations of motion are separated into (approximately) Gaussian
and collective coordinates (Section \ref{sec22}). In Section
\ref{sec23} the approximately Gaussian degrees of freedom will be
integrated out. The classical solutions of the Yang Mills
equations can be classified with respect to their topology  $N\in
Z\!\!\!Z$ and are called multi-instanton solutions. For an
introduction into the mathematically beautiful theory of
instantons the reader should consult \cite{Act}.
An elementary introduction into the classical and semiclassical theory
of solitons and instantons can be found in \cite{Raj}.
The classical one-instanton solution and its corresponding
semi-classical partition function will be presented in Section \ref{sec24}.
The multi-instanton solutions can be reduced to the one-instanton case,
in the case of well-separated instantons.
Section \ref{sec25} describes how quarks modify the partition
function. In Section \ref{sec26} the infrared problem will be
discussed and the \ILM\ to ''solve'' this problem. This model is
the basis of Chapters \ref{Ch3}-\ref{Ch7}.

\section{Separating Gaussian from non-Gaussian Degrees of
Freedom *}\label{sec21}

In the path integral formulation solving QCD or any other QFT is equivalent
to calculating the partition function $Z$ including external source terms.
Of course this is not a
simple task because one has an infinite number of degrees of freedom.
One way to tackle this problem is to reduce the number of degrees
of freedom by integrating the (nearly) Gaussian degrees
perturbatively leaving the non-Gaussian degrees, called collective
coordinates,
to be treated with other methods.

\begin{figure}[bbb]
   \epsfbox[85 0 480 550]{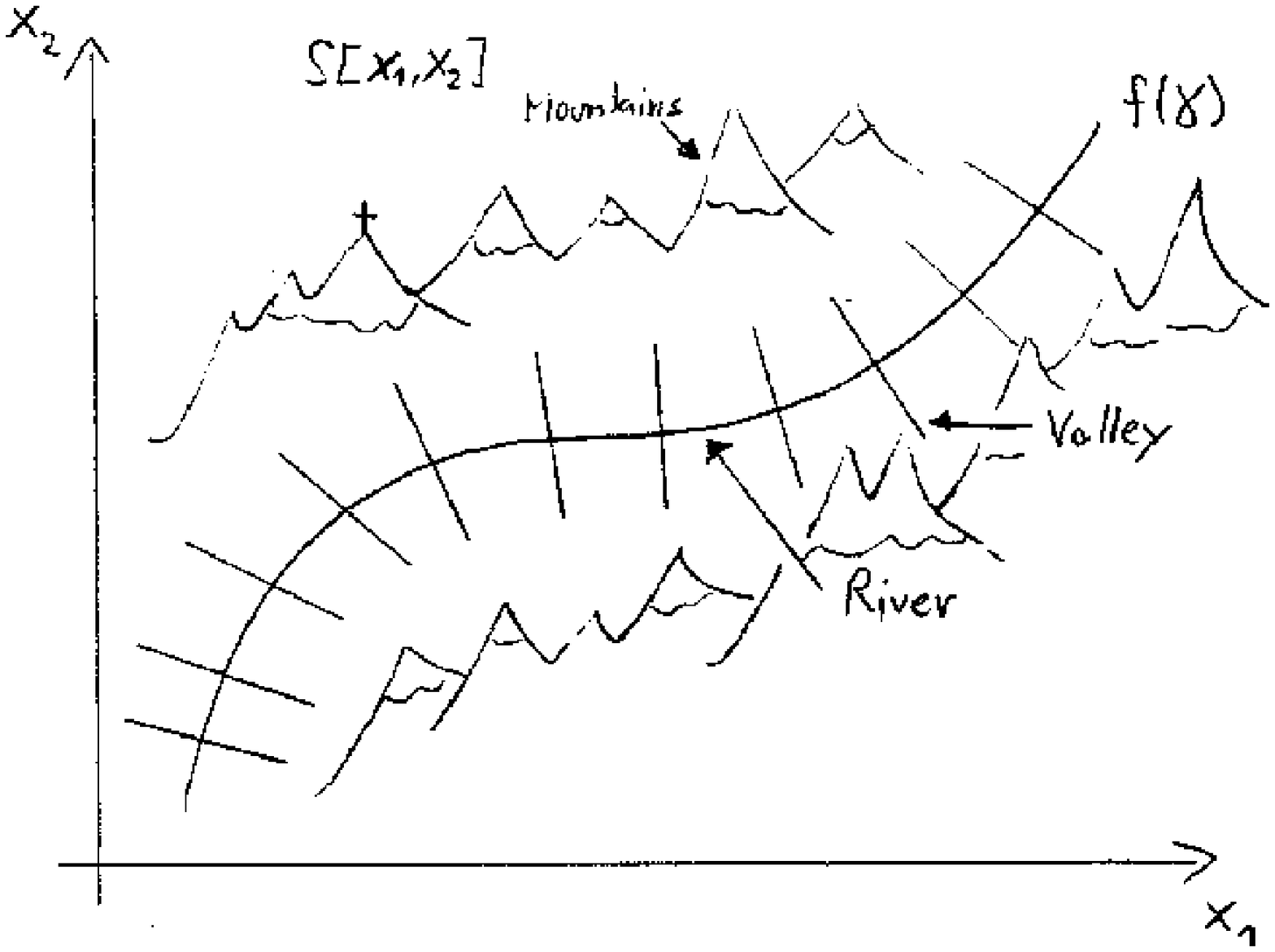}
   \caption[Panorama Function]{\label{fig31}
      \it $S$ is minimal at the river,
                   small in the valley and
                   large in the mountains,
            Therefore $Z=\int\!dx_1dx_2\,e^{-S[x_1,x_2]}$
              is dominated by the valley.
           }
\end{figure}

Let me first describe this method most generally
without referring to QFT. Our aim is to calculate
\beq\label{e31}
  Z=\int\!d^n\!x\,e^{-S[x]}
\eeq
where S is some real valued function depending on the $n$ dimensional vector
$x\in I\!\!R^n$. Let the integral be dominated by values of $x$ which lie
in the vicinity of a $k$ dimensional submanifold of $I\!\!R^n$ which may
be parameterized by $x=f(\gamma)$, $\gamma\in I\!\!R^k$.
Vectors in the tangent
space of $f(\gamma)$ are called (approximate) zeromodes, because
$S[x(\gamma)]$ is (nearly) constant in this direction. Usually
$f(\gamma)$ represents some degenerate or approximate minima of $S$.
In addition $f(\gamma)$ may contain points which do not contribute
much to the functional integral, this will cause no error, but
$f(\gamma)$ must not forget any significant points.
In figure \ref{fig31} a two dimensional example is shown containing a
river (one dimensional manifold) with steep mountains aside his
banks. So the integral will be dominated by the shaded area.
This area can be parameterized in the following way:
\beq\label{e32}
  x=f(\gamma)+y                     \quad,\quad
  x,y\in I\!\!R^n                   \quad,\quad
  \gamma\in I\!\!R^k            \quad.
\eeq
To make this representation unique, we have to demand $k$ linear extra
conditions to $y$:
\beq\label{e33}
  y\!\cdot\! g_i(\gamma)=c_i \quad,\quad 1\leq i\leq k
\eeq
E.g., $c_i=0$ and
$g_i(\gamma)=\partial f(\gamma) / \partial \gamma_i $
would fix $y$ to be orthogonal to the "river" or equally stated:
It disallows fluctuations in the non-Gaussian zeromode direction.
Now every point $x$ (at least in the vicinity of the river) can
uniquely be described by $y$ satisfying the condition
(\ref{e33}) and by $\gamma$ via (\ref{e32}).
All we have to do now is to represent $Z$ in terms of the fluctuation
vector $y$ and the collective coordinates $\gamma$.
A convenient way is to introduce a Faddeev-Popov unit
\beq\label{e34}
  1=\int\!d^k\!\gamma\,d^n\!y
  \,\delta^k\big(y\!\cdot\! g_i(\gamma)-c_i\big)
  \,\delta^n\big(f(\gamma)+y-x\big)
  \,\Phi(x)
\eeq
which serves as a definition of $\Phi$.
Inserting this into (\ref{e31}) and integrating over $x$ yields
\beq\label{e35}
  Z=\int\! d^k\!\gamma\,d^n\!y
  \,\delta^k(y\!\cdot\! g_i(\gamma)-c_i)
  \,\Phi(f(\gamma)+y)
  \,e^{-S[f(\gamma)+y]}               \quad.
\eeq
The $y$-integration in (\ref{e34}) can trivially be performed:
\beq\label{e36}
  \Phi^{-1}(x) =
  \int\! d^k\!\gamma^\prime
  \,\delta^k\big( (x-f(\gamma^\prime))g_i(\gamma^\prime)-c_i \big)
  \quad.
\eeq
For $x=f(\gamma)+y$ the $\delta$-function in the
integral only contributes for $\gamma^\prime=\gamma$.
Thus we can expand the $\delta$-argument
up to linear order in $\gamma^\prime-\gamma$
\begin{eqnarray}\label{e37}
  \Phi^{-1}(f(\gamma)+y) & = &
  \int\! d^k\!\gamma^\prime
  \delta^k
    \left( \sum_j {\textstyle
      (y\!\cdot\!{\partial g_i(\gamma)\over\partial\gamma_j}
       -g_i(\gamma)\!\cdot\!{\partial f(\gamma)\over\partial\gamma_j} ) }
      (\gamma^\prime_j-\gamma_j)
    \right)
\\
  & = & \left| \mbox{det}_{ij}
    \left(
      y\!\cdot\!\partial_j g_i
      -g_i\!\cdot\! \partial_j f
    \right) \right|^{-1}
   \quad \footnotemark
\nonumber
\end{eqnarray}
\footnotetext{In the following we will omit the absolute bars;
              when necessary they have to be reinstated. }
$$ \mbox{with}\quad y\!\cdot\! g_i(\gamma)=c_i \quad.$$
Inserting $\Phi$ into (\ref{e35}) we get
\beq\label{e38}
  Z=\int\! d^k\!\gamma\,d^n\!y
  \,\delta^k(y\!\cdot\! g_i(\gamma)-c_i)
  \,\mbox{det}_{ij}
    \left(
      y\!\cdot\!\partial_j g_i(\gamma)
      -g_i(\gamma)\!\cdot\!\partial_j f(\gamma)
    \right)
  \,e^{-S[f(\gamma)+y]} \quad.
\eeq
One may write $Z$ in a slightly different form, usually used in QFT,
because it is more suitable for semiclassical approximations.
$Z$ is independent of $c_i$ and therefore, although the r.h.s. of (\ref{e38})
explicitly contains $c_i$, it is actually independent of it.
We can smooth the $\delta$-function by a further multiplication with
\beq\label{e39}
  1=(2\pi\xi)^{-k/2}\int\! d^k\!c\,e^{-{1\over{2\xi}}\sum_i c_i^2}
  \quad.
\eeq
The determinant can be written in the form
\beq\label{e310}
  \mbox{det}A =\int\!d\eta\,d\bar\eta\,e^{\bar\eta A\eta}
\eeq
where $\eta$ are anticommuting Grassmann variables (ghosts).
Inserting (\ref{e39}) and (\ref{e310}) into (\ref{e38}) and performing
the $c$-integration we finally get
\begin{eqnarray}\label{e311}
  Z &=& (2\pi\xi)^{-k/2}\int\!d^k\!\gamma\,d^n\!y\,d\eta\,d\bar\eta
    \,e^{ - S[f(\gamma)+y] -
        S_{gf}[y,\gamma] + S_{FPG}[y,\gamma,\eta,\bar\eta]}
\quad,\nonumber \\
  S_{gf} &=& {1\over 2\xi}y^T \left( \sum_i g_i(\gamma)
    g_i^{_T}(\gamma) \right) y
\quad,\\
  S_{FPG} &=& \sum_{ij} \bar\eta_i
    \left(
      y\!\cdot\!\partial_j g_i(\gamma)
      -g_i(\gamma)\!\cdot\!\partial_j f(\gamma)
    \right) \eta_j
\quad.\nonumber
\end{eqnarray}
For a globally bijective transformation this representation of the
partition function $Z$ is still exact, for a locally bijective
transformation the representation is exact in every order
perturabtion theory, especially in semiclassical approximation. At
this point the (approximately) Gaussian degrees of freedom $y$,
$\eta$ and $\bar\eta$ could be integrated out in semiclassical
approximation. The partition function then reduces to an integral
over the collective coordinates. In Section \ref{sec23} this step
will be performed directly for the QCD case.

\private{ 2D-Example, problems if f is not bijective,
          exact to all orders pert., semiclassical approx.}

\section{Effective QCD Lagrangian in a Background Field *}\label{sec22}

Now it is time to return to QCD
\beq \label{e312}
  Z=\int\!DA_\mu\,e^{-S_{YM}[A]} \quad,\quad
  S_{YM}[A]=\int\!dx\,{1\over {4g^2}} G_{\mu\nu}^a G^{\mu\nu}_a
  \quad.
\eeq
The background configurations which (approximately) minimize $S_{YM}$
will be denoted by $\bar A_\mu(\gamma)$. A general gauge field
can be written in the form
\beq \label{e313}
  A_\mu = (\bar A_\mu(\gamma)+B_\mu)^\Omega \quad,\quad
  \gamma\in I\!\!R^k
\eeq
where $B_\mu$ are fluctuations around the background and
$A_\mu^\Omega$ is a gauge transformed field
\beq\label{e314}
  A_\mu^\Omega = S A_\mu S^\dagger + iS\partial_\mu S^\dagger
  \quad , \quad S=e^{i\Omega}\in SU(N_c) \quad.
\eeq
As in the finite dimensional case we have to make this representation
unique by introducing extra conditions,
\beq\label{e315}
  D_\mu(\bar A)B_\mu(x)=C(x) \quad,
\eeq
  to fix the gauge of the fluctuating field and
\beq \label{e316}
  \int\!d^4x\,\psi_\mu^i(x;\gamma)B_\mu(x) = c_i \quad,\quad
  1\leq i\leq k
\eeq
to avoid fluctuations in (approximately) zero mode direction.
The derivation of an effective action similar to (\ref{e311})
can now be performed in full analogy to the previous case
with only some notational complication. There is the following
correspondence:

\beq\label{e317}
  \begin{array}{lllllll}
    \mbox{finite example} & : & i   & x & y & \gamma_i           & g_i \\
    {\mbox{QCD}}            & : & i,x & A & B & \gamma_i,\Omega(x) &
\psi_\mu^i
  \end{array} \quad.
\eeq

The Faddeev-Popov unit has the form

\beq\label{e318}
  1=\int\!d^k\!\gamma\,D\Omega\,DB_\mu\,
    \delta\big( D_\mu(\bar A)B_\mu\big )
    \delta^k\big( \int\!\psi_\mu^iB_\mu d^4\!x \big)
    \delta\big( (\bar A_\mu+B_\mu)^\Omega-A_\mu \big)
    \Phi[A_\mu] \quad.
\eeq

The steps to get an expression for $\Phi$ are now:

Add primes to $\gamma,\Omega$ and $B$, insert $A=\bar A+B$,
linearize the last $\delta$-argument around $B_\mu^\prime=B_\mu$,
perform the functional $B_\mu^\prime$ integration and linearize the
remaining $\delta$ arguments around $\gamma^\prime=\gamma$ and
$\Omega^\prime=0$. Omitting the details of this calculation one gets
\cite{Dya}

\beq \label{e319}
  \Phi^{-1}(\bar A(\gamma)+B) =
  \int\!d^k\!\gamma^\prime\,D\Omega^\prime
  \delta^k(X_i)\,\delta(Y) \quad,
\eeq

\beq \label{e320}
  X_i=\int\!d^4\!x\,\sum_j
    \left(
      \psi_\mu^i(\gamma) { \partial\bar A_\mu \over\partial\gamma_j }
      - { \partial\psi^i_\mu(\gamma) \over \partial\gamma_j }B_\mu
    \right) (\gamma_j^\prime-\gamma_j) +
    \psi_\mu^i D_\mu(\bar A+B)\Omega^\prime \quad,
\eeq

\beq \label{e321}
  Y = \sum_j D_\mu(\bar A+B)
        { \partial\bar A_\mu \over\partial\gamma_j }
        (\gamma_j^\prime-\gamma_j) +
      D_\mu(\bar A)D_\mu(\bar A+B)\Omega^\prime \quad.
\eeq

From (\ref{e315}), (\ref{e316}), (\ref{e320}) and (\ref{e321}) one can
read off the form of the partition function $Z$:

\beq \label{e322}
  Z=N(\xi) \int\!d^k\!\gamma\,DB_\mu\,D\eta\,D\bar\eta
    \delta^k\big( \int\!\psi_\mu^iB_\mu d^4\!x \big)
    e^{-S_{QCD}[\bar A,B,\eta,\bar\eta]} \quad,
\eeq

\begin{eqnarray} \label{e3225}
  S_{QCD} &=& S_{YM}[\bar A+B]-S_{gf}[\bar A,B]+
        S_{FPG}[\bar A,B,\eta,\bar\eta]                  \quad,\nonumber  \\
  S_{YM} &=& \int\!d^4\!x
        {1\over 4g^2}G_{\mu\nu}^a G^{\mu\nu}_a(\bar A+B) \quad,\nonumber  \\
  S_{gf}&=& {1\over 2\xi} \int\!d^4\!x\,(D_\mu(\bar A)B_\mu)^2 \quad,     \\
  S_{FPG} &=& \sum_{ij}\bar\eta_i
    \left[ \int\!d^4\!x\,
        \psi_\mu^i(\gamma) { \partial\bar A_\mu \over\partial\gamma_j }
      - { \partial\psi^i_\mu(\gamma) \over \partial\gamma_j }B_\mu
    \right] \eta_j                                           \nonumber  \\
    &+&
    \sum_i\int\!d^4\!x\,\bar\eta_i\psi_\mu^i D_\mu(\bar A+B)\eta(x)
    +
    \int\!d^4\!x\sum_j \bar\eta(x) D_\mu(\bar A+B)
      { \partial\bar A_\mu \over\partial\gamma_j }          \nonumber  \\
    &+&
    \int\!d^4\!x\, \bar\eta(x) D_\mu(\bar A)D_\mu(\bar A+B) \eta(x)
    \quad.\nonumber
\end{eqnarray}
$S_{QCD}$ does not depend on the gauge parameter $\Omega$.
For this reason the $\Omega$ integration can be
absorbed in the normalization factor $N(\xi)$.
$\eta(x)$ are the usual ghost fields originating from the gauge fixing.
For every extra condition (\ref{e316}) one gets an additional ghost variable
$\eta_i$. For $\bar A=0$ and no extra condition ($k=0$) the action
given above just reduces to the usual QCD action including Faddeev-Popov
ghosts in
$R_\xi$ gauge
\beq
  S_{gf} = {1\over 2\xi} \int\!d^4\!x\,(\partial_\mu B_\mu)^2 \quad,\quad
  S_{FPG} = \int\!d^4\!x\, \bar\eta(x)\partial_\mu D_\mu (B) \eta(x)
  \quad.
\eeq

Note that the action (\ref{e322}) is still exact
with the non-harmonic degrees of freedom $\gamma_i$ now separated
from the hopefully more Gaussian ones, $B_\mu$ and $\eta$.

For small coupling $g$ it is now possible to establish Feynman rules
from (\ref{e3225}) in analogy to the case with no background. For this
one has to know the "free" gluon, ghost and quark propagator in a given
background $\bar A$. If $\bar A$ is a non constant
field even this is a very complicated task in contrast to usual
perturbation theory around $\bar A=0$. For a multi-instanton
configuration explicit expressions for the gluon and ghost propagator
are derived in \cite{Bro}.

\private{ In the rest of this chapter we want to develop the
Feynman rules of (\ref{e3225}) and give formal expressions for the
propagators in a multi pseudoparticle background. To a large extend
the rules and formulas are independent of whether
the pseudoparticles are instantons or some other configurations.
}
\section{The Semiclassical Limit *}\label{sec23}

Before developing perturbation theory to all orders it is wise
to study the semiclassical limit where one keeps only terms
up to quadratic order in the fields. In QCD (and many other field theories)
this is equivalent to lowest order perturbation theory,
but around a very nontrivial background!

Up to now we have not specified $\psi^i_\mu$. A natural choice
would be $\psi^i_\mu= \partial\bar A_\mu / \partial\gamma_i$
to fix the fluctuations to be orthogonal to the zero modes.
Somewhat more convenient is to bring $\psi_\mu^i$ in background
gauge:
\beq
  \psi^i_\mu=\left( {\partial\bar A_\mu \over \partial\gamma_j} \right)
             ^\Omega
  \quad \mbox{with }\Omega\mbox{ such that} \quad
  D_\mu(\bar A)\psi_\mu^i=0 \quad.
\eeq
Furthermore we assume that $\bar A$ minimizes the gauge action
$S_{YM}$ which is true for widely separated instantons thus
neglecting linear terms $S_{QCD}$.

Up to quadratic order in the fields one has
\begin{eqnarray}
  S_{QCD} &=& S_{YM}[\bar A]
  + \int\!d^4\!x\, {1\over 2g^2}B_\mu K_{\mu\nu}(\bar A) B_\nu
  + \int\!d^4\!x\, \bar\eta(x) D^2(\bar A)\eta(x)          \nonumber  \\
  &+& \sum_{ij} \bar\eta_i
       \psi_\mu^i { \partial\bar A_\mu \over\partial\gamma_j }\eta_j
  + \int\!d^4\!x\sum_j \bar\eta(x)D_\mu(\bar A)
      { \partial\bar A_\mu \over\partial\gamma_j }\eta_j
  + O(\mbox{field}^3) \quad,
\end{eqnarray}
\beq
  K_{\mu\nu} = -D^2\delta_{\mu\nu}+2iG_{\mu\nu}+
     (1-{1\over\xi})D_\mu D_\nu \quad,\quad
  G_{\mu\nu}=F^c G_{\mu\nu}^c  \quad,\quad
  (F^c)_{ab}=if_{acb} \quad.
\eeq
Performing the integration over gauge fields and ghosts one gets
an effective action depending only on the collective coordinates
$\gamma_i$:
\beq
  Z=\int\!d^k\!\gamma\, e^{-S_{eff}[\gamma]}
\eeq
\beq \label{e330}
  e^{-S_{eff}[\gamma]} =
  \mbox{det}_{ij}
  \left(
     \psi_\mu^i(\gamma) { \partial\bar A_\mu \over\partial\gamma_j }
  \right)
  {\mbox{Det}(-D^2(\bar A)) \over
     (\mbox{Det}^\prime K_{\mu\nu}^\prime(\bar A))^{1/2} }
  e^{-S_{YM}[\bar A]}
\eeq
The $\delta$-function in (\ref{e322}) causes a restriction of the gauge
field
fluctuation to be orthogonal to $\psi_\mu^i$. $K_{\mu\nu}^\prime$ is
defined as $K_{\mu\nu}$ projected to the space orthogonal to
$\psi_\mu^i$, $\mbox{Det}^\prime$ takes into account all eigenvalues
of $K_{\mu\nu}^\prime$ except the $k$ zeromodes caused by the projection.

\section{Instantons in QCD}\label{sec24}

The solutions of the classical Yang-Mills equations of motion can
be classified w.r.t.\ their topology $N\in Z\!\!\!Z$ and are
called $N$ instantons\footnote{ To simplify notations we will
treat instantons and anti-instantons on the same footing. Both
will be called instantons and are distinguished by their
topological charge $Q_I=\pm$ if necessary.} solutions \cite{Act}.
The 1-instanton solution has the well known form
\bqa\nonumber
  A_{I\mu}^a(x) &=& O_I^{ab}\eta_{b\mu\nu}^{Q_I}
                  { (x-z_I)_\nu \over (x-z_I)^2 }
                  { 2\rho^2 \over (x-z_I)^2+\rho^2 }
          \quad\mbox{in singular gauge}  \\ \nonumber
  A_{I\mu}^a(x) &=& O_I^{ab}\eta_{b\mu\nu}^{-Q_I}
                  { 2(x-z_I)_\nu \over (x-z_I)^2+\rho^2 }
          \quad\quad\quad\quad\quad\mbox{in regular gauge}
                         \\ \nonumber
\gamma_I=(z_I,O_I,\rho_I,Q_I)&=& {\mbox{(location, orientation, radius,
topological charge)}} \nonumber
\eqa
The instanton parameters $\gamma_I$ reflect the symmetries of the
Lagrangian (translation, rotation, scale invariance, parity).
It was a great achievement of 't Hooft to compute
the functional determinant in the 1-instanton background.
To get a finite results $Z_I$ has to be normalized to the
$\bar A_\mu^a=0$ case, regularized and renormalized.
The final result of the very complicated calculation \cite{tHo} for
the partition function in semiclassical approximation is
\begin{eqnarray}
  \left({Z_I\over Z_0}\right)_{reg} &=&
    \int\!d\gamma_I\,D(\rho_I) \;=\;
    {1\over 2}\sum_{Q_I=\pm 1} \int\!d^4\!z_I dO_I d\rho_I
    \, D(\rho_I) \;=\; V_4\int_0^\infty\nq d\rho\,D(\rho)
    \;=\; V_4 \bar D
  \nonumber\\
  D(\rho) &=& {C_{N_c}\over\rho^5}S_0^{2N_c}e^{-S_1(\rho) }
    \;=\; C_{N_c}\rho^{-5}S_0^{2N_c}(\rho\Lambda)^b
  \nonumber\\
    C_{N_c} &=& {4.6e^{-1.679N_c} \over\pi^2(N_c-1)!(N_c-2)! }
  \nonumber\\
  S_0 &=& {8\pi^2\over g_0^2} \quad,\quad
  S_1(\rho) \;=\; {8\pi^2\over g_1^2(\rho)} \;=\;
    b\ln{1\over\rho\Lambda} \quad,\quad
  \Lambda=\Lambda_{PV}
  \nonumber\\
  S_2(\rho) &=& {8\pi^2\over g_2^2(\rho)} \;=\;
    b\ln{1\over\rho\Lambda} + {b^\prime\over b}\ln\ln{1\over\rho\Lambda}
    + O({1\over\ln{1\over\rho\Lambda}}) \quad,\quad
  \nonumber
\end{eqnarray}
$D(\rho)$ is the density of instantons of radius $\rho$, $g(\rho)$
the running coupling constant in 1/2-loop approximation,
$b={11\over 3}N_c$ and $b^\prime={17\over 3}N_c^2$. $S_0$ is the
classical action of an instanton, and $g_0$ is the bare
(unrenormalized) tree-level coupling constant. One can obtain an
estimate for $g_0$ by replacing $g_0$ with the running coupling
constant at a suitable scale. This unlucky situation can be
avoided by using the 2-loop expression for $D(\rho)$, i.e.\ by
replacing $S_0$ with $S_1$ and $S_1$ with $S_2$. But this
replacement is an improvement only for small coupling. When $\rho$
reaches the QCD scale $\Lambda$ one should confine oneself to the
tree-level or 1-loop approximation in order to obtain useful
results, since the perturbation series is only asymptotically
convergent.

\section{Quarks}\label{sec25}

Additional fields coupled in a gauge invariant way to the gluon field
can simply be incorporated by adding the appropriate Lagrangian
with gauge field $A$ replaced by $\bar A +B$  and performing the
functional integration over the new fields.
So every quark contributes an extra factor
\beq
  \int\!D\Psi D\bar\Psi e^{-\int\!dx\,\bar\Psi(iD\!\!\!\!/+im)\Psi}
  = \mbox{Det}(iD\!\!\!\!/+im)
\eeq
to the partition function $Z$ where
\beq
  iD_\mu=i\partial_\mu + \bar A_\mu + B_\mu
\eeq
is the covariant derivative. In the semiclassical approximation $B_\mu$
can be set to zero. $D(\rho)$ has to be multiplied by the fermionic
factor
\beq
  F(m\rho)=\left\{
  \begin{array}{l@{\quad\mbox{for}\quad}l}
    1.34m\rho(1+m^2\rho^2\ln(m\rho)+\ldots) & m\rho\ll 1 \\
    1-{2\over 75m^2\rho^2}+\ldots           & m\rho\gg 1
  \end{array}\right.
\eeq
and $b$ and $b^\prime$ are now
\beq
  b = {11\over 3}N_c - {2\over 3}N_f \quad,\quad
  b^\prime = {17\over 3}N_c^2 - {13\over 3}N_cN_f
           + {1\over 2}{N_f\over N_c} \quad.
\eeq

\section{The Instanton Liquid Model}\label{sec26}

The sum of well-separated instantons ($Q_I=\pm 1$)
is also an approximate solution of the YM equations:
$$
  A=\sum_{I=1}^N A_I \quad,\quad S[A]\approx NS_0
$$
The partition function of this so called instanton gas is
$$
 Z=\sum_{N=0}^\infty Z_N \quad,\quad Z_N\approx {1\over N!}(V_4\bar D)^N
$$
The sum is dominated by instanton configuration with density
$N/V=\bar D$. Unfortunately $\bar D$ is infinite and the
assumption of a diluted gas turns out to be wrong.
The probability of small size instantons is low
because $D(\rho)$ vanishes rapidly for small distances.
On the other hand for large distances $D(\rho)$ blows up and soon
gets large. This is the origin of the infrared problem
which made a lot of people no longer believing
in instanton physics. Those who were not deterred by that have thought
of the following outcome \cite{Shu}.
For larger and larger distances, the
vacuum gets more and more filled with instantons of increasing size.
At some scale the instanton gas approximation breaks down and one
has to consider the interaction between instantons which might be
repulsive to stabilize the medium. The stabilization might occur
at distances at which a semiclassical treatment is still possible
and at densities at which the various instantons are still well
separated objects - say - not much deformed through their interaction.
So there is a narrow region of allowed values for the instanton radius.
This picture of the vacuum is called the instanton liquid model.
The idea has been confirmed in the course of years by very
different approaches:
{}
\begin{itemize}\parskip=0ex\parsep=0ex\itemsep=0ex
\item Infrarot-Cutoff  \cite{CDG}
\item Hardcore assumption \cite{Ilg}
\item Variational Approach \cite{Dya}
\item Numerical studies \cite{Shu}
\item Phenomenological success \cite{ShV}
\end{itemize}
{}
The simplest suggestion is to introduces a cutoff $\rho_c$
and to ignore large instantons \cite{CDG}:
$$
  \bar D_{\rho_c}=\int_0^{\rho_c}\!\! d\rho\,D(\rho)
$$
The cutoff has to be chosen sufficiently small such that
the space time fraction $f$ filled with instantons is smaller than
1 in order to justify the model of a diluted gas.
$$
  f = {2\over N_c}\int_0^{\rho_c}\!\! d\rho\,{1\over 2}\pi^2\rho^4 D(\rho)
      \quad < \quad 1
$$
This simple cutoff procedure can be improved by introducing a
scale invariant (hardcore) repulsion between instantons, which
effectively suppresses large instantons \cite{Ilg}.
This procedure has the advantage of respecting the scaling Ward
identities which are otherwise violated by the simple cutoff
ansatz
In \cite{Dya} such an repulsion has been found leading to
a phenomenologically welcomed packing fraction.
Unfortunately this repulsion is an artifact of the sum-ansatz as
has been shown by \cite{Ver2}. Therefore the infrared problem is
still unsolved.

Nevertheless it is possible to make successful prediction by
simply assuming a certain instanton density and some average
radius.
It seems that the vacuum can be described  by
effectively independent instantons of size $\rho=600\,\mbox{MeV}^{-1}$
and mean
distance $L_0=200\,$MeV. The integral instanton density is fixed
by the experimentally known gluon condensate \cite{SVZ2}:
\beq
  n = N/V_4 = 1/L_0^4
    = {1\over 32\pi^2}<G_{\mu\nu}^aG^{\mu\nu}_a>
    = (200\,\mbox{MeV})^4_{exp.}
\eeq
In Chapter \ref{Ch3} it is shown that this relation is not
modified by light quarks.
The ratio $L_0/\rho$ is estimated in different works to be
\beq
  (L_0/\rho)_{theor.} = 3.0 \ldots 3.2
\eeq


The \ILM\ is therefore defined by the following assumption on $D(\rho_I)$:
\beq
 D(\rho_I) = n\delta(\rho_I-\rho)   \quad,\quad
 n=(200\;\mbox{MeV})^4              \quad,\quad
 \rho = 600\;\mbox{MeV}^{-1}        \quad.
\eeq
The model describes very successfully the physics of light
hadrons. Extensive numerical studies can be found in
\cite{Shu,ShV}.

In high energy processes with momentum transfer $p$ of $1-10$ GeV,
$D(\rho)$ is often multiplied with function, which is sharply
peaked around $\rho\sim p^{-1}$. The integral over $\rho$ is then
dominated by small instantons and is infrared convergent. The
results are, hence, independent of the infrared cutoff. No
additional assumptions have to be made. A similar phenomenon
allows in Chapter \ref{Ch8} to derive a relation between the quark
condensate and the QCD scale $\Lambda$ without model assumptions.
All other chapters are based on the \ILM.



\chapter{Light Quark Propagator}\label{Ch3}
\unitlength=1mm

In this chapter the average quark propagator in
the multiinstanton background will be calculated. The methods
developed in \cite{Dya} will be extended by taking into account
dynamical quark loops. In the first section the instanton
background is treated as a classical external perturbation, but
the background field is not small (e.g. in the coupling $g$) and
therefore we have to sum up {\it all} Feynman graphs. This is
possible in the case of one quark flavor within the so-called
zeromode approximation. This is possible due to an exact
cancellation of certain graphs and by renormalizing the instanton
density. As a byproduct we answer positively the question whether
it is allowed to identify the gluon condensate with the instanton
density in the presence of dynamical quarks. The quark condensate
and a constituent quark mass are extracted from the quark
propagator in section \ref{sec38}. In the last section it is shown
that the case of two or more quark flavors can be reduced to the
one flavor case in the limit of $N_c\to\infty$. The results
obtained in the one flavor case are therefore still valid when
making this further approximation.

\section{Perturbation Theory in the Multi-instanton Background
*}\label{sec31}

It is well known how to calculate correlators in the presence
of an external classical gauge field at least as perturbation series
in powers of the external field $A^a_\mu(x)$. In the case of QCD
(or more accurately in classical chromodynamics) within the
instanton liquid model the external field
is a sum of well separated scatterers $A=\sum_I A_I$ called
instantons with a fixed radius $\rho$ and distributed randomly and
independently in Euclidian space.

For a while we will restrict ourselves to the  case of one quark flavor
and ignore gluon loops.
The Euclidian Feynman rules have the following form:
\beq
\begin{picture}(57,22)
\put(16,17){\vector(-1,0){8}}
\put(8,17){\line(-1,0){8}}
\put(20,17){\makebox(0,0)[lc]
  {$\displaystyle =\quad {1\over p\!\!/+im}\quad=\quad S_0\quad,$}}
\put(13,2){\vector(-1,0){3}}
\put(10,2){\vector(-1,0){5}}
\put(5,2){\line(-1,0){2}}
\put(8,2){\line(0,1){5}}
\put(13,7){\makebox(0,0)[cc]{$A_I$}}
\put(20,2){\makebox(0,0)[lc]{$ \displaystyle =\quad A\!\!\!/_I\quad.$}}
\put(8,7){\makebox(0,0)[cc]{$\sf x$}}
\end{picture}
\eeq
In operator notation,
\beq
  \langle p|S_0|q\rangle = {1\over p\!\!/+im}\deltabar(p-q)
  \quad,\quad
  \langle x|A\!\!\!/_I|y\rangle = A\!\!\!/_I(x)\delta(x-y)
  \quad,
\eeq
\beqn
  \deltabar^d(\cdots):=(2\pi)^d\delta(\cdots)
  \quad,\quad
  \int\!\dbar^d\!p:=\int\!{d^d\!p\over(2\pi)^d}
  \quad,
\eeqn
graphs are simply alternating chains of $S_0$ and $A_I$.
To average a graph over the instanton parameters one has to
perform the following integration for each instanton:
\beq
  \langle \ldots \rangle_I =
  {N\over V_4}\int d\gamma_I\ldots =
     {N\over 2}\sum_{Q_I=\pm}{1\over V_4}\int d^4\!z_I\int dO_I\ldots
  \quad.
\eeq
For example
\begin{center}
\special{em:linewidth 0.4pt}
\linethickness{0.4pt}
\begin{picture}(60,25)
\put(2,5){\makebox(0,0)[lc]{$\displaystyle\bigg\langle$}}
\put(45,5){\vector(-1,0){5}}
\put(40,5){\vector(-1,0){10}}
\put(30,5){\vector(-1,0){10}}
\put(20,5){\vector(-1,0){10}}
\put(10,5){\line(-1,0){5}}
\put(15,5){\line(3,4){5.33}}
\put(20.33,12){\line(2,-3){4.67}}
\put(35,5){\line(0,1){7}}
\put(25,12){\makebox(0,0)[cc]{$A_I$}}
\put(40,12){\makebox(0,0)[cc]{$A_J$}}
\put(48,5){\makebox(0,0)[lc]{$\displaystyle\bigg\rangle_{I\neq J}=$}}
\put(27,20){\oval(12,10)[]}
\put(27,15){\vector(1,0){1}}
\put(27,25){\vector(-1,0){1}}
\put(35,12){\line(-3,5){3}}
\put(22,17){\line(-2,-5){2}}
\put(20,12){\line(0,0){0}}
\put(20,12){\line(0,0){0}}
\put(20.25,12){\makebox(0,0)[cc]{\sf x}}
\put(35,12){\makebox(0,0)[cc]{\sf x}}
\end{picture}
\end{center}
\beq
  = -{N^2\over V_4^2}\int d\gamma_I d\gamma_J
  \langle p|S_0 A\!\!\!/_I S_0 A\!\!\!/_I S_0 A\!\!\!/_J S_0|q\rangle
  \mbox{Tr}(S_0 A\!\!\!/_J S_0 A\!\!\!/_I) \quad.
\eeq
The origin of the factor $N^2$ is the summation over all
pairs of different instantons $(I,J)$ yielding a factor
$N(N-1)\approx N^2$. The quark loop is the origin of the minus
sign and of the functional trace ''Tr''.

\section{Exact Scattering Amplitude in the one Instanton
Background *}\label{sec32}

Perturbation theory is suitable to study scattering processes.
To achieve chiral symmetry breaking or bound states one
has to sum up infinite series of a subclass of graphs or solve
Schwinger Dyson or Bethe Selpeter equations.

The first thing we can do is to sum up successive scatterings
at one instanton
\begin{center}
  \begin{picture}(96,7.85)
  \put(5,2.85){\vector(-1,0){3}}
  \put(2,2.85){\line(-1,0){2}}
  \put(8,2.85){\circle{6}}
  \put(16,2.85){\vector(-1,0){3}}
  \put(13,2.85){\line(-1,0){2}}
  \put(8,2.85){\makebox(0,0)[cc]{$V_I$}}
  \put(21,2.85){\makebox(0,0)[cc]{$:=$}}
  \put(36,2.85){\vector(-1,0){3}}
  \put(33,2.85){\vector(-1,0){5}}
  \put(28,2.85){\line(-1,0){2}}
  \put(31,2.85){\line(0,1){5}}
  \put(36,7.85){\makebox(0,0)[cc]{$A_I$}}
  \put(41,2.85){\makebox(0,0)[cc]{$+$}}
  \put(61,2.85){\vector(-1,0){3}}
  \put(58,2.85){\vector(-1,0){5}}
  \put(53,2.85){\vector(-1,0){5}}
  \put(48,2.85){\line(-1,0){2}}
  \put(51,2.85){\line(2,5){2}}
  \put(53,7.85){\line(2,-5){2}}
  \put(58,7.85){\makebox(0,0)[cc]{$A_I$}}
  \put(66,2.85){\makebox(0,0)[cc]{$+$}}
  \put(91,2.85){\vector(-1,0){3}}
  \put(88,2.85){\vector(-1,0){5}}
  \put(83,2.85){\vector(-1,0){5}}
  \put(78,2.85){\vector(-1,0){5}}
  \put(73,2.85){\line(-1,0){2}}
  \put(76,2.85){\line(1,1){5}}
  \put(81,7.85){\line(1,-1){5}}
  \put(81,2.85){\line(0,1){5}}
  \put(86,7.85){\makebox(0,0)[cc]{$A_I$}}
  \put(96,2.85){\makebox(0,0)[lc]{$+\ldots\quad,$}}
  \put(31,7.85){\makebox(0,0)[cc]{$\sf x$}}
  \put(53,7.85){\makebox(0,0)[cc]{$\sf x$}}
  \put(81,7.85){\makebox(0,0)[cc]{$\sf x$}}
  \end{picture}
\end{center}
\beq
  V_I := A\!\!\!/_I + A\!\!\!/_I S_0 A\!\!\!/_I + A\!\!\!/_I S_0
  A\!\!\!/_I S_0 A\!\!\!/_I + \ldots
      = S_0^{-1}(S_I-S_0)S_0^{-1}
\eeq
where
\beq
  S_I = (S_0^{-1}-A\!\!\!/_I)^{-1}
\eeq
is the quark propagator in the one instanton background.

\section{Zero Mode Approximation}\label{sec33}

The quark propagator in the one instanton background
can be represented in terms of the eigenvalues $\lambda_i$ and
eigenfunctions $\psi_i$ of the Dirac operator
\beq
  (i\partial\!\!\!/-A\!\!\!/_I)\psi_i=\lambda_i\psi_i \quad\ff\quad
  \langle x|S_I|y\rangle = \sum_i
    { \psi_i(x)\psi_i^\dagger(y) \over \lambda_i+im } \quad.
\eeq
There is one zero eigenvalue $iD\!\!\!/\psi_I=0$
(\look Appendix \ref{appa2})
which makes the propagator singular
in the chiral limit:
\beq
  S_I={|\psi_I\rangle\langle\psi_I| \over im} + S_I^{NZM} \quad.
\eeq
In the so called zero mode approximation one replaces the non
zero mode part by the free propagator
\beq
\makebox(24,6){
\begin{picture}(24,13)
\put(11,5){\circle{6}}
\put(11,5){\makebox(0,0)[cc]{$V_I$}}
\put(22,5){\vector(-1,0){4}}
\put(18,5){\line(-1,0){4}}
\put(8,5){\vector(-1,0){4}}
\put(4,5){\line(-1,0){4}}
\end{picture}
} =
   S_I-S_0 \approx {|\psi_I\rangle\langle\psi_I| \over im} \quad.
\eeq
\private{wrong for $m\neq 0$}

Although this approximation is good for large as well as for small
momenta it may be bad for intermediate ones, but what is more important
is the fact that it is a wild approximation
and so might violate general theorems like Ward identities.
In contrast, all other approximations we make
are of systematic nature respecting all known symmetries of QCD.
\begin{itemize}\parskip=0ex\parsep=0ex\itemsep=0ex
  \item semiclassical approximation (systematic)
  \item multi-instanton background ("systematic")
  \item large $N_c$ expansion (systematic) (see next section)
  \item zero mode approximation (wild)
\end{itemize}
Note that every choice of a background gauge field is "systematic"
in the sense of respecting the symmetries of QCD as long as the
background satisfies these symmetries on average.

In the following sections we will see that the advantage of
the zeromode approximation is so great that we cannot disregard this
simplification.

\private{systematic in leading density but too trivial}

\section{Effective Vertex in the Multi-instanton Background *}\label{sec34}

In Section \ref{sec32} the exact scattering amplitude $V_I$ of the
1-instanton vacuum has been computed. In this Section a formal
expression for the exact scattering amplitude $M_I$ at one instanton
in a multi-instanton bath will be derived.

Let us consider a quark line with two scatterings at $V_I$ and
insert in between a number of instantons which differ from $I$ and
from all other instantons occurring elsewhere in the graph. This
enables us to average over these enclosed instantons independently
from the rest of the graph. Summation over all possible insertions
with at least one instanton just yields the exact quark propagator
minus the free propagator. Remember that direct repeated
scattering at $A_I$ is already included in $V_I$.

Let us define
\beq\label{e544}
\begin{picture}(129,23)
\put(40,20){\vector(-1,0){3}}
\put(37,20){\line(-1,0){2}}
\put(32,20){\circle{6}}
\put(32,20){\makebox(0,0)[cc]{$V_I$}}
\put(29,20){\vector(-1,0){3}}
\put(26,20){\line(-1,0){2}}
\put(20,20){\makebox(0,0)[cc]{$:=$}}
\put(16,20){\vector(-1,0){3}}
\put(13,20){\line(-1,0){2}}
\put(8,20){\circle{6}}
\put(8,20){\makebox(0,0)[cc]{$M_I$}}
\put(5,20){\vector(-1,0){3}}
\put(2,20){\line(-1,0){2}}
\put(44,20){\makebox(0,0)[cc]{$+$}}
\put(56,20){\circle{6}}
\put(56,20){\makebox(0,0)[cc]{$V_I$}}
\put(53,20){\vector(-1,0){3}}
\put(50,20){\line(-1,0){2}}
\put(75,20){\vector(-1,0){3}}
\put(72,20){\line(-1,0){2}}
\put(67,20){\circle{6}}
\put(67,20){\makebox(0,0)[cc]{$V_I$}}
\put(61,21){\makebox(0,0)[cb]{$.$}}
\put(79,20){\makebox(0,0)[cc]{$+$}}
\put(91,20){\circle{6}}
\put(91,20){\makebox(0,0)[cc]{$V_I$}}
\put(88,20){\vector(-1,0){3}}
\put(85,20){\line(-1,0){2}}
\put(102,20){\circle{6}}
\put(102,20){\makebox(0,0)[cc]{$V_I$}}
\put(96,21){\makebox(0,0)[cb]{$.$}}
\put(121,20){\vector(-1,0){3}}
\put(118,20){\line(-1,0){2}}
\put(113,20){\circle{6}}
\put(113,20){\makebox(0,0)[cc]{$V_I$}}
\put(107,21){\makebox(0,0)[cb]{$.$}}
\put(124,20){\makebox(0,0)[lc]{$+\ldots$}}
\put(8,6){\makebox(0,0)[cb]{$.$}}
\put(20,5){\makebox(0,0)[cc]{$:=$}}
\put(44,5){\makebox(0,0)[cc]{$-$}}
\put(64,5){\vector(-1,0){8}}
\put(56,5){\line(-1,0){8}}
\put(68,5){\makebox(0,0)[lc]{$=\quad S-S_0$}}
\thicklines
\put(64,20){\line(-1,0){5}}
\put(61,20){\makebox(0,0)[cc]{$<$}}
\put(99,20){\line(-1,0){5}}
\put(96,20){\makebox(0,0)[cc]{$<$}}
\put(110,20){\line(-1,0){5}}
\put(107,20){\makebox(0,0)[cc]{$<$}}
\put(16,5){\line(-1,0){16}}
\put(8,5){\makebox(0,0)[cc]{$<$}}
\put(40,5){\line(-1,0){16}}
\put(32,5){\makebox(0,0)[cc]{$<$}}
\thinlines
\end{picture}
\eeq
\bqan
  M_I&=& V_I+V_I(S-S_0)V_I+V_I(S-S_0)V_I(S-S_0)V_I+\ldots \\
     &=& V_I+V_I(S-S_0)M_I
\eqan

This equation can be solved for $M_I$ with the following ansatz:

\beq
  M_I={1\over i\mu}S_0^{-1}|\psi_I\rangle\langle\psi_I|S_0^{-1}
\eeq
Inserting $M_I$ and $V_I$ into (\ref{e544}) we get
\beq
  {1\over i\mu}S_0^{-1}|\psi_I\rangle\langle\psi_I|S_0^{-1}
  = {1\over im}(1+{1\over i\mu}
         \langle\psi_I|S_0^{-1}(S-S_0)S_0^{-1}|\psi_I\rangle )
  S_0^{-1}|\psi_I\rangle\langle\psi_I|S_0^{-1}
\eeq
\beq\label{e550}
  \ff\quad
  \mu = m + i\langle\psi_I|S_0^{-1}(S-S_0)S_0^{-1}|\psi_I\rangle
  \quad.
\eeq
\private{$\mu$=eff. mass, $\mu=m-<\psi\psi>$}

The 1-instanton bath causes a replacement of the current mass $m$
with an effective mass $\mu$.

\section{A Nice Cancellation *}\label{sec35}

It is possible to arrange the graphs in
such a way that every $M_I$ occurs only once.
Consider a graph containing two scattering processes $M_I$
at the same instanton. The interesting part of the graph has the following
form
\beq
  \begin{picture}(35,10)(0,10)
  \put(2,5){\makebox(0,0)[cc]{$s$}}
  \put(5,5){\vector(1,0){4}}
  \put(9,5){\line(1,0){4}}
  \put(16,5){\circle{6}}
  \put(16,5){\makebox(0,0)[cc]{$M_I$}}
  \put(19,5){\vector(1,0){4}}
  \put(23,5){\line(1,0){4}}
  \put(16,17){\circle{6}}
  \put(16,17){\makebox(0,0)[cc]{$M_I$}}
  \put(27,17){\vector(-1,0){4}}
  \put(23,17){\line(-1,0){4}}
  \put(13,17){\vector(-1,0){4}}
  \put(9,17){\line(-1,0){4}}
  \put(2,17){\makebox(0,0)[cc]{$p$}}
  \put(30,17){\makebox(0,0)[cc]{$q$}}
  \put(30,5){\makebox(0,0)[cc]{$r$}}
  \put(16,12){\makebox(0,0)[cc]{$\vdots$}}
  \end{picture}
           =\quad
                \bigg[^{\nq\!\!\alpha_p}_{\nq i_p}
                { \psi_I(p)\psi_I^\dagger(q) \over i\mu }\bigg]
                ^{\alpha_q}_{i_q}
            \quad
                \bigg[^{\nq\!\!\alpha_r}_{\nq i_r}
                { \psi_I(r)\psi_I^\dagger(s) \over i\mu }\bigg]
                ^{\alpha_s}_{i_s}
\eeq
$\alpha_p$/$i_p$ are the color/Dirac indices at the quark leg
with momentum $p$.
In a physical process in addition to the graph containing the above subgraph
there
exists another graph with only two quark lines interchanged.
\beq
\makebox(35,0)
{\begin{picture}(36,20)
 \put(2,5){\makebox(0,0)[cc]{$s$}}
 \put(5,5){\vector(1,0){3}}
 \put(8,5){\line(1,0){2}}
 \put(10,5){\line(0,1){3}}
 \put(10,11){\circle{6}}
 \put(10,11){\makebox(0,0)[cc]{$M_I$}}
 \put(10,14){\line(0,1){3}}
 \put(10,17){\vector(-1,0){3}}
 \put(7,17){\line(-1,0){2}}
 \put(2,17){\makebox(0,0)[cc]{$p$}}
 \put(27,17){\vector(-1,0){3}}
 \put(24,17){\line(-1,0){2}}
 \put(22,17){\line(0,-1){3}}
 \put(22,11){\circle{6}}
 \put(22,11){\makebox(0,0)[cc]{$M_I$}}
 \put(22,8){\line(0,-1){3}}
 \put(22,5){\vector(1,0){3}}
 \put(25,5){\line(1,0){2}}
 \put(30,17){\makebox(0,0)[cc]{$q$}}
 \put(30,5){\makebox(0,0)[cc]{$r$}}
 \put(16,11){\makebox(0,0)[cc]{$\cdots$}}
 \end{picture}
 }
        = - \quad
                \bigg[^{\nq\!\!\alpha_p}_{\nq i_p}
                { \psi_I(p)\psi_I^\dagger(s) \over i\mu }\bigg]
                ^{\alpha_s}_{i_s}
            \quad
                \bigg[^{\nq\!\!\alpha_r}_{\nq i_r}
                { \psi_I(r)\psi_I^\dagger(q) \over i\mu }\bigg]
                ^{\alpha_q}_{i_q}
\eeq
As usual, the interchange of two quark lines causes a minus sign in
the amplitude. Inspecting the two expressions, we see that they
coincide except for the sign, thus there exists a complete cancellation
\beq\label{e552}
  \begin{picture}(84,17.85)
  \put(0,2.85){\vector(1,0){4}}
  \put(4,2.85){\line(1,0){4}}
  \put(11,2.85){\circle{6}}
  \put(11,2.85){\makebox(0,0)[cc]{$M_I$}}
  \put(14,2.85){\vector(1,0){4}}
  \put(18,2.85){\line(1,0){4}}
  \put(11,14.85){\circle{6}}
  \put(11,14.85){\makebox(0,0)[cc]{$M_I$}}
  \put(22,14.85){\vector(-1,0){4}}
  \put(18,14.85){\line(-1,0){4}}
  \put(8,14.85){\vector(-1,0){4}}
  \put(4,14.85){\line(-1,0){4}}
  \put(42,2.85){\vector(1,0){3}}
  \put(45,2.85){\line(1,0){2}}
  \put(47,2.85){\line(0,1){3}}
  \put(47,8.85){\circle{6}}
  \put(47,8.85){\makebox(0,0)[cc]{$M_I$}}
  \put(47,11.85){\line(0,1){3}}
  \put(47,14.85){\vector(-1,0){3}}
  \put(44,14.85){\line(-1,0){2}}
  \put(64,14.85){\vector(-1,0){3}}
  \put(61,14.85){\line(-1,0){2}}
  \put(59,14.85){\line(0,-1){3}}
  \put(59,8.85){\circle{6}}
  \put(59,8.85){\makebox(0,0)[cc]{$M_I$}}
  \put(59,5.85){\line(0,-1){3}}
  \put(59,2.85){\vector(1,0){3}}
  \put(62,2.85){\line(1,0){2}}
  \put(32,8.85){\makebox(0,0)[cc]{$+$}}
  \put(74,8.85){\makebox(0,0)[cc]{$=$}}
  \put(84,8.85){\makebox(0,0)[cc]{$0$}}
  \put(53,8.85){\makebox(0,0)[cc]{$\cdots$}}
  \put(11,9.71){\makebox(0,0)[cc]{$\vdots$}}
\end{picture}
\eeq
Whenever an $M_I$ occurs twice or more than twice in a graph
there exists another graph with opposite sign. Both contributions
cancel each other and can be ignored.
So a quark can scatter only once at every
instanton. This can be seen in another way: Because of Fermi statistics
every state can be occupied only once, and there is only one state
for each quark in the zero mode approximation, namely
the zeromode.

There are two equivalent descriptions of Feynman graphs:
\begin{enumerate}
\item Draw all topologically distinct graphs with non-numbered vertices
  and assign a symmetry factor to each graph,
\item Draw all topologically distinct graphs with numbered vertices
  and assign a factor $1/V$, where $V$ is the number of vertices.
\end{enumerate}
If all possible graphs containing $M_I$ are allowed it is not difficult
to see within the second description that they can really be paired
as stated above.

\section{Renormalization of the Instanton-Density *}\label{sec36}

Up to now the cancellation is incomplete because not all graphs are allowed.
Consider e.g.
\beqn
  \begin{picture}(85,30)
  \put(27,9){\circle{6}}
  \put(27,9){\makebox(0,0)[cc]{$M_I$}}
  \put(38,9){\vector(-1,0){4}}
  \put(34,9){\line(-1,0){4}}
  \put(8,9){\vector(-1,0){4}}
  \put(4,9){\line(-1,0){4}}
  \put(11,9){\circle{6}}
  \put(11,9){\makebox(0,0)[cc]{$M_I$}}
  \put(24,9){\vector(-1,0){5}}
  \put(19,9){\line(-1,0){5}}
  \put(19,1){\makebox(0,0)[cc]{\it not allowed}}
  \put(59,9){\circle{6}}
  \put(59,9){\makebox(0,0)[cc]{$M_I$}}
  \put(70,9){\vector(-1,0){4}}
  \put(66,9){\line(-1,0){4}}
  \put(56,9){\vector(-1,0){4}}
  \put(52,9){\line(-1,0){4}}
  \put(43,9){\makebox(0,0)[cc]{$+$}}
  \put(85,9){\makebox(0,0)[rc]{$=\quad 0$}}
  \put(59,16.3){\makebox(0,0)[cc]{$\vdots$}}
  \put(59,21){\circle{6}}
  \put(59,21){\makebox(0,0)[cc]{$M_I$}}
  \put(55,28){\vector(1,0){4}}
  \put(59,28){\line(1,0){4}}
  \put(59,1){\makebox(0,0)[cc]{\it not allowed}}
  \put(55.50,24.5){\oval(10,7)[l]}
  \put(62.50,24.5){\oval(10,7)[r]}
 \end{picture}
\eeqn
As in the case of $V_I$ both graphs are not allowed. Another example is
\beqn
 \begin{picture}(85,30)
 \put(27,9){\circle{6}}
 \put(27,9){\makebox(0,0)[cc]{$M_I$}}
 \put(38,9){\vector(-1,0){4}}
 \put(34,9){\line(-1,0){4}}
 \put(8,9){\vector(-1,0){4}}
 \put(4,9){\line(-1,0){4}}
 \put(11,9){\circle{6}}
 \put(11,9){\makebox(0,0)[cc]{$M_I$}}
 \put(19,1){\makebox(0,0)[cc]{\it not allowed}}
 \put(59,9){\circle{6}}
 \put(59,9){\makebox(0,0)[cc]{$M_I$}}
 \put(70,9){\vector(-1,0){4}}
 \put(66,9){\line(-1,0){4}}
 \put(56,9){\vector(-1,0){4}}
 \put(52,9){\line(-1,0){4}}
 \put(43,9){\makebox(0,0)[cc]{$+$}}
 \put(85,9){\makebox(0,0)[rc]{$=\quad 0$}}
 \put(59,16.3){\makebox(0,0)[cc]{$\vdots$}}
 \put(59,21){\circle{6}}
 \put(59,21){\makebox(0,0)[cc]{$M_I$}}
 \thicklines
 \put(62.50,24.5){\oval(10,7)[r]}
 \put(55.50,24.5){\oval(10,7)[l]}
 \put(55,28){\line(1,0){8}}
 \put(59,28){\makebox(0,0)[cc]{$>$}}
 \put(24,9){\line(-1,0){10}}
 \put(19,9){\makebox(0,0)[cc]{$<$}}
 \thinlines
 \put(59,1){\makebox(0,0)[cc]{\it most general tadpole}}
 \put(19,10){\makebox(0,0)[cc]{$\cdot$}}
 \put(59,29){\makebox(0,0)[cc]{$\cdot$}}
 \end{picture}
\eeqn
It is a general fact that all disallowed graphs can be paired
with other disallowed graphs or with tadpoles and vice versa.

What we have to do is to "disallow" all tadpole graphs.
Every $M_I$ can be surrounded by tadpoles which contribute
with a universal multiplicative factor which can be absorbed in a
redefinition of the instanton density $n_R$. Using this renormalized
density $n_R$ the pairing is now perfect and the statement
"every $M_I$ occurs only once" becomes true.

One further can show that in the presence of dynamical quarks this
renormalized density has to be identified with the gluon
condensate instead of the "bare" density because the same tadpoles
contribute to the gluon condensate too. The often asked question,
whether in the presence of light quarks the gluon condensate can
be identified with the instanton density, can be answered, in the
above sense, with yes!

\private{relation between $n_R = n_G = $ gluon condensate $\neq n/N_c$}

\section{Selfconsistency Equation for the Quark Propagator *}\label{sec37}

Quark loops are no longer possible because they cannot be connected
to another part of the graph via a common instanton. All graphs
which can contribute to the propagator are chains of different $M_I's$.
\begin{center}
 \unitlength=0.9 mm
 \begin{picture}(165,8)
 \put(70,5){\vector(-1,0){3}}
 \put(67,5){\line(-1,0){2}}
 \put(62,5){\circle{6}}
 \put(62,5){\makebox(0,0)[cc]{$M_I$}}
 \put(59,5){\vector(-1,0){3}}
 \put(56,5){\line(-1,0){2}}
 \put(74,5){\makebox(0,0)[cc]{$+$}}
 \put(86,5){\circle{6}}
 \put(86,5){\makebox(0,0)[cc]{$M_I$}}
 \put(83,5){\vector(-1,0){3}}
 \put(80,5){\line(-1,0){2}}
 \put(105,5){\vector(-1,0){3}}
 \put(102,5){\line(-1,0){2}}
 \put(97,5){\circle{6}}
 \put(97,5){\makebox(0,0)[cc]{$M_J$}}
 \put(109,5){\makebox(0,0)[cc]{$+$}}
 \put(121,5){\circle{6}}
 \put(121,5){\makebox(0,0)[cc]{$M_I$}}
 \put(118,5){\vector(-1,0){3}}
 \put(115,5){\line(-1,0){2}}
 \put(132,5){\circle{6}}
 \put(132,5){\makebox(0,0)[cc]{$M_J$}}
 \put(151,5){\vector(-1,0){3}}
 \put(148,5){\line(-1,0){2}}
 \put(143,5){\circle{6}}
 \put(143,5){\makebox(0,0)[cc]{$M_K$}}
 \put(151,5){\makebox(0,0)[lc]{$\Bigg\rangle_{I\neq J\neq\ldots}$}}
 \put(23.5,5){\makebox(0,0)[cc]{$=\,\Bigg\langle$}}
 \put(44,5){\vector(-1,0){8}}
 \put(36,5){\line(-1,0){8}}
 \thicklines
 \put(16,5){\line(-1,0){16}}
 \put(8,5){\makebox(0,0)[cc]{$<$}}
 \thinlines
 \put(140,5){\vector(-1,0){3}}
 \put(137,5){\line(-1,0){2}}
 \put(129,5){\vector(-1,0){3}}
 \put(126,5){\line(-1,0){2}}
 \put(94,5){\vector(-1,0){3}}
 \put(91,5){\line(-1,0){2}}
 \put(49,5){\makebox(0,0)[cc]{$+$}}
 \end{picture}
\unitlength= 1mm
\end{center}
The $M_I's$ can be averaged independently

\beq\label{e553}
  M(p) := i \langle M_I \rangle =
         {n_R \over 2\mu}p^2\varphi^{\prime 2}(p) \quad,
\eeq
where $\varphi^\prime$ is defined in appendix \ref{ChA}.
The resulting expression for the propagator now has the form
\beq\label{e554}
  S = S_0 + S_0{M\over i}S_0 + S_0{M\over i}S_0{M\over i}S_0
      +\ldots
    = (S_0^{-1}+M)^{-1}
\eeq
where $M=M(p)$ is the momentum dependent mass defined above. There is
just one thing to do: We have to solve the circular dependence
\begin{center}
 \begin{picture}(28,15.15)
 \put(1,9.14){\makebox(0,0)[cc]{$n_R$}}
 \put(14,9.14){\makebox(0,0)[cc]{$M$}}
 \put(22,4.14){\makebox(0,0)[cc]{$\mu$}}
 \put(22,14.14){\makebox(0,0)[cc]{$S$}}
 \put(16,10.14){\vector(1,1){4}}
 \put(22,12.14){\vector(0,-1){6}}
 \put(20,4.14){\vector(-1,1){4}}
 \put(13,15.15){\makebox(0,0)[cc]{$\scriptstyle (\ref{e553})$}}
 \put(28,9.14){\makebox(0,0)[cc]{$\scriptstyle (\ref{e550})$}}
 \put(13,2.99){\makebox(0,0)[cc]{$\scriptstyle (\ref{e554})$}}
 \put(5,9.03){\vector(1,0){6}}
 \end{picture}
\end{center}
but $\mu$ is just a number, which makes the solution very simple.
Inserting (\ref{e553}) and (\ref{e554}) into (\ref{e550}) one gets the
following equation for $\mu$
\beq\label{e555}
  \mu = m + \int\dbar^4\!p\,
   {2\varphi^{\prime 2}(p)M_p \over p^2+(m+M_p)^2}
   (p^2+m(m+M_p))
\eeq
\private{$M_p$ \& $\mu$ definition such that not poles in $M_p$}

which may be solved numerically for different current masses $m$.

\section{Some Phenomenological Results}\label{sec38}

In the chiral limit (\ref{e555}) reduces to
\beq\label{e556}
  \mu^2 = n_R\int\dbar^4\!p\,
   {p^4\varphi^{\prime 4}(p) \over p^2+M_p^2} =
   \alpha n_R\rho^2 + O(n_R^2)
\eeq
$\mu^2$ is proportional to $n_R$ and thus $M$ is proportional to
$\sqrt{n_R}$ in contrast to a linear dependence on $n_R$ obtained
from a naive density expansion.

\private{ failure of density expansion $\sqrt{n_R+m}$ }

In the last expression the denominator
has been expanded in the density and
\beq
  \alpha = \rho^{-2}\int\dbar^4\!p\,
   p^2\varphi^{\prime 4}(p) = 6.6
\eeq
is a universal number.
For the standard values of $n_R$ and $\rho$ one gets
\bqa
  \mu_0^2 &=& 6.6n_R\rho^2 = (100\mbox{MeV})^2   \quad,\nonumber\\
  M(p=0)  &=& 7.7\rho\sqrt{n_R} = 300\mbox{MeV}  \quad.
\eqa
The exact solution of (\ref{e556}) which has been obtained numerically by
iteration,
differs from the leading density value by 15\%:
\beq
  M(0) = 345\mbox{MeV} \quad.
\eeq
The momentum dependence of the quark mass is shown in figure \ref{fig51}.
The mass $m+M(p)$ may be interpreted as the mass of a constituent
quark. At high energies it tends to the current mass, at low momentum
chiral symmetry breaking occurs and the quark gets its constituent
mass M(0). Note that this is not a pole mass but a virtual mass at
zero momentum squared.

Let us now take into account a small current mass $m$ formally of
the order $\sqrt{n_R}$. The selfconsistency equation now reads
\beq
  1={m\over\mu}+{\mu_0^2\over\mu^2}+O(n_R) \quad.
\eeq
Solving it for $\mu$ leads to
\beq
  \mu_0 \quad\leq\quad \mu={1\over 2}m+\sqrt{{1\over 4}m^2+\mu_0^2}
        \quad\leq\quad m+\mu_0
\eeq
For the strange quark $\mu$ is increased by a factor of 2:
\beq
  \mu(m_s=150\mbox{MeV}) = 200\mbox{MeV}
\eeq
It is interesting that $m_s+M(0)$ remains to be $300$MeV.
For zero momentum the increase of the current mass is just
compensated by an equal decrease of the dynamical mass $M(0)$.

From the propagator one can obtain the quark condensate
\beq
  \langle \bar\psi\psi\rangle :=
  \lim_{x\to 0}\mbox{tr}_{CD}(S(x)-S_0(x)) =
  N_c\int\!\dbar^4\!p\mbox{tr}_D(S(p)-S_0(p)) \quad.
\eeq
In leading order in the density one gets
\beq
  i\langle \bar\psi\psi\rangle = {n_RN_c\over\mu} =
  \langle G_{\mu\nu}^aG^{\mu\nu}_a\rangle /32\pi^2\mu \quad.
\eeq
This leads to the following condensates for $u$,$d$ and $s$ quarks:
\beq
  i\langle \bar uu\rangle = i\langle \bar dd\rangle =
    (250\mbox{MeV})^3 \quad,\quad
  \langle \bar ss\rangle = 0.5\langle\bar uu\rangle \quad.
\eeq
For heavy quarks there exists a similar relation
\beq
  i\langle \bar\psi\psi\rangle =
  \langle G_{\mu\nu}^aG^{\mu\nu}_a\rangle /48\pi^2m + O(m^{-3})
  \quad,
\eeq
which leads within 10\% to the same value for the strange quark
condensate. This nicely confirms
the hypothesis that the strange quark can be treated as a
light quark as well as a heavy quark. This hypothesis is used in heavy
to light quark matching formulas.

\private{discussion}

\section{Large $N_c$ expansion *}\label{sec39}



Consider now the case of $N_f$ light quark flavors $u,d,s,\ldots$.
The discussion of the one flavor case in the previous sections
can be copied up to the pairing and cancelation of graphs
which contain more than one $M_I$ (\ref{e552}). This is still true in
the multiple flavor case but now both quark lines in (\ref{e552})
must have the same
flavor because $M_I$ always connects quarks of the same flavor.
So we have the theorem: "every $M_I$ occurs only once for each flavor".
From this point on the discussion of the one flavor case breaks
down because there are now graphs contributing to the
propagator containing quark loops. The simplest new contribution
has the form
\begin{center}
 \begin{picture}(38,18)
 \put(27,3){\circle{6}}
 \put(27,3){\makebox(0,0)[cc]{$M_J$}}
 \put(38,3){\vector(-1,0){4}}
 \put(34,3){\line(-1,0){4}}
 \put(8,3){\vector(-1,0){4}}
 \put(4,3){\line(-1,0){4}}
 \put(11,3){\circle{6}}
 \put(11,3){\makebox(0,0)[cc]{$M_I$}}
 \put(24,3){\vector(-1,0){5}}
 \put(19,3){\line(-1,0){5}}
 \put(11,9.86){\makebox(0,0)[cc]{$\vdots$}}
 \put(11,15){\circle{6}}
 \put(11,15){\makebox(0,0)[cc]{$M_I$}}
 \put(27,9.86){\makebox(0,0)[cc]{$\vdots$}}
 \put(27,15){\circle{6}}
 \put(27,15){\makebox(0,0)[cc]{$M_J$}}
 \put(4,5){\makebox(0,0)[cc]{$u$}}
 \put(19,5){\makebox(0,0)[cc]{$u$}}
 \put(34,5){\makebox(0,0)[cc]{$u$}}
 \put(25,17){\vector(-1,0){6}}
 \put(19,17){\line(-1,0){6}}
 \put(25,13){\line(0,0){0}}
 \put(13,13){\vector(1,0){6}}
 \put(19,13){\line(1,0){6}}
 \end{picture}
\end{center}
Is this contribution small in some sense ? Yes it is !
Quark loops are suppressed by a factor $1/N_c$. In perturbative
context this is extensively discussed in \cite{Wit2}, in instanton physics
it was first used by \cite{Dya}.
Although $1/3$ is not a very small number
the large $N_c$ expansion seems to be a good approximation in
various cases.
\private{$N_c\to\infty$ is formal in I-Physics because
         no I in $N_c=\infty$}

Consider a graph and add to it a new quark loop consisting of
\begin{center}
\begin{tabular}{l l}
  $N$ & new instantons        \\
  $S$ & instanton scatterings \\
\end{tabular}$\quad\quad S\geq N \quad.$
\end{center}
This multiplies the graph (see table \ref{tab2}) by a factor
$\mbox{loop} = O(N_c^{1+N-S})$.
The following cases are possible:
\begin{center}
\begin{tabular}{c l}
  $\underline{1+N-S}$ &                                      \\
  $=1$    & $\gdw M=N \gdw$ all instantons are new            \\
          & $\gdw$ the loop is disconnected                   \\
  $=0$    & $\gdw$ one old instanton $\gdw$ the loop is a tadpole \\
  $<0$    & the loop is suppressed by at least one $1/N_c$     \\
\end{tabular}
\end{center}
Disconnected graphs are cancelled by the denominator and tadpoles
have been absorbed in $n_R$. So quark loops are indeed suppressed
in the limit $N_c\to\infty$. The same is true for gluon loops.
This can be seen by the same argument using the $N_c$ dependencies
from table \ref{tab2}.

\begin{table}
  \begin{center}\begin{tabular}{|l|l|} \hline
  Parameters           & $n_R,\rho,N_c,(g_R)$     \\\hline\hline
  Instanton density    & $n = N/V_4 = n_R N_c \approx
                                           (200\mbox{MeV})^4$ \\
  Instanton radius     & $\rho \approx (600\mbox{MeV})^{-1}$  \\
  Number of colors     & $N_c=3$                              \\
  Coupling constant    & $g=g_R/N_c$
                         (not used in semicl. limit)          \\
  \hline
  Gluon condensate     & $\langle GG\rangle =
                          \langle G_{\mu\nu}^a G^{\mu\nu}_a \rangle
                          /32\pi^2 =: n_R N_c $               \\
  Quark condensate     & $\langle\bar\psi\psi\rangle \approx
                          0.39\,N_c\rho^{-1}\sqrt{n_R}
                          \approx (253\mbox{MeV})^3 $         \\
  Constituent mass     & $M(p)\sim\rho\sqrt{n_R}$             \\
  Quark mass           & $M_{quark}(0)\approx 7.7\,\rho n_R^{1/2}\approx
                          300\mbox{MeV}$                      \\
  Gluon mass           & $M_{gluon}(0)\approx 10.9\,\rho n_R^{1/2}\approx
                          420\mbox{MeV}$\cite{Hut1}           \\
  Ghost mass           & $M_{ghost}(0)\approx 7.7\,\rho n_R^{1/2}\approx
                          300\mbox{MeV}$\cite{Hut1}           \\
  Meson correlator     & $ \langle \bar\psi\Gamma\psi(x)\bar\psi
                         \Gamma\psi(0) \rangle_{trunc.} \sim N_c $ \\
  \hline
  Quark loop           & \makebox(20,0) {\unitlength=0.7mm
                         \begin{picture}(22,8)
                         \put(11,5){\circle{6}}
                         \put(12,8){\vector(-1,0){2}}
                         \end{picture}}
                         $\sim N_c$                 \\
  Instanton scattering & \makebox(20,0) {\unitlength=0.7mm
                         \begin{picture}(22,8)
                         \put(11,5){\circle{6}}
                         \put(11,5){\makebox(0,0)[cc]{$I$}}
                         \put(22,5){\line(-1,0){8}}
                         \put(8,5){\line(-1,0){8}}
                         \end{picture}}
                        $\sim N_c^{-1}\rho n_R^{-1/2}$ \\
  Instanton occurrence  &\hspace{20mm} $\sim N_c n_R$    \\
  Gluon loop           & \makebox(20,0) {\unitlength=0.7mm
                         \begin{picture}(22,8)
                         \put(11,5){\circle{6}}
                         \put(12,8){\vector(-1,0){2}}
                         \end{picture}}
                       $\sim N_c^2-1$               \\
  Instanton scattering & \makebox(20,0) {\unitlength=0.7mm
                         \begin{picture}(22,8)
                         \put(11,5){\circle{6}}
                         \put(11,5){\makebox(0,0)[cc]{$I$}}
                         \put(22,5){\line(-1,0){8}}
                         \put(8,5){\line(-1,0){8}}
                         \end{picture}}
                      $\sim (N_c^2-1)^{-1}$             \\
  \hline
\end{tabular}\end{center}
\vspace{-3ex}
\caption{\label{tab2}
  Dependence of various quantities on the parameters of the
  instanton liquid model $n_R,\rho,N_c,(g_R)$.
}\end{table}

\section{Summary}\label{sec310}

The quark propagator has been computed in the \ILM\ in zeromode
approximation and in $1/N_c$ expansion. It was possible to absorb
dynamical quark loops into a renormalized instanton density, which
should be identified with the gluon condensate. In the case of one
quark flavor the $1/N_c$ expansion is exact. The values for the
dynamical quark masses of the up, down and strange quark have been
computed. Dynamical quark loops are suppressed in every order
perturabtion theory in the limit $N_c\to\infty$. For the quark
propagator this fact remains valid beyond perturbation theory. In
the next chapter we will see that in meson correlators there are
non-suppressed quark loops.


\input dise.t4


\chapter{Correlators of Light Mesons}\label{Ch5}

Meson correlators, also called polarization functions, contain
information about the meson spectrum of QCD. Poles at $p^2=m^2$
show the existence of mesons with mass $m$. With the 4 point
functions, calculated in the \ILM\ (\look  Chapter \ref{Ch4}), we
also possess the meson correlators analytically, besides
integration (Section \ref{sec51}). In Section \ref{sec51} we will
discuss them briefly, whereby the pseudoscalar channel is of
special interest. Due to chiral symmetry breaking there are
massless Goldstone bosons in the triplet channel (but not in the
singlet channel), because instantons explicitly break the $U(1)_A$
symmetry. Both phenomenons have analogies in superconductivity. In
order to compute the meson masses we make an ansatz for the meson
spectrum in Section \ref{sec53}. Comparing the theoretical curve
with this approach in Section \ref{sec54} allows to determine the
masses of the lightest mesons in the various channels. The
evaluation of the integrals and the fit will be done numerically.

\section{Analytical Expressions}\label{sec51}

In the last chapter we have calculated various quark 4 point functions
(\ref{e662}). The meson correlators or polarization functions are just
local versions of these vertices and can be obtained by simply
setting $x=y$ and $z=w$. In momentum space the meson correlators
have the form

\beq
  \Pi_{\Gamma\Gamma^\prime}(t)  =
  \Pi^{disc}(t) + \Pi^{conn}(t) = \makebox(17,2)
  {\unitlength=0.5mm
   \begin{picture}(31,19)
   \put(14,1){\framebox(8,18)[cc]{$K$}}
   \thicklines
   \put(22,10){\oval(14,12)[r]}
   \put(14,10){\oval(14,12)[l]}
   \put(11,4){\makebox(0,0)[cc]{$\scriptstyle >$}}
   \put(11,15.14){\makebox(0,0)[cc]{$\scriptstyle <$}}
   \put(25,16){\makebox(0,0)[cc]{$\scriptstyle <$}}
   \put(25,4.86){\makebox(0,0)[cc]{$\scriptstyle >$}}
   \put(30,10){\makebox(0,0)[lc]{$\sim\!\!\!\!\prime$}}
   \put(6,10){\makebox(0,0)[rc]{$\sim\!\!\!\!\prime$}}
   \put(5,14.15){\makebox(0,0)[cc]{$\scriptstyle \Gamma$}}
   \put(31,14.15){\makebox(0,0)[cc]{$\scriptstyle \Gamma^\prime$}}
   \thinlines
   \end{picture}
   } =
\eeq
\beqn
  = \int\!(dpds)\int\!(dqdr)\Pi(p,s,q,r) =
  - \int\!dx\,e^{itx}
    \langle 0|{\cal T}j_\Gamma(x)j_{\Gamma^\prime}(0)|0\rangle \quad,
\eeqn
\beqn
  j_\Gamma(x)^{s/t} = {1\over\sqrt{2}}
    (\bar u\Gamma u(x)-\bar d\Gamma d(x)) \quad,\quad
  j_{\Gamma^\prime}(0)^{s/t} = {1\over\sqrt{2}}
    (\bar u\Gamma^\prime u(0)-\bar d\Gamma^\prime d(0)) \quad,
\eeqn
\beqn
  \int\!(dpds) = \int\!\dbar p\dbar s\,\deltabar(p-s-t) \quad,\quad
  \int\!(dqdr) = \int\!\dbar q\dbar r\,\deltabar(q-r-t) \quad,\quad
\eeqn
\beqn
  t=p-s=q-r \quad.
\eeqn

From the explicit expressions of the 4 point functions obtained in the
last chapter
one can get, up to integration, analytical expressions for
the meson correlators. The following list is a complete
summary of all formulas needed to evaluate the meson correlators:
\bqa
  \Pi_{\Gamma\Gamma^\prime}^{disc}(t) &=&
  N_c\int\!(dpds)\,\mbox{tr}_D(\Gamma S(p)\Gamma^\prime S(s))
  \quad,\nonumber\\
  \Pi_{\Gamma\Gamma^\prime}^{conn}(t) &=&
  -N_c(C_0(t)\Gamma_\Gamma^0(t)\Gamma_{\Gamma^\prime}^0(t) +
       C_5(t)\Gamma_\Gamma^5(t)\Gamma_{\Gamma^\prime}^5(t))
  \quad,\label{e705}\\
  C_{0/5}(t) &=& - {1\over F_{0/5}(t)\pm 1} \quad
    {+\; \mbox{for singlet correlator} \atop
     -\; \mbox{for triplet correlator} }
  \quad,\nonumber\\
  \Gamma_\Gamma^0(t) &=& {1\over\sqrt{n_R}}
    \int\!(dpds)\,\sqrt{M_pM_s}\mbox{tr}_D(S(p)\Gamma S(s))
  \quad,\nonumber\\
  \Gamma_\Gamma^5(t) &=& {1\over\sqrt{n_R}}
    \int\!(dpds)\,\sqrt{M_pM_s}\mbox{tr}_D(S(p)\Gamma S(s)\gamma_5)
  \quad,\nonumber\\
  F_0(t) &=& {-1\over n_R}
    \int\!(dpds)\,M_pM_s\mbox{tr}_D(S(p)S(s))
  \quad,\nonumber\\
  F_5(t) &=& {-1\over n_R}
    \int\!(dpds)\,M_pM_s\mbox{tr}_D(S(p)\gamma_5 S(s)\gamma_5)
  \quad,\nonumber\\
  S(p) &=& {1\over p\!\!/+i(m+M_p)} \quad,\quad
    M_p ={n_R\over 2\mu}p^2\varphi^{\prime 2}(p) \quad,
  \quad,\nonumber\\
  p\varphi^\prime(p) &=& 2\pi\rho z {\partial\over\partial z}
    [I_0(z)K_0(z)-I_1(z)K_1(z)]_{z=p\rho/2}
  \quad,\nonumber\\
  \mu &=& m + \int\dbar^4\!p\,
     {2\varphi^{\prime 2}(p)M_p \over p^2+(m+M_p)^2}
     (p^2+m(m+M_p))
  \quad,\label{e710}\\
  n_RN_c &=& (200\mbox{MeV})^4 \quad,\quad
    \rho = (600\mbox{MeV})^{-1}
  \quad.\nonumber
\eqa

\section{Analytical Results}\label{sec52}

Performing the Dirac traces leads to the following expressions:
\bqa
  F_{0/5}(t) &=& -{4\over n_R}\int\!(dpds)
    {M_pM_s(\pm (ps)-\tilde M_p\tilde M_s) \over
     (p^2+\tilde M_p^2)(s^2+\tilde M_s^2)}
  \quad,\nonumber\\
  \Gamma_{1/5}^{0/5}(t) &=& {4\over\sqrt{n_R}}\int\!(dpds)
    {\sqrt{M_pM_s}(\pm (ps)-\tilde M_p\tilde M_s) \over
     (p^2+\tilde M_p^2)(s^2+\tilde M_s^2)}
  \quad,\\
  \Gamma_{\mu 5}^5(t) &=& {4i\over\sqrt{n_R}}\int\!(dpds)
    {\sqrt{M_pM_s}(\tilde M_p s_\mu - \tilde M_s p_\mu) \over
     (p^2+\tilde M_p^2)(s^2+\tilde M_s^2)}
  \quad,\nonumber\\
  \tilde M_p &=& m + M_p
  \nonumber
\eqa
All other vertices $\Gamma$ are zero.
Consider the one instanton vertex (\ref{e670}) (the kernel).
It contributes
only to the scalar and pseudoscalar correlator. From this observation
one may have predicted that the connected part of all other channels is
small because a contribution has to be a multi-instanton effect.
Indeed, they are all zero as seen above except for the axial correlator.
Due to an extra factor $M\sim\sqrt{n_R}$ in the numerator
of $\Gamma_{\mu5}^5$ the connected part of the axial correlator is
suppressed by $O(n_R)$ therefore it is small as expected and will
be neglected in the following.
\private{ In addition it is purely longitudinal. }

Furthermore we will restrict
ourself to the {\it chiral limit}, taking $m=0$.
Using the selfconsistency equation (\ref{e710}) one can see that
\beq
  F_5(t=0)=1
\eeq
is leading to a pole at $t=0$ in the pseudoscalar triplet correlator
due to the $F_5(t)-1$ denominator in (\ref{e705}).
This is the massless Goldstone pion one expects in the chiral limit.
A more extensive discussion can be found in \cite{Dya}.
On the other side in the singlet correlator the minus sign is
replaced by a positive sign and there is no Goldstone boson in this case.
Thus, the (two flavor) $\eta^\prime$ meson is massive ! Unfortunately we
cannot make any reliable prediction of the $\eta^\prime$ mass because
the kernal is very repulsive in this channel and no boundstate is
formed. There have to be other attractive forces, e.g.\ confinement
forces, to built an $\eta^\prime$ boundstate. Similar things
happen in the scalar triplet channel (compare figure \ref{fig73}
and \ref{fig76}). But the most important thing is
that there is no massless pseudoscalar singlet meson
which is an important step towards discussing the $U(1)_A$-problem.

Both phenomenons have their analogon in the BCS-theory of
superconductors, namely massless excitons and massive plasma
oscillations. The phenomenons as well as the structure of the
calculation can be nearly 1:1 mapped onto, of course with
differences in details. The 't Hooft vertex corresponds to the
phonon-electron interaction. Further parallel can be read from
table \ref{supra}. An introduction to the BCS-theory can be found
in \cite{Schrieffer}.

\unitlength=1mm
\begin{table}
  \begin{center}\begin{tabular}{|l|c|c|} \hline
                                &
    {\bf Instantons in QCD}            &
    {\bf BCS-Theory}
  \\\hline\hline
    Ph"nomen                                        &
    Chiral Symm.-breaking                         &
    Superconductivity
  \\\hline
    Boundstates \dots                          &
    $\bar qq$-pairs=$\pi$-mesons                   &
    $e^-e^-$-pairs=Cooper-pairs
  \\\hline
    bounded by attractive \dots                     &
    instanton-induced                            &
    phonon-electron-
  \\
                                                    &
    't Hooft interaction                        &
    interaction
  \\
                                                    &
  \makebox(22,18){
  \begin{picture}(22,18)
    \put(11,3){\circle{6}}
    \put(11,3){\makebox(0,0)[cc]{$I$}}
    \put(11,15){\circle{6}}
    \put(11,15){\makebox(0,0)[cc]{$I$}}
    \put(11,9.94){\makebox(0,0)[cc]{$\vdots$}}
    \put(8,15){\line(-1,0){8}}
    \put(4,15){\makebox(0,0)[cc]{$<$}}
    \put(0,3){\line(1,0){8}}
    \put(4,3){\makebox(0,0)[cc]{$>$}}
    \put(22,15){\line(-1,0){8}}
    \put(18,15){\makebox(0,0)[cc]{$<$}}
    \put(14,3){\line(1,0){8}}
    \put(18,3){\makebox(0,0)[cc]{$>$}}
  \end{picture}}
                                &
  \makebox(22,14){
  \begin{picture}(22,14)
    \put(22,2){\line(-1,0){22}}
    \put(0,14){\line(1,0){22}}
    \put(11,11.5){\makebox(0,0)[cc]{$\vdots$}}
    \put(11,6.5){\makebox(0,0)[cc]{$\vdots$}}
    \put(5,14){\makebox(0,0)[cc]{$<$}}
    \put(17,14){\makebox(0,0)[cc]{$<$}}
    \put(5,2){\makebox(0,0)[cc]{$>$}}
    \put(17,2){\makebox(0,0)[cc]{$>$}}
  \end{picture}}
  \\\hline
    in the \dots                        &
    pseudo-scalar channel                           &
    scalar channel
  \\\hline
    Orderparameter                               &
    Quark condensate                                 &
    Density of superconducting
  \\
                                            &
    $\langle\bar\psi\psi\rangle$                    &
    electrons (Cooper-pairs) $\langle\rho\rangle$
  \\\hline
    Spontaneous breaking of               &
    chiral symmetry                     &
    gauge symmetry
  \\\hline
    leads to \dots                                  &
    Goldstone bosons =                             &
    Excitons =
  \\
                                &
    massless pions                    &
    massless density fluct.
  \\\hline
    Vacuum polarization                &
                                                    &

  \\
    diagrams make the                       &
  \makebox(26,0){
  \begin{picture}(26,13.5)
    \put(8,13.5){\line(0,-1){3}}
    \put(8,7.5){\circle{6}}
    \put(8,7.5){\makebox(0,0)[cc]{$I$}}
    \put(8,4.5){\line(0,-1){3}}
    \put(2,7.5){\makebox(0,0)[cc]{$\cdots$}}
    \put(8,13.5){\line(1,0){12}}
    \put(14,13.5){\makebox(0,0)[cc]{$<$}}
    \put(8,1.5){\line(1,0){12}}
    \put(14,1.5){\makebox(0,0)[cc]{$>$}}
    \put(20,1.5){\line(0,1){3}}
    \put(20,7.5){\circle{6}}
    \put(20,7.5){\makebox(0,0)[cc]{$J$}}
    \put(20,10.5){\line(0,1){3}}
    \put(26,7.5){\makebox(0,0)[cc]{$\cdots$}}
  \end{picture}}
                            &
  \makebox(26,0){
  \begin{picture}(24,14)
    \put(15,8){\circle{12}}
    \put(6,8){\makebox(0,0)[cc]{$\cdots$}}
    \put(24,8){\makebox(0,0)[cc]{$\cdots$}}
    \put(15,14.5){\makebox(0,0)[cc]{$<$}}
    \put(15,2){\makebox(0,0)[cc]{$>$}}
  \end{picture}}
  \\
    massless excitations                          &
                            &

  \\
    massive                          &
    in the flavor singlet case                     &
    for the Coulomb interaction
  \\
                                &
    $\ff$ massive $\eta'$                          &
    $\ff$ massive plasma osz.
  \\\hline
  \end{tabular}\end{center}
  \vspace{-3ex}
  \caption[Chiral Symmetry Breaking and Superconductivity]{\label{supra}
      \it Analogy between chiral symmetry breaking in QCD and the BCS theory
      of superconductors \cite{Schrieffer}}
\end{table}

It is interesting to see that in leading order in the instanton
density $F_0(t)=-F_5(t)$, which leads to a massless pole in
the scalar singlet correlator. Numerically the $\sigma$-meson
indeed turns out to be very light. Numerically a light scalar singlet
meson cannot be excluded. 

\section{Spectral Representation}\label{sec53}

To extract phenomenological information from the meson correlators
we make use of the spectral representation
\beq\label{e725}
  \Pi(p)= \int\!dx\,e^{ipx}
    \langle 0|{\cal T}j_\Gamma(x)j_{\Gamma^\prime}(0)|0\rangle =
    \int_0^\infty\!d\sigma^2\,D(\sigma,x)\rho(\sigma^2)
\eeq
where
\beq
  \rho(p^2) = (2\pi)^3\sum_n\delta(p-q_n)
    \langle 0|j_\Gamma(0)|n\rangle
    \langle n|j_\Gamma^\prime(0)|0\rangle
\eeq
is the spectral density and
\beq
  D(m,x) = \int\!\dbar^4 p{e^{-ipx}\over p^2+m^2}
         = {1\over 4\pi^2 x^2}(mx)K_1(mx)
\eeq
is the free propagator of mass $m$ in coordinate representation.
We have chosen the coordinate representation of $\Pi$ to be
able to compare the plots directly with lattice calculations \private{[??]}
and with numerical studies of the instanton liquid \cite{ShV}.

The spectrum consists of mesonic resonances and the continuum
contributions. If one is only interested in the properties
of the first resonance one might approximate the rest of the
spectrum by the perturbatively calculated continuum.

One might think that the disconnected part only contributes
to the continuum and the connected part will yield the boundstates.
But this is not the case. On one hand, Bethe Salpeter equations
have bound as well as continuum solutions. On the other hand
consider a theory with weak attraction between particles of mass m.
It is clear that there is only a cut above $2m$ and no
boundstate pole in the free loop. However in the exact polarization
function only a small portion of the continuum will be used
to form a pole just below the threshold because the attraction
is only weak. The Euclidian correlator will hardly be changed.
Therefore, assuming weak attraction, we can already estimate
the boundstate mass from the disconnected part. Of course in this
example we need not calculate anything because we know
that the bounstate mass is approximately $2m$ with errors of the
order of the strength of the interaction.

Assuming that all other forces neglected in QCD so far,
especially perturbative corrections, are small and attractive
in the vector and axialvector channel, we can obtain boundstate
masses although in these channels up to our approximation
there is no connected part. But things are less trivial than in
the example above because the quarks do not posses a definite mass
and we have to inspect the correlator to extract the meson masses.

Let us start with the scalar and pseudoscalar correlator.
The lowest resonance of mass $m_*$ is coupled to the current
with strength
\beq
  \lambda_*=\langle 0|j_{1/5}(0)|p\rangle \quad.
\eeq
The rest of the spectrum is approximated by the continuum
starting at the threshold $E_*$.
* means $\pi$, $\eta$, $\delta$ or $\sigma$ (see table \ref{tab3}).
{}
\begin{table}
  \begin{center}\begin{tabular}{|lcc|cc|} \hline
  Correlator & & $\Gamma=\Gamma^\prime$ & I=1 & I=0 \\
  \hline
  Pseudoscalar & $\Pi_5=\langle j_5j_5\rangle$
    & $i\gamma_5$         & $\pi$    & $\eta^\prime$              \\
  Scalar       & $\Pi_1=\langle j_1j_1\rangle$
    & $1\!\!1_D$          & $\delta$ & $\sigma$                   \\
  Vector       & $\Pi_ {\mu\mu}=\langle j_\mu j_\mu\rangle$
    & $\gamma_\mu$        & $\rho$   & $\omega$                   \\
  Axialvector  & $\Pi_{\mu\mu}^5=\langle j_\mu^5j_\mu^5\rangle$
    & $\gamma_\mu\gamma_5$ & $a_1$   & $f_1$                \\\hline
\end{tabular}\end{center}
\vspace{-3ex}
\caption{\label{tab3}
  Mesonic correlators
}\end{table}
{}
$E_*$ is typically of the order
$1.5$ GeV and therefore the continuum can be calculated perturbatively.
The spectrum thus has the form
\beq
  \rho_{1/5}(s) = \lambda^2_*\delta(s-m_*^2)+
    {3s\over 8\pi^2}\Theta(s-E_*^2) \quad.
\eeq
Inserting $\rho$ into (\ref{e725}) one gets
\beq
  \Pi^{fit}_{1/5}(x) = \lambda_*^2 D(m_*,x) + E(E_*,x) \quad,
\eeq
\beq
  E(E_*,x)= {3\over\pi^4x^6} {(E_*x)^3\over 16} (2K_3(E_*x)+(E_*x)K_2(E_*x))
  \stackrel{x\to 0}\longrightarrow {3\over\pi^4x^6} \quad.
\eeq
In the next section $m_*$, $\lambda_*$ and $E_*$ are obtained by
fitting the
phenomenological ansatz $\Pi^{fit}(x)$ to the theoretical curve
$\Pi^{sum}(x)$ in the Euclidian region where the theoretical
calculation is reliable.

Consider now the vector and axial vector correlator. The vector
current is conserved, thus the correlator is transverse and only
the vector meson can contribute. In the chiral limit the same
holds true for the axial current. In the singlet channel one
has to be careful because there are two currents.
A conserved one and a gauge invariant one which contains an anomaly.
Up to now we have only calculated the correlator of the conserved
current. Nevertheless to leading order in the instanton density
the two correlators coincide and should be both conserved.
\private{
  Current conservation and Ward identities will be discussed in the
  next chapter.}

For conserved vector and axial currents the spectral function is transverse:
\beq
  \rho_{\mu\nu}^{(5)}(p^2) =
    (-\delta_{\mu\nu}+{p_\mu p_\nu \over p^2}) \rho_T^{(5)}(p^2) \quad.
\eeq
The coupling of the vector and axial meson to the current is given by
\beq
  i\lambda_*\epsilon_\mu = \langle 0|j_\mu^{(5)}(0)|p\rangle
\eeq
where $\epsilon_\mu$ is the meson polarization.
The spectral and polarization functions have the form
\bqa
  -\rho_{\mu\mu}^{(5)}(s) &=& 3\lambda^2_*\delta(s-m_*^2)+
    {3s\over 4\pi^2}\Theta(s-E_*^2) \quad,\\
  -\Pi^{fit(5)}_{\mu\mu}(x) &=& 3\lambda_*^2 D(m_*,x) + 2E(x)
  \quad.\nonumber
\eqa
Here * means $\rho$, $\omega$, $a_1$ or $f_1$ (see table \ref{tab3}).

\section{Plot \& Fit of Meson Correlators}\label{sec54}

The meson correlators are shown in figure \ref{fig71} - \ref{fig76},
normalized to the correlators in lowest order perturbation theory:
\beq
  \Pi^0_{1/5}(x) = {3\over\pi^4 x^6} \quad,\quad
  \Pi^{0(5)}_{\mu\mu} = - {6\over\pi^4 x^6} \quad.
\eeq
The diagrams therefore show the deviation from the
perturbative behaviour. The numerical evaluation of the integrals
are discussed in Appendices \ref{appa4} and \ref{appa5}. The meson
parameters obtained by fitting the parameter ansatz to the
theoretical curve are summarized in table \ref{tab4}. Shuryak \&
Verbarshot \cite{ShV} have obtained the same parameters from a
numerical investigation of the instanton liquid model. Their
values are also shown in table \ref{tab4}. The parameters of the
vector channel coincide extremely well with ours. For the $\pi$
and $\sigma$ meson there is a large discrepancy in the masses but
this is not surprising: We are working in the chiral limit thus
the pion mass has to be zero. A similar argument holds for the
$\sigma$ meson as discussed above. The couplings fit very well.

The discrepancy in the axial channels can have various origins
which are under investigation. Alternatively one may directly
compare the graphs. They coincide very well even in cases where a
spectral fit does not work very well like in the $\delta$ and
$\eta^\prime$ channel. The conclusion is that the terms neglected
in our analytical treatment, but included in the numerical study
\cite{ShV}, are small and usually give an correction less than
10\%. These are contributions from nonzero modes and higher order
corrections in $1/N_c$. This is again an example for the
surprisingly high accuracy of the $1/N_c$ expansion. In the case
of strange quarks the nonzero mode contributions will become more
important.

Finally, one should compare the numbers with
experiment. As far as known, these numbers are also listed in table
\ref{tab4}.
A general discussion of the meson correlators and comparison with
experimental results can be found in \cite{ShV}.
\begin{table}
  \begin{center}\begin{tabular}{|cl|lllc|} \hline
  Meson & $I^G(J^{PC})$ & $m_*$[MeV] & $\sqrt{\lambda_*}$[MeV]
                                                    & $E_*$[MeV] & source
  \\\hline\hline
        &         &  0         & 508$\pm$1  & 1276$\pm$33 & $1/N_c$      \\
  $\pi$ & $1^-(0^{-+})$ & 142$\pm$14 & 510$\pm$20 & 1360$\pm$100& simulation
\\
        &         & 138        & 480        & ---         & experiment\\
  \hline
        &         & $\neq$ 0   & ?          & ?           & $1/N_c$      \\
  $\eta^\prime$
        & $0^+(0^{-+})$ & $\neq$ 0   & ?          & ?           & simulation
\\
        &         & 960        & ?          & ---         & experiment\\
  \hline
        &         & $\neq 0$   & ?          & ?           & $1/N_c$      \\
$\delta$& $1^-(0^{++})$ & $\neq 0$   & ?          & ?           & simulation
\\
        &         & 970        & ?          & ---         & experiment\\
  \hline
        &         & 433$\pm$3  & 506$\pm$3  & 1446$\pm$20 & $1/N_c$      \\
$\sigma$& $0^+(0^{++})$ & 543        & 500        & 1160        & simulation
\\
        &         & ?          & ?          & ---         & experiment\\
  \hline
        &         & 930$\pm$5  & 408$\pm$4  & 1455$\pm$33 & $1/N_c$      \\
  $\rho$& $1^+(1^{--})$ & 950$\pm$100& 390$\pm$20 & 1500$\pm$100& simulation
\\
        &         & 780        & 409$\pm$5  & ---         & experiment\\
  \hline
        &         & 930$\pm$5  & 408$\pm$4  & 1455$\pm$33 & $1/N_c$      \\
$\omega$& $0^-(1^{--})$ & ?          & ?          & ?           & simulation
\\
        &         & 780        & 390$\pm$5  & ---         & experiment\\
  \hline
        &         &1350$\pm$200& 370$\pm$30 & 1050$\pm$80 & $1/N_c$ \\
  $a_1$ & $1^-(1^{++})$ & 1132$\pm$50& 305$\pm$20 & 1100$\pm$50 & simulation
\\
        &         & 1260       & 400        & ---         & experiment\\
  \hline
        &         &1350$\pm$200& 370$\pm$30 & 1050$\pm$80 & $1/N_c$ \\
  $f_1$ & $0^+(1^{++})$ & 1210$\pm$50& 293$\pm$20 & 1200$\pm$50 & simulation
\\
        &         & 1285       & ?          & ---         & experiment\\
  \hline
\end{tabular}\end{center}
\vspace{-3ex}
\caption{\label{tab4}
  Meson mass $m_*$, coupling constant $\lambda_*$ and continuum threshold
  $E_*$ obtained within the instanton liquid model in this work
  ($1/N_c$ expansion), from numerical simulation and from
  experiment.
}\end{table}



\chapter{The Axial Anomaly}\label{Ch6}
\unitlength=1.00mm

A variety of predictions concerning chiral symmetry breaking
can be made within the instanton liquid model. Although the
't Hooft interaction \cite{tHo} explicitly breaks the $U(1)$
axial symmetry, instanton models are up to now not too successful
in describing quantitatively the axial singlet channel.
The most interesting quantities are the $\eta'$ mass
and the spin of the proton.

In Section \ref{sec61} we compute $m_{\eta'}$
by combining various techniques. The result does not depend
on the specifics of instanton physics.

Sections \ref{sec62}, \ref{sec63} and \ref{sec64} are an introduction
to the proton spin problem. In section \ref{sec65} the
proton form factors are reduced to vacuum correlators of 4 quark
fields by assuming independent constituent quarks.
The axial singlet quark and gluonic form factors are calculated
in section \ref{sec66}, \ref{sec67} and \ref{sec68} by using
the propagator and 4 point functions of the instanton liquid model.
Gauge(in)dependence is examined. A discussion
of the results and a comparison with \cite{For} is given in
section \ref{sec69}.

In Sections \ref{sec62}-\ref{sec65} we exceptionally use
Minkowski-Notation.

\section{The Mass of the $\eta'$ Meson}\label{sec61}

In many channels a direct calculation of the meson correlators
in the instanton liquid model and a spectral fit lead to
reasonable results for the masses of the lightest mesons \cite{ShV}.
This method even works in the axial triplet
channel because the model correctly describes spontaneous breaking
of chiral symmetry. In the axial
singlet channel a strong repulsion prevents the formation
of a meson \cite{Hut2,Geshkenbein}.
The conclusion is, that there is no massless Goldstone
boson in this channel, but the mass of the $\eta'$ remains undetermined.
In this letter I want to calculate the mass of the $\eta'$
by combining quite different techniques.
With the help of
\begin{itemize}\parskip=0ex\parsep=0ex\itemsep=0ex
\item current algebra theorems for the $\eta'$,
\item $1/N_c$ expansion,
\item instanton model,
\item scale anomaly
\end{itemize}
we are able to relate the $\eta'$ mass to the pion coupling constant
$f_\pi$ and the physical gluon condensate.

In leading order in $1/N_c$ it is possible to relate the $\eta'$ mass
to the $\Theta$ dependence of the topological susceptibility\footnote{
  $\langle AB\rangle_{conn} = \langle AB\rangle -
   \langle A\rangle\langle B\rangle$}
$d^2E/d\Theta^2$ of QCD without quarks \cite{Wit}:
\beq\label{pro55}
  m_{\eta'}^2 = {4N_f\over f_\pi^2}
    \left({d^2E\over d\Theta^2}\right)_{\Theta=0}^{no\;quarks}
  \quad,\quad
  {d^2E\over d\Theta^2} = \int\!d^4\!x\;
    \langle 0|{\cal T}Q(x)Q(0)|0\rangle_{conn}
\eeq
$$
  Q(x) =  {\alpha_s\over 4\pi}\mbox{tr}_c G\tilde G(x) \quad,\quad
  Q = \int\!d^4\!x\,Q(x) \in Z\!\!\!Z
$$
$Q(x)$ is the topological charge density and $Q$ the total charge.
This formula is derived by arguing, that for large
$N_c$ the topological susceptibility is dominated by the $\eta'$
state, utilizing the axial anomaly \cite{ABJ}
and the relation $f_\pi=f_{\eta'}$,
which is exact for $N_c\to\infty$.

The next step is to relate the topological susceptibility to the
gluon condensate $\langle 0|N(0)|0\rangle$:
$$
  N(x) =  {\alpha_s\over 4\pi}\mbox{tr}_c GG(x) \quad,\quad
  N = \int\!d^4\!x\,N(x)
$$
In instanton models the gluon field consists of
instantons of charge $Q=\pm 1$. The
exact N instanton solutions $(Q=N)$
are selfdual ($G_{\mu\nu}\!=\!\tilde G_{\mu\nu}$) and
\beq\label{pro80}
    \langle 0|{\cal T}Q(x)Q(0)|0\rangle_{conn} =
    \langle 0|{\cal T}N(x)N(0)|0\rangle_{conn} \quad.
\eeq
The exact anti-instanton solutions $(Q=-N)$
are anti-selfdual ($G_{\mu\nu}\!=\!-\tilde G_{\mu\nu}$) and
(\ref{pro80}) holds too, because the two minus signs cancel.
Unfortunately these exact solutions are not the most
important contributions to the partition function.

The dominating configurations are instantons and anti-instantons
in mixed combination. The simplest model is a dilute sum
$A=\sum_I A_I$ of instantons of mixed charge.
$G_{\mu\nu}$
is then approximately selfdual near the instanton centers,
approximately anti-selfdual near the anti-instanton centers and
small far away from any instanton.
In leading order in the instanton density we have
\beq
   G_{\mu\nu}(x)=\pm\tilde G_{\mu\nu}(x)
\eeq
where the sign now depends on $x$!
Let us define $N^\pm(x)$ in the following way:
$$
  N(x)=N_+(x)+N_-(x) \quad,\quad Q(x)=N_+(x)-N_-(x)
$$
$N_+(x)$ is a sum of bumps near the centers of instantons and
$N_-(x)$ near the centers of anti-instantons.
(\ref{pro80}) has to be replaced by the relation
\beq\label{pro81}
    \langle 0|{\cal T}Q(x)Q(0)|0\rangle_{conn} =
    \langle 0|{\cal T}N(x)N(0)|0\rangle_{conn} -
    4\langle 0|{\cal T}N_+(x)N_-(0)|0\rangle_{conn}
\eeq
Assuming independence of instantons and anti-instantons
$(\langle N_+N_-\rangle=\langle N_+\rangle\langle N_-\rangle)$
the equation reduces again to (\ref{pro80}). We will see that
this is a crucial assumption.

The next ingredient is the scaling behaviour of QCD.
Classical chromodynamics is scale invariant and the Noether theorem
leads to a conserved scale current. In quantum theory the scale
invariance is anomalously broken (like the axial singlet current).
Ward identities can be derived, especially \cite{Shi}
\beq\label{pro82}
  \int\!d^4\!x\;\langle 0|{\cal T}N(x)N(0)|0\rangle_{conn} =
  {4\over b}\langle 0|N(0)|0\rangle \quad,\quad
  b = {11\over 3}N_c
\eeq
Therefore in a self(anti)dual background the topological susceptibility
$d^2E/d\Theta^2$ is proportional to the gluon condensate:
\beq\label{pro83}
  {d^2E\over d\Theta^2} = {4\over b}\langle 0|N(0)|0\rangle
\eeq
The relation is still valid, when there are statistically
independent regions
of selfduality and selfantiduality, as discussed above.
It is in fact sufficient to assure independence of the
total instanton/anti-instanton number $N_\pm$.
I have checked (\ref{pro83}) by using the theoretical one loop instanton
density $D(\rho)$ calculated in \cite{tHo}. Only the $\rho$ dependence
$D(\rho)\sim\rho^{-5}(\rho\Lambda)^b$ is important.
Due to the infrared divergence
it is necessary to introduce an infrared cutoff, but
one has to assure not to break scale invariance.
A minimal change is to introduce two cutoffs $f_\pm$ in the total
instanton/anti-instanton packing fraction.
The packing fraction is the spacetime volume
occupied by the instantons and is a dimensionless quantity.
Scale invariance and independence of instantons and anti-instantons
are ensured. The partition function $Z$ is
\beq\label{pro84}
  Z=\sum_{N_+N_-}Z_{N_+}^+Z_{N_-}^-
\eeq
$$
  Z_{N_\pm}^\pm = {V_4^{N_\pm}\over N_\pm!} \int_0^\infty\nq
  d\rho_1\ldots d\rho_{N_\pm}\,D(\rho_1)\ldots D(\rho_{N_\pm})\,
  \Theta\left(f_\pm-{1\over V_4}\sum_{i=1}^{N_\pm}\rho_i^4\right)
$$
A lengthy, but quite standard calculation of statistical physics,
leads to
$$
  Z_{N_\pm}^\pm = \left({c_\pm N_\pm\over V_4\Lambda^4}\right)^{
                  -{bN_\pm\over 4}}
$$
where $c_\pm=c_\pm(f_\pm,b)$ are constants independent of $N_\pm$.
Differentiation of $\ln Z$ w.r.t. $\ln c_\pm$ two times leads to
(\ref{pro83}). The result is independent of $f_\pm$.
An attractive interaction would lower the susceptibility
(compared to the density). This can be seen in the following way:
In the extreme case of a very attractive interaction, all instantons
will be
bound to instanton-anti-instanton molecules, thus $N_+=N_-$ and
$Q\equiv 0$. On the other hand a repulsive interaction
would increase the susceptibility:
\bqa\label{pro85}
  {d^2E\over d\Theta^2} &<& {4\over b}\langle 0|N(0)|0\rangle
  \quad\mbox{for attractive $I\overline{I}$ interaction} \\
  \label{pro86}
  {d^2E\over d\Theta^2} &>& {4\over b}\langle 0|N(0)|0\rangle
  \quad\mbox{for repulsive $I\overline{I}$ interaction}
\eqa
Therefore the violation of (\ref{pro83}) is a measure of
the $I\overline{I}$ interaction.

All this should be compared to Dyakonov \cite{Dya},
where the relation
\beq\label{pro57}
  d^2E/d\Theta^2=\langle 0|N(0)|0\rangle
\eeq
has been derived, which is similar to (\ref{pro83}). In this work
a simple sum ansatz $A=\sum A_I$ has been made. This ansatz
leads to a strong repulsion between close instantons,
which is the origin of the missing factor $4/b$.
The result on its own and the comparison with
(\ref{pro86}) shows that some repulsive interaction is
at work. Verbaarshot \cite{Ver2} has shown that this repulsion
is an artifact of the simple sum ansatz. Using the much more
accurate and elaborate streamline ansatz he showed, that
the interaction strongly depends on the orientation and
the average interaction is about 14 times smaller than those
obtained in \cite{Dya}.
Therefore the best thing one can do
today is to assume no interaction at all and cut the packing
fraction at some small value. There is also a more general argument that
the relation derived in \cite{Dya} must be wrong.
The topological susceptibility is of $O(N_c^0)$, whereas the
gluon condensate is of $O(N_c)$. (\ref{pro83})
is consistent with $N_c\to\infty$ considerations
only due to the presence of the $4/b$ factor.

Let us now continue with the calculation of $m_{\eta'}$.
The physical gluon condensate in the presence of light quarks is
\beq
  \langle 0|N(0)|0\rangle_{phys} = (200\mbox{MeV})^4 \quad.
\eeq
Due to the presence of light quarks it is reduced by a factor $\alpha<1$.
\beq\label{pro56}
   \langle 0|N(0)|0\rangle_{phys} =
   \alpha\langle 0|N(0)|0\rangle_{\Theta=0}^{no\;quarks}
\eeq
Combining (\ref{pro55}), (\ref{pro80}), (\ref{pro82}) and (\ref{pro56}),
we get the final formula for the $\eta'$ mass is
\beq\label{pro2}
  m_{\eta'}^2 = {4N_f\over f_\pi^2}
    {12\over 11N_c\alpha}n_R
\eeq
One can see that $m_{\eta'}^2\sim 1/N_c$ because $f_\pi^2$ and
$n_R$ and $\alpha$ are independent of $N_c$.
The largest uncertainty lies in the determination of
$\alpha$. Using again the instanton model, $\alpha$ is the
determinant of the Dirac operator with the current quark masses replaced
by effective masses:
\beq
  \alpha = \prod_{i=u,d,s}1.34m_i^{eff}\rho \approx 0.4\ldots 0.7
\eeq
We have set the effective masses to the constituent masses
\beq
  m_u^{eff}=m_d^{eff}=300\ldots 350\;\mbox{MeV} \quad,\quad
  m_s^{eff}=400\ldots 500\;\mbox{MeV}
\eeq
and $\rho$ to the value of the instanton liquid model
($\rho=600\;\mbox{MeV}^{-1}$). This estimate is consistent with the
estimate of \cite{SVZ3}.
Inserting $N_f=3$ and $f_\pi=132\;$MeV into (\ref{pro2}) we get
\beq\label{pro1}
  m_{\eta'} = 884\pm 116\mbox{MeV}
\eeq
which is in good agreement to the experimental value of 958 MeV.
This result in turn confirms the assumption, that the interaction between
selfdual and anti-selfdual regions is small. The large uncertainty
in $\alpha$ prevents more accurate statements, but (\ref{pro57})
can definitely be excluded.

Using $m_{\eta'}$ as an input
we can determine the gluon condensate in pure QCD
\beq
  {\alpha_s\over 4\pi}\langle\mbox{tr}_c GG\rangle^{no\;quarks}
  = (246\;\mbox{MeV})^4
\eeq
where we have again set $N_f$ to 3.

The factor $4/b$ in (\ref{pro83}) is the essential term to get
the correct $N_c$
dependence of $m_{\eta'}$ and agreement with the experimental mass.
The discussion has shown, that the $\eta'$ channel can be
an experimental device for testing the independence of
selfdual and antiselfdual regions in QCD, which is an assumption
in the simplest instanton models.
It might turn out some day that the details of instanton models
are wrong but the assumption of independent self(anti)dual
regions remain valid.

\section{Measurement of the Axial Form Factors}\label{sec62}

The forward matrix elements of the axial currents
$$
  s_\mu\Delta\psi = \langle ps|\bar\psi\gamma_\mu\gamma_5\psi
  |ps\rangle \quad,\quad \psi = u,d,s
$$
can be interpreted as the quark spin content of the proton,
in a sense defined more accurately in the following sections.
Three independent linear combinations of
$\Delta\psi$ have been measured, thus allowing to extract their
individual values.

{}From the neutron $\beta$-decay, using isospin invariance,
one gets \cite{Sehgal,Jaf}
$$
  a_3 = g_A = \Delta u-\Delta d = F+D = 1.254\pm 0.06 \quad.
$$
{}From the octet hyperon $\beta$-decay, using $SU(3)_F$ symmetry,
one gets \cite{Bourquin,Jaf}
$$
  \sqrt{3}a_8 = \Delta u+\Delta d -2\Delta s = 3F-D = 0.688\pm 0.0035
  \quad.
$$
{}From the spin dependent structure function $g_1^p$ of the proton,
which has been measured by EMC \cite{EMC} and SMC \cite{SMC}, one
can extract
\beq
  \Gamma_p=\int_0^1\!g_1^p(x)dx =
  {4\over 9}\Delta u+{1\over 9}\Delta d+{1\over 9}\Delta s+O(\alpha_s)
  = 0.142\pm0.014
\eeq
where we have given the world average value.

Of special interest is the quarkspin sum, which can be extracted
from the values given above,
\beq
  \Delta\Sigma^{GI}=\sqrt{3\over 2}a_0 = \Delta u+\Delta d+\Delta s =
  0.27\pm 0.13
\eeq
where the $O(\alpha_s)$ corrections have been included. It deviates
significantly from the naive quark model value $\Delta\Sigma_{qm}=1$.
This deviation is the origin of the so called spin problem.
Further the large polarization of strange quarks in the proton
$$
  \Delta s = -0.1\pm 0.05
$$
is counter intuitive because this indicates a large strange quark
content of the proton.

Much more could be said about proton spin phenomenology and
the experiments. For an introduction and further references see
\cite{Baas,Reya,Roberts,EMC}.
We will now give a more thorough definition and
interpretation of $\Delta\Sigma^{GI}$ and other
quantities, which we want to calculate within the instanton model.

\section{Axial Singlet Currents \& Anomaly}\label{sec63}

It is well known that products of operators at the same spacetime
point are very singular objects. In order to make the expressions
well defined one has to regularize and renormalize the operator
products. An anomaly appears, if this procedure breaks a symmetry
of the theory. The most important ones are the breakdown of
the scale invariance and the breakdown of the axial symmetry \cite{Shi}.
In the following we are interested in the axial anomaly \cite{ABJ}.
The operator product which has to be regularized is the axial
singlet current,
\beq
  J_{\mu 5}(x) = \sum_{q\in\{u,d,s,\ldots\}}
                 \bar{q}(x)\gamma_\mu\gamma_5 q(x)
\eeq
which seems to be local, gauge invariant and conserved\footnote{
We will use the term 'conserved' even for $m_q\neq 0$. Sometimes
this current is called the symmetric current in the literature.}.
Unfortunately after regularization one of the three properties
is unavoidably lost. Therefore we can define two different local
currents, a conserved (c) one and a gauge invariant (GI) one.
The third GI, conserved and
non-local current is discussed in \cite{Kogut} in connection with
the $U(1)$ problem. We will suppress the summation over quark flavors
and write $\psi$ for the quark field operator:
\bqa
  J_{\mu 5}^{GI}(x) &=& \lim_{\eps\to 0}
    \bar\psi(x+\eps)\gamma_\mu\gamma_5
    P\exp\left(i\int_x^{x+\eps}\!\! dz\!\cdot\!A(z)\right)
    \psi(x) \nonumber\\
  J_{\mu 5}^c(x) &=& \lim_{\eps\to 0}
    \bar\psi(x+\eps)\gamma_\mu\gamma_5\psi(x)
\eqa
The difference between the two currents is described by the anomaly
current $K_\mu$:
\bqa\label{pro3}
  K_\mu(x) &=& {N_f\alpha_s\over 2\pi}\eps_{\mu\nu\rho\sigma}
    \mbox{tr}_c A^\nu(G^{\rho\sigma}-{2\over 3}A^\rho A^\sigma)
    \quad,\quad
    J_{\mu 5}^{GI} = J_{\mu 5}^c + K_\mu \nonumber\\
  \partial^\mu K_\mu(x) &=& {N_f\alpha_s\over 2\pi}\mbox{tr}_c G\tilde G(x)
    = a(x) \\
  \partial^\mu J_{\mu 5}^c(x) &=& 2mJ_5(x) \quad\quad\quad,\quad
    J_5 = i\bar\psi\gamma_5\psi \quad. \nonumber
\eqa
$m$ is the current quark mass and $N_f$ is the number of quark flavors.
Note, that the splitting of
$J_{\mu 5}$ in a conserved and an anomaly part is gauge dependent.
There are attempts to define both uniquely on physical
grounds \cite{Reya}. The intention is to define $J_{\mu 5}^c$ as the
naive parton model spin and $K_\mu$ as some gluonic contribution.
The proton matrix elements of the various currents can be expressed
in terms of real form factors $G_i, K_i, A and J$:
\bqa\label{pro4}
  \langle p' s'|J_{\mu 5}^{GI}(0)|ps\rangle &=&
    \bar u_{s'}(p')\left[\gamma_\mu\gamma_5
    G_1^{GI}(q^2)-q_\mu\gamma_5 G_2^{GI}(q^2)\right] u_s(p) \nonumber\\
  \langle p' s'|J_{\mu 5}^c(0)|ps\rangle &=&
    \bar u_{s'}(p')\left[\gamma_\mu\gamma_5
    G_1^c(q^2)-q_\mu\gamma_5 G_2^c(q^2)\right] u_s(p) \nonumber\\
  \langle p' s'|K_\mu(0)|ps\rangle &=&
    \bar u_{s'}(p')\left[\gamma_\mu\gamma_5
    K_1(q^2)-q_\mu\gamma_5 K_2(q^2)\right] u_s(p)              \\
  \langle p' s'|a(0)|ps\rangle &=&
    2MiA(q^2)\bar u_{s'}(p')\gamma_5 u_s(p) \nonumber\\
  \langle p' s'|J_5(0)|ps\rangle &=&
    iJ(q^2)\bar u_{s'}(p')\gamma_5 u_s(p) \nonumber
\eqa
$M$ is the proton mass and $q=p'-p$.
{}From (\ref{pro3}) one can derive the following
relations between the form factors:
$$
  G_1^{GI}=G_1^c+K_1 \quad,\quad G_2^{GI}=G_2^c+K_2
$$
\beq\label{pro5}
  G_1^c-{q^2\over 2M}G_2^c={m\over M}J \quad,\quad
  K_1-{q^2\over 2M}K_2 = A
\eeq
$$
  G_1^{GI}-{q^2\over 2M}G_2^{GI} = {m\over M}J+A \quad,
$$
where all form factors are evaluated at $q$.
The last equation relates only GI quantities.
In the next section we show that the form factor $G_1^{GI}$
at zero momentum transfer can
be connected with the proton spin.

\section{The Proton Spin and its Interpretation}\label{sec64}

The stress tensor $T_{\mu\nu}$ is conserved
$(\partial_\mu T^{\mu\nu}=0)$ symmetric and GI and can be
constructed from the Noether theorem. The angular momentum density
tensor $M^{\mu\nu\rho}$
associated with Lorentz transformations can be expressed
in terms of $T_{\mu\nu}$:
\beq
  M^{\mu\nu\rho} = x^\nu T^{\mu\rho} - x^\rho T^{\mu\nu}
\eeq
$M$ can be decomposed in spin and orbital contribution of quarks
and gluons \cite{Jaf}:
\bqa
  M^{\mu\nu\rho} &=&
    M^{\mu\nu\rho\nq\nq}_{q,orb} +
    M^{\mu\nu\rho\nq\nq}_{q,spin} +
    M^{\mu\nu\rho\nq\nq}_{g,orb} +
    M^{\mu\nu\rho\nq\nq}_{g,spin}
  - {1\over 4}G^2(x^\nu g^{\mu\rho}-x^\rho g^{\mu\nu})
  + \partial(\cdots)
  \nonumber\\
  M^{\mu\nu\rho\nq\nq}_{q,orb}  &=& {1\over 2}i\bar\psi
    \gamma^\mu(x^\nu\partial^\rho-x^\rho\partial^\nu)\psi
  \quad,\quad
  M^{\mu\nu\rho\nq\nq}_{q,spin} = {1\over 2}\eps^{\mu\nu\rho\sigma}
    \bar\psi\gamma_\sigma\gamma_5\psi = {1\over 2}J^{GI}\nq_{\sigma 5}
  \\
  M^{\mu\nu\rho\nq\nq}_{g,orb}  &=& -G^{\mu\sigma}
    (x^\nu\partial^\rho-x^\rho\partial^\nu)A_\sigma
  \quad,\quad
  M^{\mu\nu\rho\nq\nq}_{g,spin} =
    G^{\mu\rho}A^\nu-G^{\mu\nu}A^\rho
  \nonumber
\eqa
The last two terms in $M^{\mu\nu\rho}$ do not contribute to the
angular momentum operator
\beq\label{pro9}
  J^i = {1\over 2}\eps^{ijk}\int\!d^3\!x\; M^{0jk}(x) \quad.
\eeq
Taking the matrix element of $J_z$ in a proton state,
where the proton is
aligned in $z$-direction and at rest we
get the spin of the proton
\beq
  \Delta J = {1\over{\cal N}}\langle ps|J_z|ps\rangle
    = {1\over 2}\eps^{3jk}\langle ps|M^{0jk}(0)|ps\rangle
  \quad,\quad
    {\cal N} = \langle p,s|p,s\rangle = \deltabar^3(0)
\eeq
The total spin of the proton is with no doubt $1/2$ and we get
the sum rule
\beq
  \Delta J = \Delta L_q + {1\over 2}\Delta\Sigma^{GI} +
             \Delta L_g + \Delta g = {1\over 2}
\eeq
where $(\Delta L_q,{1\over 2}\Delta\Sigma^{GI},\Delta L_g,\Delta g)$
are the (quark-orbital, quark-spin, gluon-orbital, gluon-spin)
contribution to the proton spin, defined as matrix elements
of the various parts of $M$ given above.
Therefore The GI axial current measures the quark spin
contribution to the proton spin. The space integral in \ref{pro9} can be
cancelt with the state normalization and we get in covariant notation:
\beq\label{pro7}
  s_\mu\Delta\Sigma^{GI} = \langle ps|J_{\mu 5}^{GI}(0)|ps\rangle
  \quad\ff\quad \Delta\Sigma^{GI} = G_1^{GI}(0)
\eeq
In the naive quark model the proton consists of three quarks
at rest. There is no orbital and no gluonic contribution to
the proton spin. This leads to the Ellis-Jaffe sum rule
$\Delta J$ = ${1\over 2}\Delta\Sigma^{GI}$ = $1/2$.
In the real world the identification of $\Delta\Sigma^{GI}$ with
the proton spin is not correct, because $J_{\mu 5}^{GI}$
measures the spin of the (nearly massless) current quarks
whereas the proton consists of three massive ($\approx 300\;$MeV)
constituent quarks. Further in a model of non-interacting
constituent quarks the axial current which measures the
constituent quark spin should be anomaly free because
the anomaly is due to the interaction with gluons.
Therefore the conserved current $J_{\mu 5}^c$ might be identified
with the constituent quark spin operator.
\beq\label{pro11}
  s_\mu\Delta\Sigma^c = \langle ps|J_{\mu 5}^c(0)|ps\rangle
  \quad\ff\quad \Delta\Sigma^c = G_1^c(0) \stackrel{?}{=}1 \quad.
\eeq
{}From (\ref{pro5}) we get
\beq\label{pro12}
  \Delta\Sigma^{GI} = \Delta\Sigma^c + K_1(0)
\eeq
which can now be interpreted in the following way: The spin
of the constituent quarks $\Delta\Sigma^c$ are formed by the
spin of the current quarks $\Delta\Sigma^{GI}$ and a rest
$-K_1(0)$, which contains orbital and gluonic contributions.
The origin of these contributions is {\it not} the motion
and interaction of the constituent quarks inside the proton,
because the constituent quarks are noninteracting and at rest
in the naive quark model, but
due to the formation of massive quarks from massless quarks.
Therefore (\ref{pro12}) may be discussed for an individual
''constituent'' quark.
Further the gluonic configurations which are responsible
for the generation  of the quarkmass also determine the value of
$K_1(0)$.

E.g.\ in a BAG model a massive quark is formed by confining a
massless quark to a sphere. The spin of the massive constituent quark
is the sum of the spin $({1\over 2}\Delta\Sigma^{GI})$
and the oribtal ${1\over 2}(1-\Delta\Sigma^{GI})$ contribution of
the current quark. The BAG, which might be formed by nonperturbative
gluonic configurations, is responsible for the mass generation
and indirectly for the orbital contribution. From analytical and
numerical calculations we know, that in the BAG model
the constituent spin
is splitted into $70\%$ spin and $30\%$ orbital contribution when
starting with massless quarks. 

Whereas (\ref{pro12}) is rigorously true, the interpretation of ${1\over
2}\Delta\Sigma^c$
as the spin of a constituent quark and its value ${1\over 2}$
is questionable. One reason is, that
an axial current which describes massive constituent quarks
is by no means conserved in contradiction to $J_{\mu 5}^c$.

There exists another relation between $\Delta\Sigma^{GI}$ and
the form factor $A$ at zero momentum transfer. Before deriving
this relation we have to give a short discussion about
the order of limits and massless poles.
The following limits are taken: The spacetime volume goes to
infinity $(V_4\to\infty)$, because the universe is actually very large,
the current quark masses go to zero ($m\to 0)$, because the
up and down masses are very small and $q\to 0$, because we
are interested in the forward matrix elements. In principle the
results can depend on the order of the limits and therefore they
have to be chosen consistent with the physical situation. This
means, that if in the real world e.g. $q\ll m$ we first have to
take $q\to 0$ and then $m\to 0$. Actually we are interested in
the forward matrix element ($q\equiv 0$) and $m\neq 0$ in the real
world and the order of limits just stated applies. Through the
cluster theorem connected correlators in coordinate space
have to decay  to zero
when the separation of two arguments tends to infinity. Therefore
there are no $\delta(q)$-peaks in momentum space and the order of limits
$q\to 0$ and $V_4\to\infty$ can be taken at will. Because $m^4V_4\gg 1$
we have to take first $V_4\to\infty$ and then $m\to 0$.
In statistical physics this is a well known fact, that
a spontaneous breakdown of a symmetry only occurs, when there is
a small explicit symmetry breaking term and the system volume tends to
infinity. In the final end one may remove the symmetry breaking term.
In QCD chiral symmetry is spontaneously broken (SBCS)
and the small current quark
mass is the explicit breaking of the chiral symmetry. Therefore
it is mandatory first to take $V_4\to\infty$ and then $m\to 0$
\cite{Smilga}. Therefore we can use the following order of limits
\beq
  \lim_{m\to 0}\{\lim_{q\to 0}[\lim_{V_4\to\infty}(\ldots)]\}.
\eeq
This justifies the usage of the infinite volume formulation from
the very beginning.

In real QCD there are no massless particles
$(m_\pi\neq 0)$. Therefore GI form factors have no massless poles
especially
\beq\label{pro13}
   q^2G_2^{GI}(q^2)\stackrel{q\to\ 0}{\longrightarrow}0
\eeq
{}From (\ref{pro5}), (\ref{pro7}) and (\ref{pro13}) we get
\beq\label{pro14}
  \Delta\Sigma^{GI} = {m\over M}J(0)+A(0)
\eeq
This relation is true whether there are Goldstone bosons in the
axial singlet channel or not. Experimentally we know that the lightest
particle in this channel is the $\eta'$ with a  mass
of 958 MeV much too large to be a Goldstone boson. Therefore $J(0)$
remains finite in the chiral limit and we obtain
\beq\label{pro20}
  \Delta\Sigma^{GI} = A(0) \quad\mbox{for}\quad m\to 0
\eeq
Assuming the non-existence of the axial singlet Goldstone boson
from the very beginning the order of limits is of no importance
in deriving (\ref{pro20}).
Note, that (\ref{pro20}) is only true,
if we take $m=m_u=m_d=m_s$, although all three masses tend to zero.
Otherwise additional nonsinglet currents on the r.h.s. of (\ref{pro14})
would survive the chiral limit \cite{Fri}.

Combining (\ref{pro5}), (\ref{pro13}) and (\ref{pro12})
we can conclude that
$$
  2M\Delta\Sigma^c = q^2G_2^c(q^2)_{|q^2=0} =
    -q^2K_2(q^2)_{|q^2=0}
$$
$\Delta\Sigma^c$ is given by the pole residuum
of $G_2^c$. Because $G_2^{GI}$ has no massless pole $\Delta\Sigma^c$ is
also given by the pole of $-K_2$. These massless poles are called
ghost poles and they may truly appear, even if there are no physical
massless particles, because $G_2^c$ and $K_2$ are gauge dependent
objects. Note that all other form factors defined in (\ref{pro5}) are GI
and therefore free of massless poles.

Table \ref{tabpro3} summarizes the values for the form factors
at zero momentum transfer for the following three cases:
\begin{itemize}\parskip=0ex\parsep=0ex\itemsep=0ex
\item the naive quark model of non-interacting constituent
      quarks of mass $m=M/N_f$,
\item chiral QCD and the identification of $\Delta\Sigma^c$ with
      the naive spin value 1,
\item the instanton liquid model.
\end{itemize}
In the following sections we will calculate some of the form factors
for a single constituent quark in the instanton liquid model.

\begin{table}[tbb]\label{tabpro3}
  \begin{center}\begin{tabular}{
         |c|c@{=}c@{+}c|c@{=}c@{+}c|c@{-}c@{=}c@{+}c@{=}c@{+}c|} \hline
    & $\Delta\Sigma^c$ & ${q^2\over 2M}G_2^c$ & ${m\over M}J$ &
      $K_1$ & ${q^2\over 2M}K_2$ & $A$ &
        $\Delta\Sigma^{GI}$ & ${q^2\over 2M}G_2^{GI}$ &
          $\Delta\Sigma^c$ & $K_1$ & ${m\over M}J$ & $A$     \\\hline
    $N_fm=M$ & 1 & 0 & 1 & 0 & 0 & 0 &
               1 & 0 & 1 & 0 & 1 & 0                         \\\hline
    $m=0$    & 1 & 1 & 0 & A-1 & -1 & A &
               A & 0 & 1 & A-1 &  0 & A                      \\\hline
    $Instanton$ & ? & ? & 0 & 0 & 1 & (-1) &
                  1 & 0 & ? & 0 & 0 & (-1)                   \\\hline
\end{tabular}\end{center}
\vspace{-3ex}
\caption[Proton form factors at zero momentum transfer]{
  \it The proton form factors at zero momentum transfer $q^2=0$
  in the naive constituent quark model ($N_fm=M$),
  in chiral QCD (m=0) and
  in the instanton-liquid model (Instanton).
  Experimentally $A$ is 0.27.
  }
\end{table}

\section{Reduction of the Proton Form Factors to Vacuum
Correlators}\label{sec65}

In this section we will calculate some of the form factors defined
above in the instanton liquid model. To apply the methods developed
in \cite{Hut2} we relate the form factors to vacuum correlation functions
\beq\label{pro29}
  \langle p's'|B(0)|ps\rangle =
\eeq
$$
   = -{1\over Z_\eta}\bar u_{s'}(p')\left[\int\!d^4\!x\,d^4\!z\;
    e^{ip'x-ipz}(i\partial\!\!\!/_x-M)(-i\partial\!\!\!/_z-M)
    \langle 0|{\cal T}\eta(x)B(0)\bar\eta(z)|0\rangle\right] u_s(p)
    \quad.
$$
$M$ is the proton mass and $B(0)$ is an arbitrary local operator.
$\eta(x)$ is a local operator with the
quantum numbers of a proton e.g. a product of three quark fields
in an appropriate spin and flavor combination \cite{Ioffe}.
Assuming that $\eta(x)$ tends to a free proton field operator for
infinite times the proton states can be reduced and (\ref{pro29})
is just an LSZ reduction formula for composite fields.
For our purpose the following form is more suitable
\bqa\label{pro30}
  \langle p's'|B(0)|ps\rangle &=&
    Z_\eta\bar u_{s'}(p')[\lim_{p^2,p'^2\to M^2}
    S^{-1}(p')T_B(p',p)S^{-1}(p)]u_s(p)             \nonumber\\
  T_B(p',p) &=& \int\!d^4\!x\,d^4\!z\;
    e^{ip'x-ipz}\langle 0|{\cal T}\eta(x)B(0)\bar\eta(z)|0\rangle    \\
  S(p) &=& \int\!d^4\!x\;
    e^{ipx}\langle 0|{\cal T}\eta(x)\bar\eta(0)|0\rangle =
    {iZ_\eta\over p\!\!/-M} + continuum       \nonumber\\
  Z_\eta^{1/2}u_s(p) &=& \langle 0|\eta(0)|ps\rangle    \nonumber
\eqa
The advantage of this form is, that the explicit knowledge of the
mass $M$ is not needed. In Euclidian calculations like lattice-,
instanton- and OPE-calculations it is always difficult to
extract pole masses.

This form can also be interpreted as a spectral
representation of the 3 point function. Inserting two complete
sets of states into the 3 point function
and taking the limit $p^2=p'^2\to M^2$ to select the proton state
one can directly attain (\ref{pro30}).

If e.g. $B(0)$ is a quark current, the 3 point function is a product
of 8 quark fields, which is too complicated to be evaluated in
a multi-instanton background. Let us assume that the proton consists
of three nearly independent quarks. Then the main nonperturbative
properties of the proton come from the formation of constituent
quarks out of current quarks. The forces which confine the constituent
quarks in the proton are assumed to modify the properties
of the proton only in a minor way, except that the proton is then stable.
This assumption is justified by the success of the constituent quark
model. The form factors of the proton are therefore the sum of the
form factors of the constituent quarks. $\eta$ has to be replaced
by a single quark field $\psi$ of flavor {\it up} or {\it down}
and $M$ must
be replaced by the constituent quark mass. In this case it is even more
important to use (\ref{pro30})
because one does not expect a definite pole mass
for the quark propagator. Looking at the quark propagator in
the instanton liquid model we see, that the $p\!\!/$ term remains
unrenormalized and therefore $Z_\psi=1$. For a constant constituent
mass this argument would be rigorously true. For a running mass it
is plausible that $Z_\psi$ is still approximately one. This
fact is true in all models of chiral symmetry breaking I know.
A conservative estimate is
\beq
  0.7\le Z_\psi\le 1
\eeq
In the following we will set $Z_\psi=1$ remembering that this not
an exact statement. The results for all form factors have to
be multiplied with $Z_\psi$.

\section{The Axial Form Factors $G_{1/2}^{GI}(q)$}\label{sec66}

The form factor of the current $j_\Gamma=\bar\psi\Gamma\psi$ of
a constituent quark can be reduced with the help of (\ref{pro30})
to a 4 point function 
\beq
  tr_{CD}[T_{j_\Gamma}(p',p)\Gamma'] =
    \int\!d^4\!x\,d^4\!z\;e^{ip'x-ipz}tr_{CD}[\langle 0|
    {\cal T} \psi(x)\bar\psi(0)\Gamma\psi(0)\bar\psi(z)|
    0\rangle\Gamma'] =
\eeq
$$
  = \int\!\dbar^4\!q\;\Pi_{\Gamma\Gamma'}(q-p,q-p',p,p')
$$
The polarization functions $\Pi_{\Gamma\Gamma'}$  are calculated
are defined and calculated in the instanton liquid model in
\cite{Hut2} and other works.
For $\Gamma=\gamma_\mu\gamma_5$ the connected part of
the 4 point function is suppressed by $O(n_R^{1/2})$.
In leading order in the instanton density only the disconnected part
contributes and we get
\beq\label{pro33}
  T_{j_{\mu 5}^{GI}}(p',p) = S(p')\gamma_\mu\gamma_5 S(p)
\eeq
Inserting (\ref{pro33}) in (\ref{pro30}) and comparison with
(\ref{pro4}) leads to
\beq\label{pro34}
  \langle p's'|J_{\mu 5}^{GI}(0)|ps\rangle =
    \bar u_{s'}(p')\gamma_\mu\gamma_5 u_s(p)
\eeq
$$
    G_1^{GI}(q^2) = 1 \quad,\quad G_2^{GI}(q^2)=0
$$
Note, that $\Pi_{\Gamma\Gamma'}$ was calculated in singular gauge, but
the connected part is suppressed in any gauge and the disconnected
part only depends on the propagators, which cancel out anyway.
The form factors $G^{GI}_{1/2}(q)$ are indeed gauge invariant.
The result coincides with a model of free massive quarks.
Further we see that the current is not conserved. Conservation
demands $q^2G_2=MG_1$, which is clearly not satisfied by (\ref{pro34}).
In the one instanton approximation one can work from the very beginning
with the effective 't Hooft vertex \cite{tHo} which explicitly
breaks the $U(1)$ symmetry and therefore contains the anomaly.

The result for the GI form factors (\ref{pro34}), although not
consistent with the experimental value, is {\it up to now} at least
theoretical consistent.

\section{The Anomaly Form Factor $A(q)$ *}\label{sec67}

We will now calculate the anomaly form factor $A$.
Using again the reduction formula with insertion of the anomaly
current $B(0)=a(0)$ we have to calculate the 3 point function $T_a(p,s)$.
In the instanton model the field operator $a(0)$ is replaced by
a classical field $a_A(0)$ where $A=\sum_I A_I$ is a multi instanton
configuration inserted in $a$. In a given background $A$ the correlator
can be written in the form
\beq
  \langle 0|{\cal T}\psi(x)a(0)\bar\psi(z)|0\rangle_A =
  a_A(0)\langle 0|{\cal T}\psi(x)\bar\psi(z)|0\rangle_A =
  a_A(0)S_A(x,z)
\eeq
where $S_A(x,z)$ is the quark propagator in the multi instanton
background $A$. The r.h.s. has now to be averaged over the collective
coordinates $\gamma_I$ of all instantons. Without the factor $a_A(0)$
this is just the averaged quark propagator calculated in \cite{Hut2}.
$a_A(y)$ is $2N_f$ times the topological charge density at spacetime
point y.
In the vicinity of an instanton of charge $Q_I=\pm 1$ the charge density
has a positive/negative bump and is small elsewhere. Therefore
$a_A(y)$ is only nonzero when there is at least one instanton near y.
Let us fix exactly one instanton in the vicinity of $y=0$. The
orientation and charge of the remaining instantons can be averaged
independently, but when averaging the locations $z_I$ the domain
near $y$ has to be avoided. The next step is to assume 2 instantons
near y and so on. The relative error we make by neglecting these further
contributions and by forgetting about the restriction on $z_I$ are
both of $O(n_R)$. In leading order in the instanton density we
can therefore fix one instanton near $y=0$ and take only this contribution
to $a_A(0)$ into account. The remaining instantons can be averaged
as in the pure propagator case and the diagrams which have to be summed
and averaged are the same except for the fixing of one instanton $I$.
The propagator consists of a chain of instanton scatterings $A_J$
($J=1\ldots N$).
Repeated scattering at this vertex is allowed.
There are two cases: The first case is that all instantons
left to all occurrences of instanton $I$ are different to all instantons
right to all occurrences of instanton $I$.
In leading order in $1/N_c$ all instantons in the middle section
from the first up to the last occurrence of $A_I$ are different to the
exterior instantons. The instantons
on the left and on the right can be averaged independently leading
to averaged multi-instanton propagators. Averaging the middle section,
but fixing $I$ leads to the effective vertex $M_I$.
The free part of the correlator in momentum space is therefore
$$
  T_a^{free}(p,s) = \langle 2N_fQ_I(z_I)\quad\makebox(22,2)
{\begin{picture}(22,5.85)
  \thicklines
  \put(11,2.85){\circle{6}}
  \put(11,2.85){\makebox(0,0)[cc]{$M_I$}}
  \put(0,2.85){\line(1,0){8}}
  \put(4,2.85){\makebox(0,0)[cc]{$<$}}
  \put(14,2.85){\line(1,0){8}}
  \put(18,2.85){\makebox(0,0)[cc]{$<$}}
  \put(3,1){\makebox(0,0)[cc]{$p$}}
  \put(17,1){\makebox(0,0)[cc]{$s$}}
 \end{picture}
}
\rangle_I =
$$
\beq\label{pro36}
  = -2iN_f\hat Q(p-s)\sqrt{M_pM_s}S(p)\gamma_5 S(s)
\eeq
$$
  Q_I(z_I)={1\over 2N_f}a_{A_I}(0)=\pm{6\over\pi^2}
    \left(\rho\over z_I^2+\rho^2\right)^4
$$
$Q_I(z_I)$ is the charge density of one instanton of charge $Q_I=\pm 1$
and $\hat Q(q)={1\over 2}(q\rho)^2K_2(q\rho)$
its Fourier transform\footnote{
I apologize for the overload of the symbol $K$:
$K_2(q\rho)$ is a modified Bessel function, $K_\mu(x)$ is the
anomaly current and $K_{1/2}(q)$ its form factors.}
for $Q_I=+1$. For $p^2=s^2=M^2$
the term $\sqrt{M_pM_s}$ is just the onshell mass $M$. Inserting
(\ref{pro36}) into (\ref{pro30}) and comparison with (\ref{pro4})
we get for the free part of the anomaly form factor:
\beq
  A^{free}(q)=-N_f\hat Q(q) \quad,\quad A^{free}(0)=-N_f
\eeq
The second case is, that there are common instantons to the left
and to the right of instanton $I$. The connected part of the
correlator and the form factor are
\vspace{5mm}
\beq\label{pro38}
  T_a^{conn}(p,s) = \bigg\langle 2N_fQ_I(z_I) \makebox(25,2)
{\begin{picture}(25,21)
  \thicklines
  \put(3,5){\framebox(19,8)[cc]{$C^s$}}
  \put(12,18){\circle{6}}
  \put(12,18){\makebox(0,0)[cc]{$M_I$}}
  \put(7.50,15.50){\oval(3,5)[lt]}
  \put(17,15.50){\oval(4,5)[rt]}
  \put(6,15){\makebox(0,0)[cc]{$\vee$}}
  \put(19,15){\makebox(0,0)[cc]{$\wedge$}}
  \put(3,3.50){\oval(6,3)[rb]}
  \put(22,3.50){\oval(6,3)[lb]}
  \put(3,2){\makebox(0,0)[cc]{$<$}}
  \put(22,2){\makebox(0,0)[cc]{$<$}}
  \put(1,4){\makebox(0,0)[cc]{$p$}}
  \put(24,4){\makebox(0,0)[cc]{$s$}}
  \put(6,13){\line(0,1){3}}
  \put(9,18){\line(-1,0){2}}
  \put(19,13){\line(0,1){3}}
  \put(15,18){\line(1,0){2}}
  \put(19,5){\line(0,-1){2}}
  \put(6,5){\line(0,-1){2}}
  \put(25,2){\line(-1,0){4}}
  \put(0,2){\line(1,0){4}}
 \end{picture}
}
\bigg\rangle_I =
\eeq
\vspace{5mm}
$$
  = -4(N_f-1)i\hat Q(p-s)C_5^s(p-s)F_5(p-s)\sqrt{M_pM_s}
     S(p)\gamma_5 S(s)
$$
\beq
  A^{conn}(q) = -2(N_f-1)\hat Q(q)C_5^s(q)F_5(q) \quad,\quad
  A^{conn}(0) = N_f-1
\eeq
For one flavor the connected part is zero as it should.
For two flavors the result can easily derived by using the formulas of
\cite{Hut2}.
The total anomaly form factor for zero momentum transfer
\beq
  A(0)=A^{free}(0)+A^{conn}(0) = -1
\eeq
is independent of the number of flavors!
This result is welcomed due to the following
argument: The form factors of the axial singlet currents $j_{\mu 5}$
should not  depend on any quark flavor which is not involved in
the particle state. One expects that they are independent of $N_f$.
Due to (\ref{pro3}) matrix elements of $a(x)$ must then be
independent of $N_f$ too.
But this is not obvious because $a(x)$ is explicitly proportional to
$N_f$ and the gluonic field is not flavor sensitive. The calculation
given above shows how the quark interaction cancels the free part, which
is proportional to $N_f$, so that the total form factor is independent
of $N_f$ at least at zero momentum transfer.

\section{The Gluonic Form Factors $K_{1/2}^{GI}(q)$}\label{sec68}

Now we come to the calculation of $K_{1/2}(0)$. The previous calculation
can be copied with minor changes. $a(0)$ has to be replaced by
$K_\mu(0)$. This in turn induces the replacement
\beq
  2Q(z_I)    \leadsto G_\mu(z_I):={1\over N_f}K_{A_I}^\mu(0) \quad,\quad
  2\hat Q(q) \leadsto \hat G_\mu(q)
\eeq
$G_\mu(z_I)$ is $K_\mu(0)$ where the gauge field is an instanton
centered at $z_I$ of charge $Q_I=+1$ and $\hat G_\mu(q)$ is its
Fourier transform. In regular gauge we get
\bqa
  G_\mu^{reg}(z) &=& {1\over N_f}K_{A_I^{reg}}^\mu(0) =
  - {z_\mu(z^2+3\rho^2)\over\pi^2(z^2+\rho^2)^3} \\
  \hat G_\mu^{reg}(q) &=& -iq_\mu\rho^2K_2(q\rho)
  \stackrel{q\to 0}{\longrightarrow}-2iq_\mu/q^2 \nonumber
\eqa
With this replacement in (\ref{pro36}) and (\ref{pro38})
and comparison with (\ref{pro4}) $K_{1/2}(0)$
can be extracted:
\beq
  K_1^{reg}(q)=0 \quad,\quad
  \lim_{q^2\to 0}{q^2\over 2M}K_2^{reg}(q)=1
\eeq
In singular gauge we get
\beq\label{pro44}
  G_\mu^{sing}(z) = G_\mu^{reg}(z)+{z_\mu\over\pi^2 z^4}
  \quad,\quad
  \hat G_\mu^{sing}(q) = \hat G_\mu^{reg}(q) + 2iq_\mu/q^2
    \stackrel{q\to 0}{\longrightarrow} 0
\eeq
$$
  K_1^{sing}(q)=0 \quad,\quad
  \lim_{q^2\to 0}{q^2\over 2M}K_2^{sing}(q)=0
$$

An apparent observation is, that the anomaly form factor
$K_2(q)$ is gauge dependent and receives a massless pole in
regular gauge. The reason for this is the gauge
dependence of the anomaly current $K_\mu$ itself. One can show
that the forward matrix elements $K_{1/2}(0)$ are GI
for small gauge transformations. A gauge transformation is called
small, when it can be smoothly deformed into the unit transformation.
On the other hand the gauge transformation, which transforms
an instanton from regular gauge to one in singular gauge is large,
because the regular solution can not be smoothly deformed into a
singular one due to the singularity.

The next striking observation is that the relation
\beq\label{pro45}
  K_1(q)-{q\over 2M}K_2(q) = A(q)
\eeq
is violated in singular gauge as can be seen from (\ref{pro44})
\beq\label{pro46}
  K_1^{sing}(q)-{q\over 2M}K_2^{sing} \neq A(q)
\eeq
Surface terms are the origin of this violation. For
the derivation of (\ref{pro45}) one has assumed the vanishing of
surface terms. If one replaces the plane wave solution for the state
by a wave packet, the state and therefore the matrix elements
decrease sufficiently fast at spacial infinity and there are no
surface terms. A experimental state is always a more or less localized
wave packet rather than an exact plane wave. Therefore in regular gauge
there are no surface terms and
\beq\label{pro47}
  K_1^{reg}(q)-{q\over 2M}K_2^{reg} = A(q)
\eeq
is valid for all $q$.
In order to work in singular gauge we have to choose a space-time
manifold $I\!\!R^4\backslash\{0\}$ to exclude the unphysical singularity.
This small hole should not affect the physics at large distances.
Therefore all coordinate space integrals are integrals over the
domain $I\!\!R^4\backslash B_\eps(0)$. Partial integration
can now lead to surface terms at zero.
The surface term is non-zero in the case of $G_\mu^{sing}$ as can be seen
from (\ref{pro44}). This is the reason for the inequality (\ref{pro46}).
It is surprising that not the slowly decaying regular gauge field
causes a surface term at infinity but the strong singularity
at the instanton centers in
singular gauge leads to surface terms and to a violation of (\ref{pro45}).

The following conclusions should be drawn:
\begin{enumerate}
\item not to consider gauge dependent objects like $K_{1/2}(q)$ at all or
\item save the relation (\ref{pro45}) by using regular gauge
      although this violates the philosophy of \cite{Hut3} or
\item modify relation (\ref{pro45}) by including the surface
      terms and be careful when performing partial integrations.
\end{enumerate}
In the following discussion we take position 2.

\section{Discussion}\label{sec69}

Comparing the results for the form factors
$\Delta\Sigma^{GI}=G_1^{GI}(0)$, $G_2^{GI}(0)$, $A(0)$,
$K_1(0)=K_1^{reg}(0)$ and $K_2(0)=K_2^{reg}(0)$ summarized in
the last row of table \ref{tabpro3} we clearly see that they
are in contradiction. It is not possible to determine the
remaining form factors in a way that they are consistent
with (\ref{pro5}) and (\ref{pro13}). The most obvious contradiction
is $\Delta\Sigma^{GI}\neq A(0)$. An opposite sign of the
anomaly would at least be theoretical consistent and would lead
to the naive expectations. The only candidate for this violation
of the axial ward identities is the neglect of the non-zeromodes.
All other approximations respect the symmetries of QCD
as discussed in Chapter \ref{sec33}.

Forte \cite{For} has derived the relation $\Delta\Sigma+A(0)=0$ in
the instanton model in the case of one quark flavor in quenched
approximation and density expansion. $\Delta\Sigma$ was identified
with $\Delta\Sigma^c$ and $K_2(0)$ was assumed to be zero
(although not explicitly stated). Therefore $A(0)=K_1(0)$ and from
(\ref{pro12}) one can arrive at the welcomed result
$\Delta\Sigma^{GI}=0$.

In section \ref{sec68} I have shown that the anomaly contributes
to $K_2$ and not to $K_1$. This is the first discrepancy. Further,
in section \ref{sec66} I have shown that $\Delta\Sigma$ has to be
identified with $\Delta\Sigma^{GI}$. This is the second
discrepancy. It may turn out that the inclusion of non-zeromodes
removes the discrepancies in a way, that leads to a
phenomenological welcomed small $\Delta\Sigma^{GI}$. In the one
instanton approximation the inclusion is manageable and has been
performed by \cite{Geshkenbein} for the meson correlators. The
consistent extension to the instanton liquid and to the quark form
factors was not yet manageable.

These problems might be compared to calculations of the $\eta'$
mass. A brute force method of calculating the axial singlet meson
correlator and extracting $m_{\eta'}$ by a spectral fit is not
successful too. More elaborate arguments, given in Section
\ref{sec66} allowed a successful determination of $m_{\eta'}$. The
instanton model was only used as a motivation for a selfdual model
of QCD. Maybe the same model is able to solve the proton spin
problem rather than a brute force calculation.

It might also be possible that the spin problem can not be solved
on the level of individual constituent quarks formed out of current
quarks but is connected with a strong interaction in the axial
singlet channel between different constituent quarks.
This possibility in connection with
instantons is discussed in \cite{Dorokhov}.


\chapter{Gluon Mass}\label{Ch7}

Up to now the instanton vacuum has been treated on the classical
level. In this Chapter we compute the quantum fluctuations around
the 1-instanton vacuum for the gluon propagator.

Section \ref{sec71} gives a brief introduction to the problems and
phenomenological consequences of massive gauge bosons. In the
other Sections we will calculate the gluon propagator for
different gauges. In background $\xi=1$ gauge it has the simple
form $ S_{\mu\nu}^{ab}=\delta^{ab} g_{\mu\nu} / (p^2-M(p)^2) $. In
Section \ref{sec72} we will extract the inverse propagator from
the terms quadratic in the fluctuations around a background field.
In Section \ref{sec73} we will average this expression over
relevant background fields and expand it for large momentum in a
way to get the gluon mass $M(p)$. In Section \ref{sec74} we
explicitly calculate $M$ in the multi-instanton background. For
large momentum, however, this mass is cancelled out by terms which
have been neglected. Furthermore, the mass is gauge dependent. To
get reliable results for large as well as for small momentum we
make in Section \ref{sec75} a cluster expansion in the instanton
density. For this purpose we need the gluon propagator in the
1-instanton background. The relevant formulas to construct this
propagator are listed in Section \ref{sec76}. Because they are a bit
lengthy we will restrict ourselves in Section \ref{sec77} to the
case of small momentum.

\section{Introduction}\label{sec71}

The question, whether creation of a mass for gauge bosons is
possible without gauge symmetry breaking, is very old. The
explicit introduction of a gluon mass into the Lagrangian causes
several difficulties. To preserve gauge invariance the gauge
fields have to be coupled to massless scalar particles
\cite{Scott,Cornwall2}, which decouple from physical matrix
elements. The mass cannot be a constant, but must vanish for large
momentum in order to ensure renormalizability of the theory (soft
mass). Both phenomenons (massless scalar and soft gluon mass)
occur in every model which dynamically creates a gluon mass. It
should be mentioned that the terminology of a mass does not
necessarily stand for a pole mass, but for the selfenergy. A
dynamical mass means a non-vanishing selfenergy at $p^2=0$, i.e.\ if
the massless pole has disappeared from the propagator. In
\cite{Cor} a gauge invariant selfenergy has been defined and
computed. Solving the Schwinger Dyson equations leads to a mass of
$500\pm200 MeV$. Due to asymptotic freedom the mass vanishes
logarithmically for large momenta $M_{gluon}(p)\sim(\ln
p^2)^{-12/11}$. Nonperturbative arguments lead to a decay
${1\over p^2}(\ln p^2)^{12/11}$ \cite{Lane,Pagels}. Finally it
should be mentioned that a dynamical gluon mass is not in
contradiction to confinement \cite{Bralic,Fishbane,CDG2,Ambjorn}.

Apart from regularizing all infrared divergences the gluon mass
has a series of phenomenological consequences. A direct
consequence are gluon balls. In the simplest picture in which
glueballs consist of $N$ independent constituent quarks the mass
is of the order of $N\!\cdot\!M_{gluon}$. Hadrons with gluons can
also be constructed ($\bar qqg$, $qqqg$, \dots). Of course, the
gluons are highly virtual due to the strong interaction. It is not
even known whether this picture is qualitatively correct. One can
only refer to the quark sector and the success of the constituent
quark model. Furthermore, a gluon mass modifies the transverse
momentum distribution of gluon jets. With a theoretically computed
gluon mass one has a natural energy cutoff. A gluon mass is also
welcomed for instanton physics: In a massive gauge theory
instantons of a size $\rho\geq M_{gluon}^{-1}$ are exponentially
suppressed. The infrared problem disappears. Under the assumption
of the \ILM\ with independent instantons of radius $\rho=600
MeV^{-1}$ and density $n=(200 MeV)^4$ one gets a gluon mass of
$M_{gluon}=480$ MeV near the cutoff scale $\rho^{-1}=600 MeV$.
Note that this possible solution of the infrared problem is
different from the standard hope of finding a repulsive
interaction between instantons \cite{Ilg,Dya,Ver2}.

For a quantitative theory one has to generalize the calculation of
unphysical gluon propagators to gauge invariant matrix elements,
like glueball correlators. This is a problem in all works up to
now (including this one). The methods to achieve a gluon mass are
too complicated to be applied to physical quantities. For this
reason it is difficult to make the above mentioned physical
applications more quantitative.

In the following Sections we will compute the averaged quantum
gluon propagator for various gauges in the multi-instanton
background. Although explicit expressions are known for a long
time \cite{Bro} they have not been used in phenomenological
applications since they are rather complex. Till now the instanton
background has been treated at the classical level. Even the
1-instanton action \cite{tHo} is of relatively low practical
importance due to the infrared problem. On the other hand quantum
corrections to gluonic $n$ point functions can be large.

\section{Gluon Propagator *}\label{sec72}

The first task to obtain a formal expression for the
gluon propagator in a background field
is to expand ${\cal L}_{QCD}[\bar{A}+B]$ in the fluctuations
$B_\mu^a$ around our background $\bar{A}_\mu^a$.
The term quadratic in $B_\mu^a$ is then by definition
the inverse gluon propagator.
For $\bar{A}_\mu^a$ we will later use our instanton gas.
We will use the QCD-Lagrangian
$$
  {\cal L}_{QCD}= {1\over 4g^2} G_{\mu\nu}^a G^{\mu\nu}_a \quad,\quad
  G_{\mu\nu}^a(A)=\partial_\mu A_\nu^a -
             \partial_\nu A_\mu^a +
             f_{abc} A_\mu^b A_\nu^c
$$
where we have rescaled the fields in such a way that the
coupling-constant-dependence is in front of $\cal L$.
We will entirely work in Euclidian space with metric
$\delta_{\mu\nu}$ instead of $g_{\mu\nu}$ because instantons do
not make sense in Minkowski space. At the very end we can
simply rotate back to Minkowski space.
With these conventions
$$
  g^2{\cal L}_{QCD}(\bar{A}+B) =
    {1\over 4} G_{\mu\nu}^a(\bar{A}+B)
    G^{\mu\nu}_a(\bar{A}+B) =
$$ $$
  = {1\over 4} \overbrace{ \bar{G}_{\mu\nu}^a
                  \bar{G}^{\mu\nu}_a }^{O(B^0)} +
  \overbrace{ B_\mu^a\bar{D}_\nu^{ab}
              \bar{G}_a^{\mu\nu} }^{O(B^1)} +
  {1\over 2} \overbrace{ B_\mu^a (
    -\bar{D}_\rho^{ac}\bar{D}_\rho^{cb}\delta_{\mu\nu}
    -2f_{acb}\bar{G}_{\mu\nu}^c+
    \bar{D}_\mu^{ac}\bar{D}_\nu^{cb} ) B_\nu^b }^{O(B^2)} +
$$
\begin{equation}\label{LQCD}
  + \underbrace{ f_{abc}B_\mu^b B_\nu^c\bar{D}_\mu^{ad}
                           B_\nu^d }_{O(B^3)} +
  {1\over 4} \underbrace{ f_{abc}B_\mu^b B_\nu^c f_{ade}B_\mu^d
                          B_\nu^e }_{O(B^4)} +
  \partial_\mu(\ldots) \quad,
\end{equation}
where
$ \bar{D}_\mu^{ab}=\partial_\mu\delta_{ab}+
    f_{acb}\bar{A}_\mu^c
$
is the covariant derivative with $\bar{A}$ inserted instead of
$A$; similarly $\bar{G}_{\mu\nu}^a=G_{\mu\nu}^a(\bar{A})$.
As usual the terms which are total derivatives disappear
after integrating $\cal L$.
We see that if $\bar{A}$ solves the
equation of motion the linear term vanishes, as it should be.
If we choose background gauge $\bar{D}_\nu^{ac}B_\nu^c=0$ the last
term of the $O(B^2)$-contribution vanishes.
Now we can read the inverse gluon propagator from the terms
quadratic in $B$ (from now on we will omit the bars over $A$, $G$ and
$D$ because the unbared objects won't be needed further):
\begin{equation}
  (S^{-1})_{\mu\nu}^{ab}={1\over g^2}
  (-D_\rho^{ac}D_\rho^{cb}\delta_{\mu\nu}
  -2f_{acb}G_{\mu\nu}^c) \quad.
\end{equation}
We will also omit the $1/g^2$ in front of the propagator
which is a result of the rescaling of fields anyway.
For further manipulations some abbreviations are useful:
$$
  G_{\mu\nu}=F^c G_{\mu\nu}^c \quad,\quad
  A_\mu=F^c A_\mu^c \quad,
$$ $$
  (F^c)_{ab}=i f_{acb} \quad,\quad
  [F^a,F^b]=i f_{abc}F^c \quad,\quad
  tr_c F^a\!F^b=N_c \delta^{ab} \quad,\quad
  N_c=3 \quad,
$$ $$
  (\hat{P}_\mu)^{ab}=iD_\mu^{ab} \quad,\quad
  \hat{p}_\mu=i\partial_\mu \quad,\quad
  \hat{P}_\mu=\hat{p}_\mu+A_\mu \quad,
$$
\begin{equation}
  \hat{p}_\mu X=[\hat{p}_\mu,X]+X \hat{p}_\mu = i(\partial_\mu X)+X
\hat{p}_\mu \quad.
\end{equation}
The last equation has only been quoted to show that
$\hat{p}$ and $\hat{P}$ will be used in operator sense.
$F^a$ are the generators in adjoint representation and
$f_{abc}$ are the structure constants of the color gauge group $SU(N_c)$.
With these abbreviations we can now write
$$
  S^{-1}_{\mu\nu} = \hat{P}^2\delta_{\mu\nu}+2iG_{\mu\nu}
  = (\hat{p}^2+\hat{p}\!\cdot\!A+A\!\cdot\!\hat{p}+A^2)
    \delta_{\mu\nu}+2iG_{\mu\nu}
  = (S_0^{-1}+V)_{\mu\nu} \quad,
$$
\begin{equation}\label{Sinv}
  S^0_{\mu\nu}=\delta_{\mu\nu}/{\hat{p}^2} \quad,\quad
  V_{\mu\nu}=(A^2+\hat{p}\!\cdot\!A+A\!\cdot\!\hat{p})
  \delta_{\mu\nu}+2iG_{\mu\nu} \quad.
\end{equation}
$S_0$ is the free gluon propagator with no background and $V$
can be interpreted as an interaction potential caused
by the background.
The QCD-Lagrangian in the background \ref{LQCD} can be written
in another form more suitable for ordinary perturbation theory:
\begin{equation}
  g^2{\cal L}_{QCD}={1\over 4}G_{\mu\nu}^a(B)G^{\mu\nu}_a(B)
  +{1\over 2}B_\mu^a V_{\mu\nu}^{ab}B_\nu^b
  +f_{abc}f_{aed}B_\mu^b B_\nu^c B_v^d\bar{A}_e^\mu
\end{equation}
The terms independent and linear in $B$ have been omitted.
Constant terms are irrelevant for the dynamics and terms linear in
$B$ are zero if $A$ solves the QCD equations of motion.
The new terms due to the background create two additional
Feynman diagrams
$$
\unitlength=1mm
\SetScale{2.835}
\SetWidth{0.2}
\begin{picture}(68,42)
  \put(28,14){\circle{10}}
  \put(28,14){\makebox(0,0)[cc]{A}}
  \put(68,14){\makebox(0,0)[lc]{$=g f_{abc}f_{aed}A^e_\mu\delta_{\nu\rho}$}}
  \put(4,14){\makebox(0,0)[rc]{$\mu,b$}}
  \put(43,26){\makebox(0,0)[lc]{$\nu,c$}}
  \put(43,2){\makebox(0,0)[lc]{$\rho,d$}}
  \put(28,37){\circle{10}}
  \put(28,37){\makebox(0,0)[cc]{V}}
  \put(4,37){\makebox(0,0)[rc]{$\mu,a$}}
  \put(52,37){\makebox(0,0)[lc]{$\nu,b$}}
  \put(68,37){\makebox(0,0)[lc]{$=V_{\mu\nu}^{ab}$}}
  \Gluon(23,37)(5,37){2}{4}
  \Gluon(33,37)(51,37){2}{4}
  \Gluon(23,14)(5,14){2}{4}
  \Gluon(32,17)(41,26){2}{3}
  \Gluon(32,11)(41,2){2}{3}
\end{picture}
$$
It should be noted that for small coupling constant $g$ the
second graph can be treated perturbatively but not the first.
So our main concentration lies on the first term.

\section{Propagator in Statistical Background *}\label{sec73}

The next step is to use some approximation scheme to calculate
the propagator
\begin{equation}
  S=(S_0^{-1}+V)^{-1}=S_0(1\!\!1+T)^{-1} \quad,\quad T=VS_0 \quad .
\end{equation}
For large momentum $p$, $S_0$ and therefore $T$ are small and we
can expand $S$ in powers of $T$:
\begin{equation}
  S=S_0(1\!\!1-T+T^2-T^3+\ldots) \quad .
\end{equation}
Note that $S(x,y)=\langle x|S|y\rangle$ is generally not
invariant under translations and rotations
because the background $A_\mu$ and therefore $T$ are not.
Actually we are not interested in the propagator for a particular
background configuration, but only in the average over all relevant
configurations. Here we do not mean the functional integration
over quantum fluctuations around the empty vacuum, but fields
other than the perturbative $A_\mu\!=\!0\,$-vacuum which
minimize the action $\int{\cal L}\,dx$.
\begin{equation}\label{Sav}
  \overline{S}=S_0(1\!\!1-\overline{T}+\overline{T^2}-\overline{T^3}
                    +\ldots) \quad,
\end{equation}
where the bar denotes averaging over relevant configurations.
Details are specified below. If the background
is statistically invariant under translations then
$\overline{S}$ is translationally invariant and therefore
diagonal in momentum space
$  \overline{S}(p,q)=\langle p|\overline{S}|q\rangle =
   \overline{S}(p)\deltabar(p-q)$.
If we would evaluate the terms in the series
we would get an expansion of $\overline{S}$ in the form
\begin{equation}\label{Sa2}
  \overline{S}(p)=1/p^2+c_1/p^4+c_2/p^6+\ldots
\end{equation}
and the pole remains at zero --- and gets even worse with higher terms.
What we want is $\overline{S}$ in a form like
$\overline{S}(p)=(p^2+M(p)^2)^{-1}$ with $M(p)$ bounded for large
and small $p$ and interpreted as momentum dependent
gluon mass\footnote{
The $''+''$ in front of $M$ will change to the more familiar $''-''$ when
we rotate back from Euclidian space to Minkowski space.}.
So let's invert (\ref{Sav})
\begin{equation}
  \overline{S}^{-1}=\hat{p}^2+M(\hat{p})^2=(1\!\!1-\overline{T}+
  \overline{T^2}-\overline{T^3}+\ldots)^{-1}S_0^{-1}
\end{equation}
and expand it once again in $T$. Without averaging this would
just be a geometrical series which, expanded, would give the original
formula $S^{-1}=S_0^{-1}+V$. With averaging the different
terms are now in
no relation and expanding and sorting with respect to powers of
$T$ yields
\begin{eqnarray}\label{Sai}
  \overline{S}^{-1} &=& ( 1\!\!1+\overline{T}
   - (\overline{T^2}-\overline{T}^2)
   + (\overline{T^3}-\overline{T}\,\overline{T^2}
     -\overline{T^2}\,\overline{T}+\overline{T}^3) )S_0^{-1}
   + O(T^4)
\\
   &=& S_0^{-1}+\overline{V}
   -(\overline{VS_0V}-\overline{V}S_0\overline{V}) + \ldots
\nonumber \\
   &=& S_0^{-1}+M^2 ,\quad M^2=M_1^2-M_2^2+\ldots ,\quad
   M_1^2=\overline{V} ,\quad
   M_2^2=\overline{VS_0V}-\overline{V}S_0\overline{V} .\nonumber
\end{eqnarray}
In the next section we introduce the concepts of
instanton gas calculation to determine $M_1$, which is
actually very simple.

\section{A Naive Estimate of the Gluon Mass *}\label{sec74}

In the \ILM, in which $A=\sum_I A_I$ is a sum of instantons in
singular gauge with fixed radius $\rho=600$MeV$^{-1}$, the scatter
amplitude

$$  \overline{V}_{\mu\nu} =
    (\overline{A^2}+\hat{p}\cdot \overline{A}+\overline{A}\cdot \hat{p})
    \delta_{\mu\nu}+2i\overline{G}_{\mu\nu} \;:
$$
can be easily computed. A short calculation shows \cite{Hut1}
\begin{equation}\label{Aav}
  \overline{A_\mu}=\langle A_{I_\mu}\rangle_I=0\quad,\quad
  \overline{G_{\mu\nu}}=0
\end{equation}
For $M_1^2=\overline{V}=\overline{A^2}$ we get
\beq
   (\overline{A^2})
=
   {12\pi^2 N_c\over N_c^2-1}n\rho^2\delta_{ab}
\eeq
For $M_1^2=\overline{V}=\overline{A^2}$ this leads to
\begin{equation}\label{M1}
   M_1=\sqrt{12\pi^2 N_c\over N_c^2-1}\sqrt{n}\rho
   \approx 420\,\mbox{MeV} \quad,
\end{equation}
where we used the instanton density $n=(200\mbox{MeV})^4$. This
value coincides very well with the results of \cite{Cor} und
\cite{Hal}.

For large momentum there is a term $M_2$ which completely cancels
$M_1$. A vanishing mass for large momentum is expected due
to asymptotic freedom.
To see this in detail one can count the number of
$\hat{p}'s$ occurring in $M_2^2$ which will give us the dominant
behaviour of $M_2^2$ for large $p$.
$\overline{V}S_0\overline{V}=M_1^4/p^2$ vanishes for large $p$
but in principle V contains a $\hat{p}$ in the numerator
$(\hat{p}A)$ and $S_0=1/\hat{p}^2$ and $\overline{V S_0 V}$
could be finite for large $p$. To be more definite consider
$$
  V S_0 V= (pA+Ap)S_0(pA+Ap)+\mbox{other terms} =
  4A_\mu{p_\mu p_\nu\over p^2}A_\nu + \ldots
$$
where we have used $[p,A_I]=i\partial_\mu A_I^\mu=0$.
$$
   \langle x|V S_0 V|y\rangle =
   4A_\mu(x)\langle x|{p_\mu p_\nu \over p^2}|y\rangle A_\nu(y) + \ldots
$$
{}From $\delta_{\mu\nu}\langle x|p_\mu p_\nu / p^2|y\rangle =
\langle x|y\rangle = \delta(x-y)\mbox{we can conclude that}
$$
  \langle x|p_\mu p_\nu/p^2|y\rangle =
  {1\over 4}\delta_{\mu\nu}\delta(x-y)$ + traceless terms.
$$
   \langle x|V S_0 V|y\rangle =
   A_\mu(x)A_\mu(x)\delta(x-y)+\ldots=\langle x|A^2|y\rangle+\ldots
$$
So $M_2^2=\overline{A^2}+\ldots=M_1^2+\ldots$ contains a term
which fully cancels $M_1^2$ calculated above. Chapter \ref{Ch8}
explains why such cancellations are expected in singular gauge.

In the next section we present a systematic expansion of the
propagator in the instanton density.

\section{Expansion in the Instanton Density *}\label{sec75}

In this section we will make an
expansion of the gluon propagator in the instanton density $n$
which will be valid for all Euclidian momenta $p$ especially for
small $p$. Furthermore, the result will be Gauge invariant.

The average of an arbitrary power of the 1-instanton field is proportional
to the inverse volume
$ \overline{A_I^n} \sim \frac{1}{V} \quad \mbox {for} \quad n \geq 1$.
So, for example, the expansion of the square of the
multi-instanton configuration in the instanton density is
\begin{equation}
   \overline{A^2} = \sum_{I=1}^{N} \overline{A^2}
    + \underbrace{N \overline{A_I}}_{O(n)}
      \underbrace{(N-1) \overline{A_I}}_{O(n)} =
      \underbrace{N \overline{A_I^2}}_{O(n)} + O(n^2)
\end{equation}
and more general
\begin{equation}
   \overline{A^n} = N \overline{A_I^n} + O(n^2) \quad , \quad
   \overline{V^n} = N \overline{V_I^n} + O(n^2) \quad
   \mbox{for} \quad n \geq 1,
\end{equation}
where $V_I = V(A_I)$ defined in (\ref{Sinv}) with $A$ replaced
by $A_I$ and similar for $T$. Let us now sum up all terms
in (\ref{Sai}) linear in $n$:
\begin{eqnarray}\label{Sai3}
   \overline{S}^{-1} &=& (1\!\!1+\overline{T}-\overline{T^2}
    +\overline{T^3}-\ldots)S_0^{-1}+O(n^2)
\nonumber \\
        &=& (1\!\!1+N(\overline{T_I}-\overline{T_I^2}
          +\overline{T_I^3}-\ldots))S_0^{-1}+O(n^2)
\nonumber \\
        &=& (1\!\!1+N \overline{T_{eff}})S_0^{-1}+O(n^2)
         = S_0^{-1}+N \overline{V_{eff}}+O(n^2)
\end{eqnarray}
\begin{eqnarray}\label{Veff}
    T_{eff} &=& T_I-T_I^2+T_I^3-\ldots=T_I-T_I(T_I-T_I^2+\ldots)
             =T_I-T_I T_{eff}  \nonumber\\
    V_{eff} &=& V_I-V_I S_0 V_{eff}
    \quad \Longrightarrow \quad V_{eff}=S_0^{-1}(S_0-S_I)S_0^{-1}
    \quad \mbox{with} \quad \\
    S_I^{-1} &=& S_0^{-1}+V_I \quad
    \mbox{is the propagator in the 1-instanton background.}
\end{eqnarray}
More generally we can make a cluster expansion of an arbitrary
function of $A$ in the following form
\begin{equation}\label{fep}
   \overline{f(A_1+\ldots+A_N)} = \overline{f(A)} =
    \sum_{l=0}^N (^N_l)
    \overline{\sum_{k=0..l} (-)^{l-k}
    (^l_k) f(A_1+\ldots +A_k)} =
\end{equation}
$$ =\underbrace{f(0)}_{O(1)} +
    \underbrace{N\overline{f(A_1)-f(0)}}_{O(n)} +
    {1\over 2}\underbrace{N(N-1)\overline{
       f(A_1+A_2)-2f(A_2)+f(0)}}_{O(n^2)} + O(n^3)
$$
where the first line is an identity even without averaging.
Inserting a Taylor expansion for $f$ in the second line and
using the indistinguishability of different instantons one can see
that all monomials in the $k$-th term contain more or equal than $k$
different instantons. So the average will factorize in $k$ factors
each proportional to $n$ and therefore
the $k$-th term is indeed proportional
to $n^k$. It is easy to generalize (\ref{fep})
for two or more species of
fields. If $A$ is a field of $N_I$ instantons {\it and}
$N_{\bar{I}}$ anti-instantons we get to first order in $n$
\begin{equation}\label{fia}
   \overline{f(A)}=f(0)+N_I\overline{f(A_I)-f(0)}
   +N_{\bar{I}}\overline{f(A_{\bar{I}})-f(0)}+O(n^2).
\end{equation}
If we insert the propagator $S$ in (\ref{fia}) we get
\begin{equation}
  \overline{S}=\overline{S(A)}=S(0)+N\overline{S(A_I)-S(0)}+O(n^2)
   =S_0+N\overline{S_I-S_0\!}+O(n^2)
\end{equation}
which is after inversion up to $O(n^2)$ just (\ref{Sai3}).

\section{QCD Propagators *}\label{sec76}

In Section 5 we have seen that it is enough to know the 1-instanton
propagator to calculate $\overline{S}$ to first order in $n$. Luckily
this propagator is known, unfortunately it is a bit lengthy
and suffers from a divergence.

The gluon propagator $S_{I\mu\nu}^{ab}$ with spin $S=1$
in adjoint color\footnote{
Often called isospin \cite{Bro} }
representation $(C=1)$ can be constructed out of
the ghost propagator $\Delta_I^{ab}$ $(S=0,C=1)$
which is explicitly known in the 1-instanton background.
The general formulas how to construct a propagator of given
spin $S$ out of the corresponding scalar propagator with
same color $C$ in a selfdual background are shown below.
They are derived and more thoroughly discussed in \cite{Bro}.

{\it Notations:}
\begin{eqnarray}
   A_\mu &=& T^a A_\mu^a \quad,\quad
   [T^a,T^b] = i\epsilon_{abc} T^c \quad, \quad
   T^a = \mbox{generator of}\quad SU(2)_c
\nonumber \\
   T^a &=& \left\{ \begin{array}{rll}
           0 & \mbox{in scalar} & (C=0) \\
           \tau^a/2 & \mbox{in fundamental} & (C=\frac{1}{2}) \\
           i\epsilon_{\cdot a \cdot} & \mbox{in adjoint} & (C=1)
           \end{array} \right\} \mbox{representation}
\end{eqnarray}
$$
   P_\mu = p_\mu+A_\mu \quad,\quad p_\mu=i\partial_\mu \quad,
   \quad \tilde{G}_{\mu\nu}=\frac{1}{2}\epsilon_{\mu\nu\rho\sigma}
         G_{\rho\sigma}
   \quad, \quad \{\gamma_\mu,\gamma_\nu\}=2\delta_{\mu\nu}
$$

{\it Spin $0$ propagator $\tilde{\Delta}$:}
\begin{equation}
    \tilde{\Delta}^{-1}=P^2 \quad \Longrightarrow \quad
    \tilde{\Delta} = P^{-2}
\end{equation}

{\it Spin $\frac{1}{2}$ propagator $S$:
(not used but only stated for completeness)}
\begin{equation}
   S^{-1}= P\!\!\!\!\!\:/ = \gamma_\mu P^\mu
   \quad \Longrightarrow \quad
   S= P\!\!\!\!\!\:/ \tilde{\Delta} \frac{1+\gamma_5}{2}+
   \tilde{\Delta} P\!\!\!\!\!\:/ \frac{1-\gamma_5}{2}\quad
   \mbox{for} \quad G_{\mu\nu}= \tilde{G}_{\mu\nu}
\end{equation}

{\it Spin $1$ Propagator $S$:}
$$
   S_{\mu\nu}^{-1}= P^2 \tilde{\Delta} _{\mu\nu}+2iG_{\mu\nu}-
   (1-\frac{1}{\xi})P_\mu P_\nu \quad \Longrightarrow \quad
$$
\begin{equation}\label{glp}
   S_{\mu\nu}=q_{\mu\nu\rho\sigma}P_\rho \tilde{\Delta}^2 P_\sigma
   -(1-\xi)P_\mu \tilde{\Delta}^2 P_\nu \quad \mbox{for}
   \quad G_{\mu\nu}= \tilde{G}_{\mu\nu},
\end{equation}
$$
   q_{\mu\nu\rho\sigma} = \delta_{\mu\nu}\delta_{\rho\sigma}
   +\delta_{\mu\rho}\delta_{\nu\sigma}
   -\delta_{\mu\sigma}\delta_{\nu\rho}+\epsilon_{\mu\nu\rho\sigma}
   \quad.
$$

There are some comments in order. The above formulas are
valid for an arbitrary color representation, we will need
them only for $C=1$. For $S\neq1$
there are zeromodes and the propagator is only the inverse
of the kernel in a subspace orthogonal to the zeromodes.
The implications and problems will be discussed in Section 8
when they show up explicitly. The Spin $1$ kernel is the
quadratic term of the QCD-Langrangian (\ref{LQCD}) with
gauge fixing term $\frac{1}{2\xi}(D_\mu^{ab}B_\mu^b)^2$
in slight generalization to the $\xi=1$ case considered
previously.

With
\begin{equation}
   \Pi(x)=1+\frac{\rho^2}{x^2}\quad,\quad
   F(x,y)=1+\rho^2\frac{(\tau x)}{x^2}
    \frac{(\tau^\dagger y)}{y^2} \quad,
\end{equation}
$$
   \tau_\mu=(\vec\tau,i)        \quad,\quad
   \tau_\mu^\dagger=(\vec\tau,-i) \quad,\quad
   \tau_\mu\tau_\nu^\dagger=\delta_{\mu\nu}+i\bar\eta_{a\mu\nu}\tau_a
$$
we can write the ghost propagator for 1 instanton in the form
\cite{Bro}
\begin{eqnarray}\label{dIab}
  \Delta_I^{ab}(x,y) & = & { {1\over 2}\mbox{tr }\tau_a F(x,y)
    \tau_b F(y,x) \over 4\pi^2(x-y)^2\Pi(x)\Pi(y) } = \nonumber\\
  & = & {\delta_{ab} \over 4\pi^2(x-y)^2} -
    {\rho^2\delta_{ab} \over 4\pi^2(x^2+\rho^2)(y^2+\rho^2) } +
    {2\rho^2\epsilon_{abc}\bar\eta_{c\mu\nu}x_\mu y_\nu \over
     4\pi^2(x-y)^2(x^2+\rho^2)(y^2+\rho^2)}  \nonumber\\
  & & + {2\rho^4(((xy)^2-x^2 y^2)\delta_{ab}+
      \epsilon_{abc}\bar\eta_{c\mu\nu}x_\mu y_\nu+
      \bar\eta_{a\mu\nu}\bar\eta_{b\rho\sigma}
      x_\mu y_\nu x_\rho y_\sigma
     \over 4\pi^2(x-y)^2(x^2+\rho^2)x^2(y^2+\rho^2)y^2 }
\end{eqnarray}
For simplicity we have placed the instanton at the origin $z_I=0$
in standard orientation $O^{ab}=\delta^{ab}$.
It is possible to write down the gluon propagator
$S_{I\mu\nu}^{ab}$ explicitly in which we are finally
interested in using (\ref{glp})
but the expression will be rather lengthy.
While $\Delta_I^{ab}$ can be averaged and represented in momentum
space exactly with the help of modified Bessel functions this seems
not to be possible for $S_{I\mu\nu}^{ab}$ and we have to expand
$\overline{S}$ for large and/or small momenta (or work much harder
to solve the complicated integrals in terms of special functions).
So we will never use the full expression for $S$.

\section{Propagators for Small Momentum *}\label{sec77}

The calculation simplifies significantly if we expand
$\overline\Delta(p)$ or $\overline{S}(p)$ for small momentum.
Because $p$ always occurs in the dimensionless quantity $(p\rho)$
the lowest order in $p$ can be found by keeping only terms
of lowest order in $\rho$ or differently stated: Small $p$ corresponds
to large $x$ and for $x\gg\rho$ $\rho$ is negligible.
To warm up let us start with $\Delta$ from (\ref{dIab}):
\begin{equation}\label{ghpr}
   \Delta_I^{ab} = \Delta_0^{ab} - \rho^2 W^{ab} + O(\rho^4)
\end{equation}
$$ \Delta_0^{ab}(x,y) = {\delta_{ab} \over 4\pi^2(x-y)^2}
$$ $$ W^{ab}(x,y) = {\delta_{ab} \over 4\pi^2 x^2 y^2} +
       {2\epsilon_{abc}\bar\eta_{c\mu\nu}x_\mu y_\nu \over
        4\pi^2(x-y)^2 x^2 y^2}
$$
$\Delta_0(p)=1/p^2$ is
the free $(A_\mu^a\equiv 0)$ ghost propagator.
After reintroducing the instanton position $z$ we can now average
$\Delta_0-\Delta_I$. The $\epsilon$-term in $W$ will be killed by
averaging over the instanton orientation:
\begin{equation}
  \langle x|\overline{\Delta_0^{ab}\!-\!\Delta_I^{ab}}|y\rangle =
  4\pi^2\delta_{ab}\rho^2{1\over V_4}
  \int {d^4\!z \over 4\pi^2(x-z)^2 4\pi^2(z-y)^2 } \quad,
\end{equation}
The last integral is infrared divergent but noticing that
$1/4\pi^2(x-y)^2$ is the Fourier transformation of $1/p^2$
we can rewrite the above expression in the form
$$
  V_4 \langle x|\overline{\Delta_0\!-\!\Delta_I}|y \rangle =
  4\pi^2\rho^2\int d^4\!z \langle x|{1\over p^2}|z\rangle
  \langle z|{1\over p^2}|y\rangle =
  4\pi^2\rho^2 \langle x|{1\over p^4}| y\rangle
$$
The divergence is now hidden in the fact that the coordinate
representation of $p^{-4}$ does not exist. In fact we are not
interested in the coordinate representation but in the
momentum representation and so the divergence is spurious.
Inserting $\Delta$ in (\ref{Veff}) we get
\begin{equation}
   \overline{V}^{ghost}_{eff} =
   \Delta_0^{-1}\ \overline{\Delta_0\!-\!\Delta_I}\ \Delta_0^{-1} =
   p^2{4\pi^2\rho^2 \over V_4} {1\over p^4} p^2 =
   {1\over V_4} 4\pi^2\rho^2 = \rho^2 p^4\overline{W}
\end{equation}
and lead in the $SU(2)_c$ case to a ghost mass of
$$
   M^2_{ghost}(p=0) = N\overline{V}_{eff} = 4\pi^2\rho^2 n
$$
which has to be inserted into the ghost propagator
$\overline\Delta_{ab} = \delta_{ab}(p^2+M^2_{ghost}(p))^{-1}$.
Generalization to the case of $N_c$ colors gives
\beq\label{gluon30}
   M_{ghost}(p=0) = 2\pi\rho\sqrt{2n_R} = 8.9\rho\sqrt{n_R}
   = 340\,\mbox{MeV}
\eeq
where $n_R:=n/N_c$ is of the order of $O(N_c^0)$.
The most important insight is that $M\not=0$.
A scalar particle in adjunct color representation achieves a
dynamical mass in the instanton background.

Let us now calculate the gluon mass at zero momentum along the
same lines. To do this we must insert (\ref{ghpr}) into
(\ref{glp}). Expanding also $P_\mu$ up to order $\rho^2$
$$
  P_\mu=p_\mu+\rho^2\breve{A}_\mu+O(\rho^4) \quad,\quad
  \breve{A}_\mu = {2F_a\bar\eta_{a\mu\nu}x_\nu \over x^4}
$$
we get
$$
   P_\mu\Delta_I^2 P_\nu =
   p_\mu\Delta_0 p_\nu - \rho^2 p_\mu(\Delta_0 W+W \Delta_0)p_\nu +
     \rho^2(p_\mu\Delta_0\breve{A}_\nu+\breve{A}_\mu\Delta_0 p_\nu) +
     O(\rho^4)
$$
The free gluon propagator in $R_\xi$-gauge is well known or
can be obtained from (\ref{glp}) setting $G_{\mu\nu}=0$:
$$
   S^0_{\mu\nu} = q_{\mu\nu\rho\sigma}p_\rho\Delta_0^2 p_\sigma -
     (1-\xi)p_\mu\Delta_0^2 p_\nu =
   {1\over p^2}(\delta_{\mu\nu}-(1-\xi){p_\mu p_\nu\over p^2})
$$ $$
   S^0_{\mu\nu}-S^I_{\mu\nu} =
   \rho^2[q_{\mu\nu\rho\sigma}p_\rho(\Delta_0 W+W\Delta_0)p_\sigma
   -(1-\xi)p_\mu(\Delta_0 W+W\Delta_0)p_\mu]
$$ $$
   \overline{S^0_{\mu\nu}-S^I_{\mu\nu}} =
   2\rho^2\overline{W}(p^2\delta_{\mu\nu}-(1-\xi)p_\mu p_\nu) =
   2\rho^2 p^2\overline{W}S^0_{\mu\nu}
$$
where we have used in the last line that $p$ commutes with
averaged quantities like $\overline{W}$.
$$
   \overline{V}^{gluon}_{eff} = S_0^{-1}\overline{S_0-S_I}S_0^{-1} =
   2\rho^2 p^2 S_0^{-1}\overline{W}
$$ $$
   \overline{S}^{-1}=S_0^{-1}+N\overline{V^{gluon}_{eff}} =
   S_0^{-1}(1\!\!\!1+2 M^2_{ghost}(0)/p^2)
$$ $$
   \overline{S}={p^2 S_0\over p^2+M^2_{ghost}(0)} =
   {\delta_{\mu\nu}-(1-\xi){p_\mu p_\nu\over p^2} \over
     p^2+M^2_{gluon}(p^2)}\quad \mbox{with}
$$ $$
   M^2_{gluon}(p^2)=2 M^2_{ghost}(0) + O(p^2)
$$
We got the interesting result that for momentum transfer $0$ the
gluon mass is a factor of $\sqrt 2$ larger than the ghost mass.
Maybe the relation $M_{gluon}\approx 2 M_{ghost}$ is valid even for
larger $p$. The gluon mass for momentum transfer zero is
\beq\label{gluon32}
   M_{gluon}(p=0) = 4\pi\rho\sqrt{n_R} = 12.6\rho\sqrt{n_R}
   = 480\,\mbox{MeV}
\eeq

\section{Zeromodes *}\label{sec78}

In our whole calculation we have ignored the zeromodes.
All gluon field fluctuations must be orthogonal to these which
is achieved by using a gluon propagator orthogonal to the
zeromodes or otherwise stated by subtracting out from (\ref{glp})
the projections on the zeromode subspace. This procedure
also deletes a divergence coming from the second term in (\ref{dIab})
squared. There are some further problems to be solved because
the scalar product of the zeromodes with the propagator does not exist
\cite{Bro}. But all this concerns a term which is proportional
to $\rho^4$ which was irrelevant in our zero momentum approximation.
Also the principle mechanism of mass generation does not depend
on the zeromodes which can be seen from the ghost propagator
because in the spin 0 case there are no zeromodes.
This should be contrasted with the fermionic case where the zeromode
are the main ingredient for mass generation and the so called
zeromode approximation simplifies calculations enormously.

\section{Conclusions and further developments}\label{sec79}

We have computed the gluon propagator in the \ILM\ for small
momentum $p$ in leading order in the instanton density. A mass
independent of $N_c$ has been extracted. Furthermore, the mass is
gauge independent, at least for $o^2=0$ within the class of
$R_\xi$ gauges. As expected, for $\xi\to 0$, the gluon propagator is
transversal.

A massless pole remains for $\xi\neq 1$ due to the $(1-p)p_\mu
p_\nu/p^2$ term. We expect that the massless state, which is
associated with this pole, decouples from physical matrix
elements, as in conventional perturbation theory.

The result is in agreement with the values calculated
by Cornwall ($500\pm 200 MeV$) \cite{Cor} or
extracted from $pp$ scattering ($370 MeV$) \cite{Hal}.

The next step may be the calculation of some gauge invariant
correlation function like a glueball correlator or topological
susceptibility. As usual this involves products and integrals over
propagators but with the difference that we have to use
(\ref{glp}) instead of the simple free propagator. Without further
simplification this may cause a headache.



\chapter{Gauge Invariant Quark Propagator}\label{Ch8}

A variety of predictions concerning chiral symmetry breaking and
concerning the lightest hadrons in various channels can be made
within the instanton liquid model. Although there are various
attempts to derive this model from first principles, it is still
an open question, whether instantons melt or not. Thus the
infrared problem remains unsolved. In this chapter I attempt to
make a few predictions which do not rely on the Instanton Liquid
Model, but exploit the theoretical one-loop density $D(\rho)$.

The quark condensate will be calculated in Section \ref{sec85} and
\ref{sec86}. I get a finite result by choosing an appropriate gauge
and performing a self energy resummation without introducing an
infrared cutoff in the instanton density.

Because the finiteness essentially depends on the choice of gauge,
I give a more general discussion in Section \ref{sec81},
\ref{sec82} and \ref{sec83} of how to choose a gauge when
calculating gauge dependent quantities. The quark propagator in
the background of one instanton in the well known regular and
singular gauge is compared to a gauge invariant propagator, for
which explicit expressions are calculated in Section \ref{sec81}.

\section{Generalities On the Choice of Gauge}\label{sec81}
Gauge symmetry is a rather large symmetry, an infinite product
of $SU(N_c)$ in the case of QCD. A physicist is always happy
of having symmetries because they can be exploited to make
predictions even without solving the theory. Gauge symmetry is
necessary to get a physical vector particle spectrum.
As long as one does not make an approximation which manifestly
breaks gauge symmetry one can choose a comfortable gauge for
calculations because the result is GI.
But it is very difficult {\it not} to break GI, especially
in a non-abelian gauge theory. It is not easy to find a GI
regularization and furthermore, the gluon propagator, the primary
object in perturbation theory, is not GI. Of course it is meanwhile
well known how to perform GI calculations in every order
perturbation theory using FP-ghosts and dimensional regularization.
Every new approach beyond perturbation theory is again confronted
with the problem of GI. In lattice theory the Wilson action had
to be invented. In Schwinger-Dyson and Bethe-Salpeter type
selfconsistency equations GI is still an open problem. In
instanton physics when going beyond the one instanton approximation
the choice of gauge is also important. This will be discussed in the
next paragraph. There is a related problem when considering
non-GI objects from the very beginning like the gluon or
quark propagator. Strictly
speaking they are only {\it defined} when relying to a certain gauge.
In principle one should not give them any physical meaning at all.
Often one is tempted to do so and therefore it is necessary
to give some motivation of choosing this or that gauge.
\section{A Natural Gauge}\label{sec82}
The gauge field $A_\mu^a$ describes the connection between
neighboring vector bundles over the spacetime manifold $I\!\!R^4$.
Thus a choice of gauge is like the choice of a coordinate system
in general relativity with connection $\Gamma^\mu_{\;\;\nu\rho}$.
When choosing a crooked coordinate system, although being in a smooth
universe, there will appear fictitious accelerations towering above
the real physical accelerations
$$
  \ddot x^\mu_{phys} = \ddot x^\mu_{fict} + \Gamma^\mu_{\;\;\nu\rho}
                       \dot x^\nu \dot x^\rho \quad.
$$
When making general covariant calculations
these fictitious accelerations and the $\Gamma$ contribution
will cancel out thus leading to the correct
small result. But the slightest unsystematic approximation
will produce gross errors. The natural solution of this problem is to
use a coordinate system as smooth as possible to avoid fictitious
accelerations, e.g. to choose $\Gamma^\mu_{\;\;\nu\rho}$ as small
as possible. To make this statement more quantitative we
may try to minimize $(\ddot x^\mu_{phys}-\ddot x^\mu_{fict})^2$
simultaneously for
all curves. This is done by choosing a coordinate system which
minimizes\footnote{In Euclidian space this is a positive definite norm}
$$
  ||\Gamma||^2 \;:=\; \int \Gamma_\mu^{\;\;\nu\rho}
                           \Gamma^\mu_{\;\;\nu\rho}
                           \,d^4\!x
$$
This obviously measures the crookedness of the coordinate system.

Let us now transfer this to QCD. The analog norm for the
gauge potential is
$$
  ||A||^2 \;:=\; \int tr_c A_\mu A^\mu \,d^4\!x
$$
A stationary point is found by variating $||A||$ w.r.t.
gauge transformations
$$
  \delta A_\mu = i[A_\mu,\Omega]+\partial_\mu\Omega \quad,\quad
  \delta ||A||^2 = 2i\int tr(\partial_\mu A^\mu)\Omega d^4\!x = 0
  \quad \forall\Omega   \quad\gdw\quad
  \partial_\mu A^\mu =0
$$
Therefore in Lorentz gauge, $A_\mu^a$ contains as few pure gauge
as possible, if the stationary point is a minimum.
An expansion in $A$ is thus most rapidly convergent in Lorentz gauge.
\private{Im Pfadintegral Feynman gauge}
In applications where $A$ is not needed in total e.g. when
only a certain momentum region is probed, different norms and
different gauges may be optimal in the sense discussed above.
Especially one should include derivatives of $A$ into the
norm in order to guarantee a smooth $A$ which is important for
high energies.
\section{On the Gauge in Instanton Physics}\label{sec83}
When calculating GI quantities in the background of
one instanton in a GI way the choice of gauge is only a matter
of convenience. But one can see that there are large cancellations
between different terms in regular gauge at large distances due
to their slow decay and in singular gauge at small distances
due to the topological singularity at the instanton center.
For non-GI invariant quantities like the gluon or quark propagator,
or when making some unsystematic approximation,
the lesson is to use singular/regular gauge when dealing with
low/high energies to avoid these cancellations.
This is consistent with the discussion
given above. Singular as well as regular gauge fulfill
the Lorentz condition. $||A_{sing}||$ is finite and a minimum.
$A_{sing}$ is therefore a good choice for low energies.
For high energies it is important to have a smooth $A$ which
is obviously only satisfied by the regular gauge.

To linearly superpose instantons they have to decay
rapidly enough. Therefore one has to use singular gauge.
This argument can in principle be circumvented
by superposing two fields $A_N$ and $A_{\overline{N}}$ the
former/latter being an exact multi-instanton/anti-instanton
configuration in regular gauge.
Despite this, for low energies singular gauge is in any case
a good choice and for high energies a one instanton approximation
is already a good approximation.
\section{The Quark Propagator in Axial Gauge *}\label{sec84}
\begin{figure}
   \epsfbox[85 0 480 550]{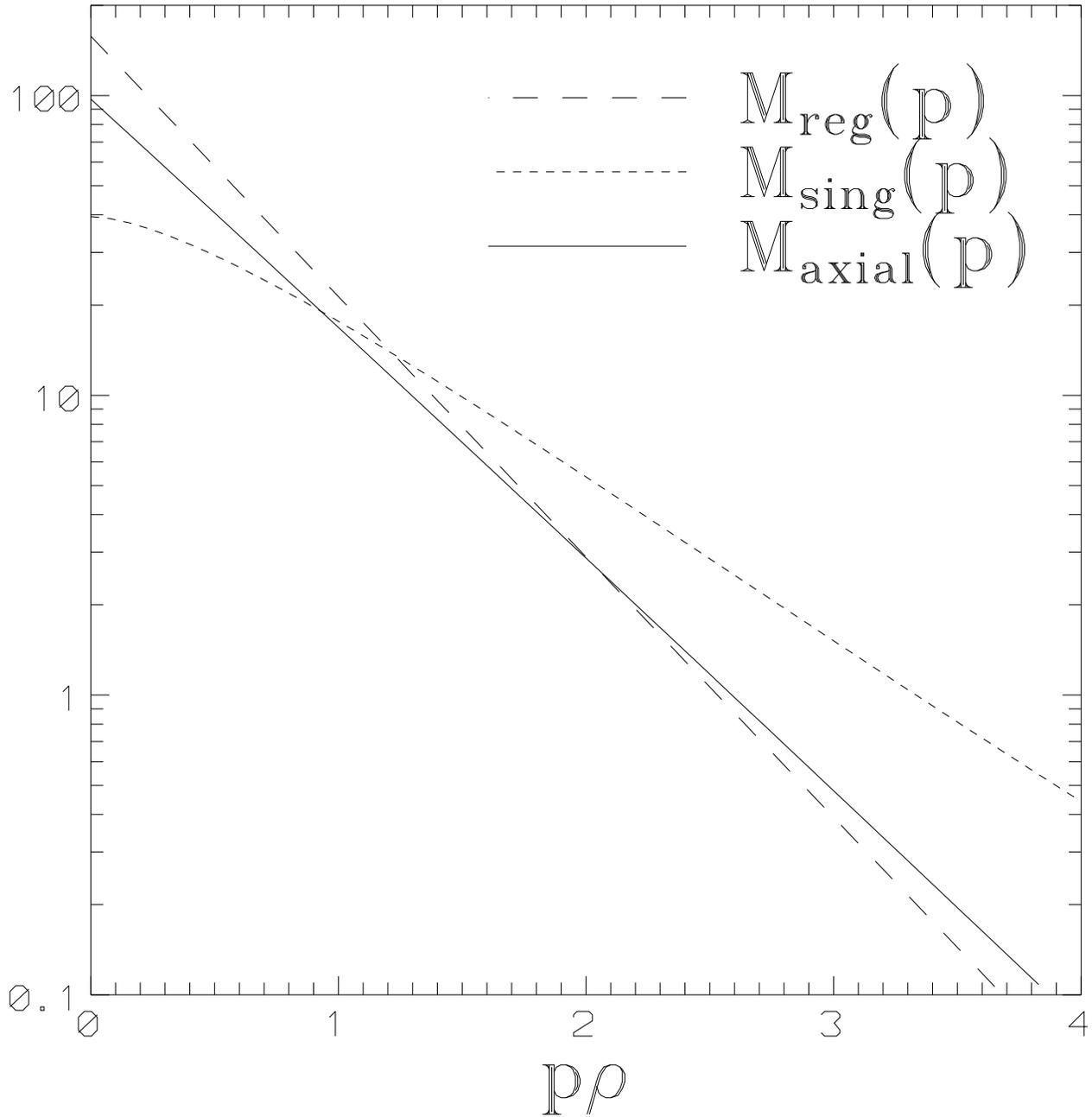}
   \caption[Constituent Quark Mass in Regular,
            Singular and Axial Gauge]{\label{fig51}\label{fig1}
            \it Constituent quark mass $M(p)\sim p^2\varphi^2(p)$
            in singular, regular
            and axial gauge for fixed instanton radius $\rho$
            in arbitrary normalization. For a given momentum
            the corresponding lowest curve may be interpreted as
            the ''most physical'' one.}
\end{figure}

\begin{figure}
  \epsfbox[85 0 480 550]{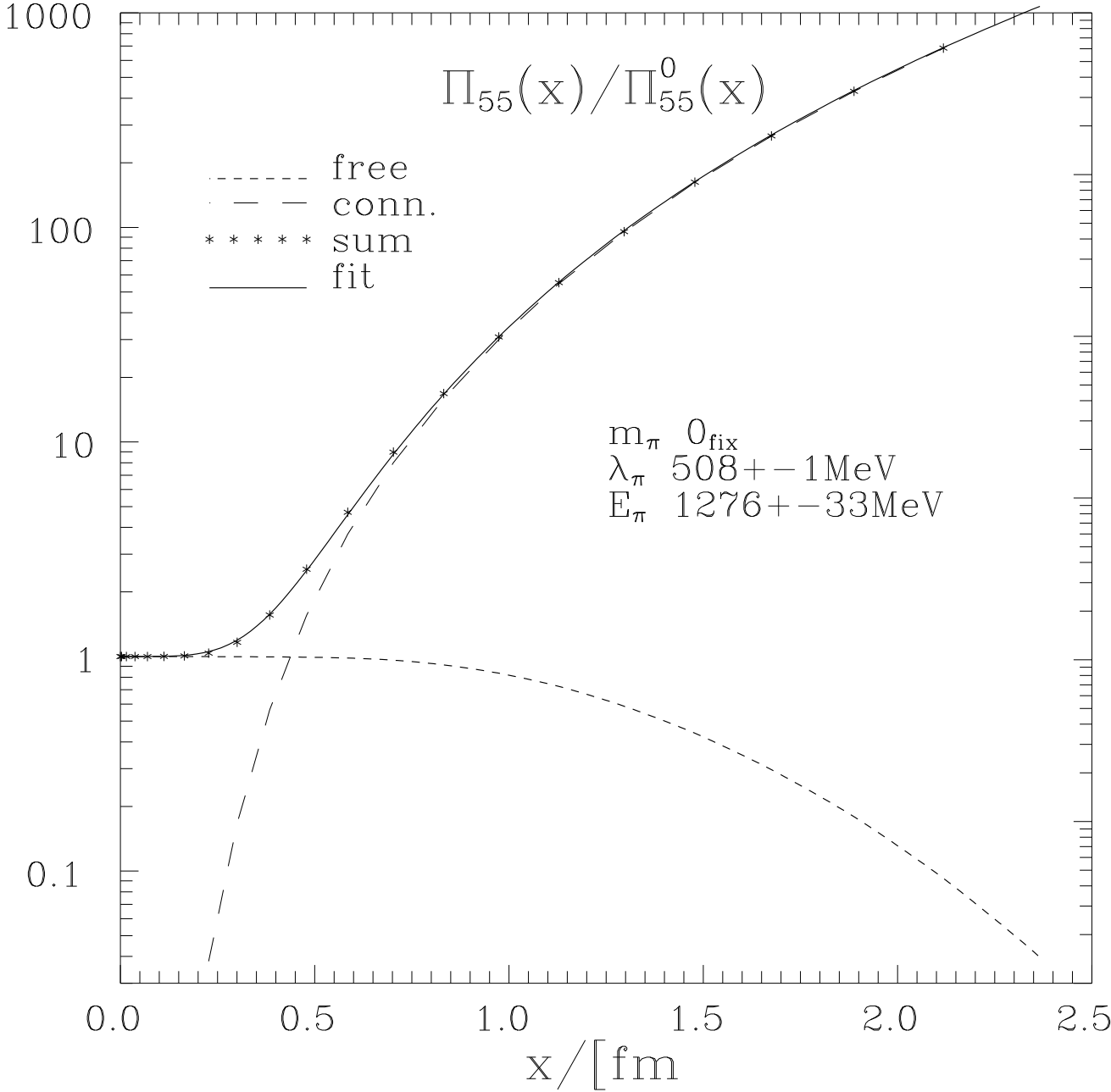}
  \caption[Pseudoscalar Triplet Correlator ($\pi$)]{\label{fig71}
     \it Pseudoscalar triplet correlator normalized to the
     free massless quark correlator. The pion coupling constant
     $\lambda_\pi$ and the continuum threshold $E_\pi$ are {\it fit}ted
     in order to match the spectral ansatz with the theoretical
     {\it sum} of the {\it free} and the {\it conn}ected part.
           }
\end{figure}

A specific example to test the gauge dependence is the quark
propagator. The contribution of one instanton of radius $\rho$
to $M(p)=ip^2\bar S_I(p)$
which usually is interpreted as a constituent quark mass is shown
in figure \ref{fig1} in regular, singular and axial gauge.
The regular graph is larger than the singular at low momentum
and the singular graph shows the slow decay (only polynomial in $1/p$)
for large momenta. The analytical expressions are well known
and are listed in Appendix \ref{appa2} together with expressions
in axial gauge which will be derived and discussed below.

A correlator containing color-non-singlet operators can be made GI
by connecting distant points with a special path-ordered exponential
containing the gauge field.
The exponential ensures the parallel transport of color
from one point to the other.
The GI quark propagator may symbolically written as
\beq\label{gi31}
  S_{ax}(x,y) = \langle 0|\Psi(x)
           P\exp\left(i\int_x^y\!\! dz\!\cdot\!A(z)\right)
           \bar\Psi(y)|0\rangle
\eeq
$P$ denotes path-ordering.
We have already defined $S_{ax}$ to be its color singlet part
because only the singlet part is GI.
$S_{ax}$ will be called the axial propagator because in axial gauge
with $n_\mu=x_\mu-y_\mu$ the exponential vanishes.
In the one instanton background in zeromode approximation we get
$$
  S_{ax}(x,y) = {1\!\!1_c\over N_c} tr_c\left[
           P\exp\left(i\int_x^y\!\! dz\!\cdot\!A(z)\right)
           \psi(x)\bar\psi(y) \right]
$$
where $A$ is now the instanton field and $\psi$
is the zeromode in any gauge.
In a coordinate system where the instanton sits at the origin and
$x-y$ is in time direction\footnote{
Although working in Euclidian space we will adopt the Minkowskian
language $(x_0, {\bf x}) =$ (time, space).}
$({\bf x}={\bf y}={\bf z})$ the path ordered exponential reduces
to an ordinary exponential. Alternatively we could have tried
to find a gauge transformation which transforms the regular gauge
in axial gauge. In both cases we get:
\bqa\label{gi33}
  S_{ax}(x,y) &=& {1\over N_c}tr_c\left[\psi_{ax}(x)
                \bar\psi_{ax}(y)\right] \quad,\quad
  \psi_{ax}(x) = R(x)\psi_{reg}(x) ,
\\ \label{gi34}
  R(x) &=& e^{\pm i\alpha(x){{\bf\tau}\!\cdot\!{\bf x} \over |{\bf x}| } }
       = \cos\alpha(x)\pm i{{\bf\tau}\!\cdot\!{\bf x} \over |{\bf x}| }
         \sin\alpha(x) =: \pm i\tau_\mu^\pm\!\cdot\! \tilde{x}(x)
\\ \label{gi35}
  \alpha(x) &=& {|{\bf x}|\over\sqrt{{\bf x}^2+\rho^2}}
              \arctan{x_0\over\sqrt{{\bf x}^2+\rho^2}}
\eqa
$\alpha(x)$ may also be written in a covariant form
$$
  \alpha(x)=\pm\left(1+{\rho^2(x-y)^2\over x^2y^2-(xy)^2}\right)^{-1/2}
  \arctan\sqrt{(x^2-(xy))^2\over x^2y^2-(xy)^2+\rho^2(x-y)^2}
$$
but now $\alpha(x)$ depends also on $y$ and the expression for the
propagator no longer factorizes. The reason for this is that the
axial gauge is not covariant, but the definition of the
propagator is. Inserting (\ref{gi34}) and (\ref{gi86}) into
(\ref{gi33}) we get
\beq\label{gi36}
  S_{ax}(x,y) = {1\over N_c} \left( (\tilde{x}\tilde{y})
     - {1\over 2}\tilde{x}_\mu\tilde{y}_\nu\sigma^{\mu\nu} \right)
       {1\pm\gamma_5\over 2}\varphi_{reg}(x)\varphi_{reg}(y)
\eeq
Inserting (\ref{gi34}), (\ref{gi35}) and (\ref{gi86})
into (\ref{gi36}) the space-time averaged propagator can
be expressed as an integral over elementary functions
$$
  \bar S(x-y) = \int_0^\infty\nq dr\int_{-\infty}^\infty\nq dt\,4\pi r^2
     \cos\left[{r\over R}
       \left(\arctan{t+|x-y|\over R} -
             \arctan{t\over R}
       \right) \right]\cdot
$$
$$
  \cdot{1\over 2N_c}{\rho^2\over
     \pi^2(R^2+(t+|x-y|)^2)^{3/2} (R^2+t^2)^{3/2} }
  \quad,\quad R^2 = r^2+\rho^2
$$
The difference between the propagator in regular and axial gauge
is the insertion of the $\cos[\ldots]$ factor. Therefore the
axial propagator is everywhere smaller than the regular propagator,
except at $x=y$ where they coincide because the path-ordered
exponential is one.
At large distances it is smaller by a factor $\pi/4$.
Instead of performing the integration in coordinate space, let us
go directly to the more interesting momentum space representation:
\beq\label{gi85}
  \bar S_I(p) = {1\over 2N_c}\varphi_{ax}^\mu(p)
    \varphi_{ax}^{\mu\dagger}(p)  \quad,\quad
  \varphi_{ax}^\mu(p) = \int \tilde{x}^\mu(x)\varphi_{reg}(x)e^{ipx}dx
\eeq
Although $\varphi_{ax}^\mu$ does not transform like a vector,
we can choose a convenient direction of $p$ because
$\varphi\varphi^\dagger$ is a Lorentz scalar. For pure spacelike $p$
the spatial components of $\varphi$ vanish because the integrand
is anti-symmetric w.r.t time reflection. Only the time component is
nontrivial
$$
  \varphi_{ax}(p) := \varphi_{ax}^0(p) = \int\! d^3\!r\int\!dt\,
    \cos\left[{r\over R}
       \arctan{t\over R} \right]
  {\rho\over\pi(R^2+t^2)^{3/2}}
  e^{i{\bf p}\cdot{\bf r}}
$$
With the following hints
\bqa
  \cos(\gamma\arctan x) &=& \mbox{Re}
    \left({1+ix\over 1-ix}\right)^{\gamma/2}           \nonumber\\
  \int_{-\infty}^\infty\!\! (R-it)^{-\alpha}(R+it)^{-\beta} dt &=&
    2\pi(2R)^{1-\alpha-\beta}
  {\Gamma(\alpha+\beta-1)\over\Gamma(\alpha)\Gamma(\beta)} \nonumber\\
    \Gamma({3\over 2}-x)\Gamma({3\over 2}+x) &=&
  {(1/4-x^2)\pi\over\cos\pi x}
  \nonumber\\
    \int d^3\!r\,e^{i{\bf p}\cdot{\bf r}}f(r) &=&
    {2\pi\over p}\int_0^\infty\!\! f(r)\sin(pr)r\,dr
\eqa
the reader should be able to perform the $t$
and the angular integration $d\Omega_r$,
\beq\label{gi86}
  \varphi_{ax}(p) = {8\over p\rho}\int_0^\infty\nq
  \cos\left({\pi r\over 2R}\right)\sin(pr)r\,dr
\eeq
I was not able to perform this last integral analytically,
but for small momenta it is easy to see that $\varphi_{ax}^0(p)$
behaves like $\pi^2\rho/p$. For large $p$ it decays like
$\sim e^{-p\rho}$ with a non-polynomial coefficient because
of an essential singularity at $r=\pm i\rho$.
Comparing $\varphi_{reg}$, $\varphi_{sing}$ and $\varphi_{ax}$ plotted
in figure \ref{fig1} we see that the axial $\varphi$ lies somewhat
in between the regular and the singular. So one may conclude
that axial gauge is a good compromise for all momenta.

The calculation of the GI propagator seems to make the discussion
of its gauge dependent partners obsolete. I will now argue that
this is not the case. The reason is that there are a huge
number of GI definitions of a quark propagator and (\ref{gi31})
is only one
possible choice. One obvious generalization is to choose a more
complicated path from $x$ to $y$ than a straight line.
The next thing one could do is not to restrict oneself to a specific
path, but to take into account all paths one is interested in
and average the results with arbitrary weights. Another
possibility is to let the path depend on the gauge
field itself, as long as this choice is made in a GI way.
Finally one can combine both generalizations. I am sure that
it is possible to produce any result for the propagator with
a suitable generalized definition.
The advantage of the standard axial propagator is, that the
definition is simple and that
the non local operator has a physical interpretation. It creates
a quark-antiquark pair connected by a thin gluon flux tube.
This might be a good choice for a non-local meson creation operator.
But it is also plausible that one of the generalizations given
above is even better. The only thing I want to point out is,
that the GI definition for the propagator given above is
nothing more than to work in axial gauge.
One still has to choose the right gauge using more sophisticated
arguments.

\section{Effective Quark Mass *}\label{sec85}
The only reason for performing the $N_c\to\infty$ limit is
to make the $N_c$ dependence of the resulting formulas simple.
The accuracy has been checked to be within the standard 10\%
for $N_c=3$ usually achieved by $1/N_c$ expansion.
Here the accuracy can simply be understood. The actual expansion
parameter is not $1/N_c=1/3$ itself but $1/b\approx 1/11$.
The following asymptotic formulas will be used
\bqa
  N_c!^{1/N_c} \eqd N_c/e \quad,\quad b \eqd {11\over 3}N_c \quad,\quad
  C_{N_c}^{1/b} \eqd 2.22b^{-6/11} \,    \nonumber\\
  \rho^5 D(\rho) \sim (2.22(S_0/b)^{6/11}\rho\Lambda)^b \quad,\quad
  S_0/b = {24\pi^2\over 11}(g_0^2 N_c)^{-1}
\eqa
Every equality in the large $N_c$ limit will be marked with a dot.
Notice that in this limit instantons of size
$\rho < {1\over 2.22}(S_0/b)^{-6/11}\Lambda^{-1}$ are completely
suppressed. Above this threshold the instanton density gets
infinite. $S_0/b$ is independent of $N_c$ because
the coupling is $g_0\sim 1/\sqrt{ N_c}$. $\Lambda$ is the QCD
scale.

In the presence of one light quark flavor the instanton density
$D(\rho)$ has to be multiplied with the functional determinant of the
Dirac operator
$$
  Det(iD\!\!\!\!/+im) \approx 1.34m\rho
$$
which is proportional to $m$ because of a zeromode of $D\!\!\!\!/$.
The quark propagator in the background of one instanton
is dominated by this zeromode
$$
  S_I(p,q)={\psi_I(p)\psi^\dagger_I(q) \over im}
$$
Averaging this expression over all collective coordinates $\gamma_I$
one gets
$$
  M(p) := ip^2\bar S_I(p) = {1.34\over{2N_c}}\int_0^\infty\nq d\rho\,
          p^2\rho D(\rho) \varphi^2(p)
$$
Summing the contribution to the propagator
of $0,1,2,3,\ldots$ instantons, which is the analog of a selfenergy
resummation in perturbation theory,
$$
  S(p) = {1\over p\!\!/} +
         {1\over p\!\!/}{M(p)\over i}{1\over p\!\!/} +
         {1\over p\!\!/}{M(p)\over i}{1\over p\!\!/}
         {M(p)\over i}{1\over p\!\!/} + \ldots
       = {1\over p\!\!/+iM(p)}
$$
justifies to call $M(p)$ a dynamical quark mass.
Expressions of $\varphi$ in various gauges are given in
appendix \ref{appa2}.
The graphs of $\varphi(p)$ in singular
and regular gauge cross over at
$$
  \varphi_{sing}(p)=\varphi_{reg}(p) \quad\gdw\quad
  2p\rho\approx 2.5
$$
Therefore one should use regular gauge for large $\rho$ and
singular gauge for small $\rho$. This choice of gauge also
makes the integral convergent for large $\rho$. At this stage we
have no infrared problem.
Using regular gauge in the whole integration interval we get
$$
  M_{reg}(p) = Bp({\Lambda\over 2p})^b \quad,\quad
  B=1.34\cdot16\pi^2(C_{N_c}b^{2N_c})(S_0/b)^{6b/11}I_b/N_c
$$
$$
  I_b = \int_0^\infty\nq dz\,z^{b-2}e^{-z} = (b-2)! \quad,\quad
  z=2p\rho
$$
The integral is sharply dominated by
$$
  z \eqd b\pm b^{1/2} >> 2.5
$$
therefore the result is independent of the choice of gauge for
$z<2.5$ justifying our use of regular gauge over the whole
integration interval. Axial gauge would lead to nearly the
same result as can be seen from figure \ref{fig1}.
Using singular gauge for large
$\rho$ would produce a divergent integral dominated by arbitrary
large instantons  inconsistent with the choice of gauge discussed
above.
The infrared "problem" shows up in the rapid raise of $M(p)$
for low $p$, which effectively suppresses the propagation
of quarks with low virtuality $p^2$.
Consider some process involving quarks at distances $x=1/p_c$.
The effective quarkmass $M(1/x)$ is dominated by instantons
of much larger size
$$
  \rho=\rho_c(1\pm b^{-1/2}) >> x \quad,\quad \rho_c={b\over 2p_c}.
$$
In other words, given
an instanton of radius $\rho$ influences the physics at a much
smaller scale $x={2\over b}\rho << \rho$. Therefore the
interior of the instanton is probed and one should avoid
the singularity at its center by using regular gauge.

\section{The Quark Condensate}\label{sec86}
Let us now calculate a real physical gauge invariant observable,
the quark condensate
$$
  \langle \bar\psi\psi\rangle :=
  \lim_{x\to 0}\mbox{tr}_{CD}(S(x)-S_0(x)) =
  -4iN_c\int {M(p)\over p^2+M^2(p)} {d^4\!p\over(2\pi)^4}
$$
Inserting M(p) and performing the angular integration we get
$$
  |\langle \bar\psi\psi\rangle| =
  {N_c\over 16\pi^2}B^{3/b}J_b\Lambda^3
$$
$$
  J_b = \int_0^\infty {z^{b+2}\over 1+z^{2b}} dz =
  {\pi\over 2b\sin({b+3\over 2b}\pi)} \eqd {\pi\over 2b}
  \quad,\quad p=B^{1/b}{\Lambda\over 2}z
$$
The integral is finite and sharply dominated by $z\eqd 1\pm b^{-1}$.
Without resummation of the selfenergies the integral $J_b$
and thus condensate
would have turned out to be infinite.
The condensate is dominated by quark wavefunctions with
momenta
$$
  p=p_c(1\pm b^{-1}) \quad,\quad p_c=\beta b{\Lambda\over 2}
  \quad,\quad
  \beta := {1\over b}B^{1/b} \eqd {2.22\over e}(S_0/b)^{6/11}
$$
and depends on $\Lambda$ and $g_0$
$$
  |\langle \bar\psi\psi\rangle|^{1/3} =
  0.139\beta b\Lambda \quad.
$$
Expressing $p_c$ and $\rho_c$ in terms of
$|\langle\bar\psi\psi\rangle|$ by eliminating $\beta$
we get our main result
\bqa
  p_c    &=& 3.59 |\langle \bar\psi\psi\rangle|^{1/3}  \nonumber \\
  \rho_c &=& {1.96\over N_c} |\langle \bar\psi\psi\rangle|^{1/3}
  \label{gi810} \\
  2.22N_c\Lambda &=& (g_0^2N_c)^{6/11}|\langle\bar\psi\psi\rangle|^{1/3} \nonumber
\eqa
A weak point is the experimental extraction of $g_0$.
It should be extracted from a reliable treelevel process at low
energies presumably of the order of $\rho_c$.
In QCD improved Bag-Models the main nonperturbative effect is
modeled by the bag and the hyperfinesplitting is caused
by a one gluon exchange. $g_0$ extracted from $\Delta-N$ splitting
is \cite{Clo}
$$
g_0^{bag} \approx 2.6
$$
Let me also give a theoretical guess of $g_0$.
The change to a two loop expression for the instanton density
$S_{0/1} \leadsto S_{1/2}$ can be effectively performed
by only replacing $S_0$ in the following way
$$
  (S_0/b)^{6/11} \leadsto (\ln{1\over \rho\Lambda})^\alpha
  \quad,\quad \alpha = {15\over 121}
$$
Because $\alpha$ is very small
$(\ln{1\over\rho\Lambda})^\alpha$ is approximately one
in a large range of values for $\rho\Lambda$.
and for
$$
  g_0^{guess} = 2.7\sqrt{3/N_c}
$$
the two 2 loop density coincides with the one loop density.
Because we do not believe that the 2 loop density is an improvement,
one should not take $g_0^{guess}$ too seriously.
At least it is not in contradiction with $g_0^{bag}$.

The condensate is well known to be
$|\langle \bar\psi\psi\rangle|^{1/3}=240\mbox{MeV}$.
Setting $N_c=3$ and taking $g_0=2.6$ for granted we get
\bqa
  p_c   \pm\Delta p   &=& (860\pm 80)\mbox{MeV}      \nonumber\\
  \rho_c\pm\Delta\rho &=& (160\pm 50)^{-1}\mbox{MeV}          \\
  \Lambda_{PV}  &\approx& 190\mbox{MeV}              \nonumber
\eqa
The most interesting thing is, that the condensate is sharply
dominated by quark field wave functions of rather large
momentum $p_c$. On the other hand the dominating instantons
have a very large radius $\rho_c$, 4 times larger than usually
assumed in instanton liquid models. Nevertheless the predicted value of
$\Lambda_{PV}$, which of course must be assigned a large error
because of the rough estimate of $g_0$, is in agreement with
experiment.

\section{Summary}\label{sec87}
Whenever one is calculating gauge dependent objects or when making
gauge breaking approximations, one is confronted with the problem
of choosing a ''good'' gauge. Specializing the general discussion
of Section \ref{sec82} to the case of instantons, we came to the
conclusion that the regular gauge is appropriate for small distances
and the singular gauge for processes involving large distances.
The GI propagator was defined, calculated and compared to
the propagator in singular and regular gauge (figure \ref{fig1}).
The conclusion was,
that the GI propagator is not a-priori a good choice,
but lies somewhat in between regular and singular gauge.

Using an appropriate gauge along the lines discussed in Section
\ref{sec82} we were able to derive a finite quark condensate
without taking an infrared cut-off for the instanton radius nor
relying on some instanton model. The linear relation between
$|\langle\bar\psi\psi\rangle|^{1/3}$ and the QCD scale $\Lambda$
is in agreement with experiment. The condensate if formed by
quark fields of high momenta $p_c=860$MeV mainly lying within
the sharp region $\Delta p=80$MeV. The dominating instantons
are very large ($\rho_c=160$MeV).



\chapter{Conclusions}\label{Ch9}

\section{Summary}\label{sec91}

Some known and various new results have been obtained in this
work. In most cases, the computation was based on the \ILM.

The following new insights were gained.
\begin{itemize}
\item For the quark propagator dynamical quark loops
can be absorbed in a renormalized instanton density, which can be
identified with the gluon condensate (Chapter \ref{Ch3}).
\item Concise Bethe-Salpeter equations have been obtained and
were solved for the quark 4-point functions. In the flavor singlet
channel, a chain of quark loops contribute, which is responsible
for the missing $U(1)$ Goldstone bosons (Chapter \ref{Ch4}).
\item From the meson correlators the masses and couplings of the
$\sigma$, $\rho$, $\omega$, $a_1$ and $f_1$ mesons have been
obtained by a spectral fit (Table \ref{tab4}, Figures
\ref{fig71}-\ref{fig76}).
\item The $\eta'$ mass has been predicted successfully (Equation \ref{pro1}).
\item The axial proton form factors of the singlet current $j_{\mu 5}$,
the gluon current $K_\mu$, and the anomaly $a$ have been computed
and their gauge (in)dependence has been discussed
(Table \ref{tabpro3}). It has been shown that $A(0)=-1$ is
independent of the number of flavors.
\item For small momentum a ghost and a gauge independent gluon
mass have been calculated (Equations \ref{gluon30} and \ref{gluon32}).
\item General rules regarding the choice of a gauge, especially
when to chose the singular and when the regular gauge
have been established (Sections \ref{sec81}, \ref{sec82} and
\ref{sec83}).
\item A gauge invariant quark propagator in the 1-instanton background
has been derived (Equations \ref{gi85} and \ref{gi86}).
\item The quark condensate has been identified with the QCD scale
$\Lambda$, where neither an infrared cutoff, nor a specific
instanton model was necessary (Equation \ref{gi810}).
\end{itemize}

\section{Outlook}\label{sec92}

The \ILM\ has proven to be successful in various sectors of QCD as
the results of this and other works show. Although this model
cannot strictly be derived from first principles, its success
proves that there is some truth to it after all. The limits of
this model also show up clearly: Confinement cannot be explained
and the axial singlet channel still causes problems, although
instantons explicitly break the $U(1)_A$ symmetry.

Future work should take into account non-zeromodes, which become
important in the case of strange quarks and probably essential in
respecting the axial Ward identities. Baryon correlators to determine
Baryon masses could be calculated. Both extensions have already
been realized numerically \cite{ShV}.

Although a direct calculation of the axial singlet correlators was
not successful, refined arguments allowed determining the $\eta'$
mass. A similar consideration may also solve the proton spin problem.

Gauge invariant glueball corrections should be calculated.

The most urgent theoretical problem is still a correctness proof
of the \ILM. This would clarify the range of validity and increase
the general acceptance of this model.

QCD is, indeed, a tough theory.  Experimental physicists do a good job in
measuring the hadron parameters, whereby the precision of these
experiments increases steadily. This progress should challenge the
theorist to keep up with them.

\section{Acknowledgements}\label{sec93}

First and foremost I would like to thank Prof. H. Fritzsch for his
willingness to take on a novice in his department. He opened the
door to the fascination of particle physic and enabled contact
with important particle physicists, especially Prof. Karliner in
Israel, Prof. S. Veneziano, S. Forte and many others at CERN. I
owe Prof. Dyakonov special thanks for many precious discussions
and his prompt willingness to answer all questions via email. I
wish to thank my good friend Michael Birkel for bringing fresh
wind to the Institute and Dr. A. Blumhofer for many long,
stimulating evenings spent at the University, during which the
borderline between physics and philosophy was often crossed.
Thanks also to all colleagues of the department, for occasional
discussions. Thank you to my father, and M. Matuschek, and D.
Holtmannsp\"otter for proof-reading my thesis. I am also very
grateful to the German "Forschungs-Gemeinschaft" for financial
support.


\private{}

\begin{appendix}

\chapter[Appendix]{}\label{ChA}

\section{Notations}\label{appa1}

Nearly all occurring factors $2\pi$ can be absorbed in the
following definitions:
$$
  \deltabar^d(\cdots):=(2\pi)^d\delta(\cdots)
  \quad,\quad
  \int\!\dbar^d\!p:=\int\!{d^d\!p\over(2\pi)^d}
  \quad,
$$
At many places vectors and operators in Dirac notation in
$I\!\!R^4$ are used:
$$
  \langle x|p\rangle = e^{ipx} \quad,\quad
  \langle x|y\rangle = \delta^4(x-y) \quad,\quad
  \langle p|q\rangle = \deltabar^4(p-q)
$$ $$
  \langle x|\psi\rangle = \psi(x) \quad,\quad
  \langle p|\psi\rangle = \hat\psi(p) \quad,\quad
  \langle p|S|q\rangle = S(p,q)
$$ $$
  \int\!\dbar^4\!p\,|p\rangle\langle p| =1\!\!1 \quad,\quad
  \int\!d^4\!x\,|x\rangle\langle x| =1\!\!1 \quad,\quad
$$ $$
  [\hat p,X] = i(\partial_\mu X)
$$
This notation should not be confused with the vacuum state
$|0\rangle$ and the proton state $|ps\rangle$ of the QCD Hilbert
space and the averaging $\langle\cdots\rangle_I$ over instantons.
\bqan
  N_c         &=& \mbox{Number of colors}        \\
  N_f         &=& \mbox{Number of quark flavors} \\
  m           &=& \mbox{Current masses}             \\
  M           &=& \mbox{Dynamic masses}        \\
  \mbox{tr}_D &=& \mbox{Trace in Dirac space}       \\
  \mbox{tr}_C &=& \mbox{Trace in color space}        \\
  \mbox{Tr}   &=& \mbox{Functional trace}          \\
  \mbox{Det}  &=& \mbox{Functional determinant}  \\
  \lambda^a/2 &=& \mbox{Generators of } SU(N_c),\quad a=1\ldots N_c^2-1 \\
  \tau^a/2    &=& \lambda^a/2 = \mbox{Generators of } SU(2)_c
                 ,\quad a=1\ldots 3
\eqan

\section{Instantons in Singular, Regular and Axial Gauge}\label{appa2}
Instantons are solutions of the classical Yang-Mills equations of
motion. The instanton at the origin in standard orientation in
singular, regular and axial gauge is given by:
\bqa
   A_\mu^{sing}(x) &=& \eta^\pm_{\mu\nu} {x_\nu\over x^2}
     {\rho^2\over x^2+\rho^2} \quad\quad,\quad\quad
     \tau_\mu^\pm\tau_\nu^\mp=\delta_{\mu\nu}+i\eta_{\mu\nu}^\pm
                                                        \nonumber\\
   A_\mu^{reg}(x)  &=& \eta^\mp_{\mu\nu}
     {x_\nu\over x^2+\rho^2} \quad\quad\quad,\quad\quad
     \tau_\mu^\pm = (\pm i,{\bf\tau})           \nonumber\\
   A_\mu^{ax}(x)   &=& R(x)A_\mu^{reg}(x)R^\dagger(x)
     + iR(x)\partial_\mu R^\dagger(x)
\eqa
The upper/lower sign denotes an instanton/antiinstanton
($Q=\pm 1$).
$$
  R(x) = \pm i\tau_\mu^\pm\tilde{x}^\mu(x) \quad,\quad
  \tilde{x}^\mu(x) = { \cos\alpha(x) \choose
    {{\bf x}\over |{\bf x}|}\sin\alpha(x) }
$$
$$
  \alpha(x) = {|{\bf x}|\over\sqrt{{\bf x}^2+\rho^2}}
              \arctan{x_0\over\sqrt{{\bf x}^2+\rho^2}}
$$
The covariant derivative $D\!\!\!\!/$ has a zeromode
$$
  iD\!\!\!\!/\psi = (i\partial\!\!\!/-A\!\!\!/)\psi = 0
$$
where the zeromode has the following form:
\bqa
  \psi_{sing}(x)   &=& \sqrt{2}\varphi(x){x\!\!/\over|x|}\chi \nonumber\\
  \psi_{reg}(x)    &=&\sqrt{2}\varphi(x)\chi \quad\quad,\quad
  \varphi(x) = {\rho\over\pi(x^2+\rho^2)^{3/2}}              \\
  \psi_{ax}(x)     &=& \sqrt{2}\varphi(x)R(x)\chi       \nonumber
\eqa
$\chi$ is a color \& Dirac spinor, given by
$$
  \chi^\pm\bar\chi^\pm = {1\over 16}\gamma_\mu\gamma_\nu
    {1\pm\gamma_5\over 2}\tau^\mp_\mu\tau^\pm_\nu \quad.
$$
For light quarks the propagator is dominated by the zeromode.
When averaged over the instanton orientation, position and charge
the propagator is diagonal in momentum space and given by
$$
   \langle\psi(p)\psi^\dagger(p)\rangle =
   {1\over 2N_c}\varphi^2(p),
$$
where
\bqa
  \psi_{sing}(p)   &=& \sqrt{2}\varphi_{sing}(p){p\!\!/\over|p|}\chi
  \quad,\quad
  \varphi_{sing}(p) = \pi\rho^2{d\over dz}
    [I_1(z)K_1(z)-I_0(z)K_0(z)]_{z=p\rho/2}              \nonumber\\
  \psi_{reg}(p)    &=&\sqrt{2}\varphi(p)\chi \quad\quad\quad\quad,\quad
  \varphi_{reg}(p)  = {4\pi\rho\over p}e^{-p\rho}           \\
  \psi_{ax}(p)     &=& FT\{\psi_{ax}\}(p)    \quad\quad\quad\quad,\quad
  \varphi_{ax}(p)   = {8\over p\rho}\int_0^\infty\nq
  \cos\left({\pi r\over 2\sqrt{r^2+\rho^2}}\right)\sin(pr)r\,dr\nonumber
\eqa
$I_\nu(z)$ and $K_\nu(z)$ are modified Bessel functions.
The asymptotics are given in the following table
\begin{table}[hhh]\label{taba1}
  \begin{center}\begin{tabular}{|c|c|c|c|}               \hline
    ${p\over\rho}\varphi(p)$ & singular & regular & axial   \\ \hline
    $p\rho\ll 1$             & $2\pi$   & $4\pi$  & $\pi^2$ \\ \hline
    $p\rho\gg 1$             & ${12\pi\over(p\rho)^3}$
      & $4\pi e^{-p\rho}$    & $\sim e^{-p\rho} $        \\ \hline
\end{tabular}\end{center}
\vspace{-3ex}
\caption{
  Asymptotic behaviour of ${p\over\rho}\varphi(p)$
}\end{table}

The constituent mass of a quark in the \ILM\ is
$$
  M_\rho(p) \sim p^2\varphi^2(p) \quad.
$$
$M_\rho(p)$ is plotted in figure \ref{fig1} in all three gauges.

\section{Averaging over the Instanton Parameter $\gamma_I$}\label{appa3}
The instanton in general orientation and location
and the corresponding zeromode have the following form
\bqan
  A_{I\mu}(x) &=& O_I^{ab}A_\mu^b(x-z_I) \\
  \psi_I(x) &=& U_I\psi(x-z_I)  \quad,\quad
    \chi_I:=U_I\chi \quad,\quad
    {1\over 2}\mbox{tr}(U\lambda^aU^\dagger\lambda^b) = O^{ab} \\
  \gamma_I=(z_I,O_I,\rho_I,Q_I) &=&
    \mbox{(location, orientation, radius, topological charge)}
\eqan
$A_\mu^b$ is and instanton in standard orientation with radius
$\rho=\rho_I$ and charge $Q_I=\pm 1$ and $\psi$ is the corresponding
zeromode, given in Appendix \ref{appa2}.
$U_I\in SU(N_c)$ and $O_I\in Ad[SU(N_c)]$ are orientation matrices
in fundamental and adjunct representation, respectively.

Át various points I have to average over the collective
coordinates $\gamma_I$:
\beq
   \langle\ldots\rangle_I \;=\;
      \int\!d\gamma_I\,D(\rho_I)\ldots \;=\;
    {1\over 2}\sum_{Q_I=\pm 1} \int\!d^4\!z_I dO_I d\rho_I
    \, D(\rho_I)\ldots
\eeq
The following assumption on $D$ defines the \ILM:
\beq
 D(\rho_I) = n\delta(\rho_I-\rho)   \quad,\quad
 n=(200\;\mbox{MeV})^4              \quad,\quad
 \rho = 600\;\mbox{MeV}^{-1}        \quad.
\eeq
The most important formulas for the Haar measure $\int dO_I$ and $\int dU_I$
are:
\beq
  \int dO\;1 = 1 \quad,\quad
  \int dO\;O^{ab}O^{cd} = {1\over N_c^2-1}\delta^{ac}\delta^{bd}
\eeq
$$
  \int dU\;1 = 1 \quad,\quad
  \int dU\; U^i_kU^{\dagger j}_l = {1\over N_c}\delta^i_l\delta^j_k
$$
Integrals over an odd number of matrices are zero. The following
formulas are also useful
\beq
  N_C\langle\chi_I^\pm\bar\chi_I^\pm\rangle_I =
  \mbox{tr}_C\chi_I^\pm\bar\chi_I^\pm =
  {1\over 2}({1\pm\gamma_5\over 2}) \nonumber
\eeq
where the averaging over $Q_I$ has not been performed yet.
\beq
  \bar\chi\chi = \mbox{tr}_{CD}\chi\bar\chi = 1
\eeq
Collecting space, color and Dirac terms yields
\beq
  \psi_I^\dagger(p)\psi_I(p) = 2\varphi^2(p) \quad,\quad
  \int\!\dbar^4\!p\,\psi_I^\dagger(p)\psi_I(p) = 1
\eeq

\section{Numerical Evaluation of Integrals}\label{appa4}

The integral expressions for the meson correlators have been
evaluated numerically. Two types of operations have to be performed:
{}
\begin{enumerate}\parskip=0ex\parsep=0ex\itemsep=0ex
  \item{Convolution of Lorentz covariant functions
    ($F_{0/5},\Gamma_\Gamma$)}
  \item{Fourier transformation (FT) of
    the correlators to coordinate space}
\end{enumerate}

Let us first consider the FT generalized to $d$ dimensions:
\beq
  \hat f_{\mu_1\ldots\mu_n}(x) =
  F_d\{f_{\mu_1\ldots\mu_n}\}(x) =
  \int\!\dbar^d\!p\,e^{-ipx}f_{\mu_1\ldots\mu_n}(p)
\eeq
For a scalar spherically symmetric function $f=f(|p|)$ the FT
reduces to a one dimensional integral
\beq\label{eb10}
  F_d\{f(|p|)\}(x) = \int_0^\infty \left({m\over 2\pi|x|}\right)^{d/2}
  f(m) J_{d/2-1}(m|x|)|x|\,dm
\eeq
where $J_\nu$ are Bessel functions.

If $x$ is not too large and $f$ decays rapidly, the integration
can be performed with Gaussian (or other) integration methods.
If the decay is too slow one has to subtract the asymptotic
part from $f$ thus improving convergence. The FT of the asymptotic
part can be performed analytically and has to be added to the numerical
FT of the reduced function.

The FT of a general Lorentz covariant function can also be reduced to
(\ref{eb10}) with (formally) an increased dimension $d$:
\bqa
  F_d\{p_\mu f(|p|)\}(x) &=& 2\pi ix_\mu F_{d+2}\{f\}(x)
  \\
  F_d\{p_\mu p_\nu f(|p|)\}(x) &=&
    {1\over d-1}\Big[ (\delta_{\mu\nu}-{x_\mu x_\nu\over x^2})
      F_d\{p^2f(|p|)\}(x) -
  \nonumber\\
  & & (\delta_{\mu\nu}-d{x_\mu x_\nu\over x^2})
     (4\pi F_{d+2}\{f\}(x)-4\pi^2x^2F_{d+4}\{f\}(x)) \Big]
  \nonumber\\
  &\ldots&
  \nonumber
\eqa

\section{Numerical Evaluation of the Convolution}\label{appa5}

The convolution integral is defined as
\beq
  f\!*\!g(p) = \int\!\dbar^d\!q\, f(q)\!\cdot\!g(p-q)
\eeq
This type of integral can be reduced to the FT discussed above:
\beq
  f\!*\!g(p) = F_d^{-1}\{F_d\{f\}\!\cdot\!F_d\{g\}\}
\eeq
This is a quick and easy method for evaluating convolution integrals.
For the disconnected part of the correlators it has the advantage
that in coordinate representation $F_d^{-1}$ can be dropped.
The disadvantage of this formula is that the FTs involve oscillating
integrals which are numerically problematic.
If the back-transformation $F_d^{-1}$ is needed as in the case of
$F_{0/5}$ and $\Gamma_\Gamma$ it is better to perform the convolution
directly.
Similar to the FT the convolution can be reduced to the scalar case. The
convolution
of two scalar functions can further be reduced to a two dimensional
integral:
\beq\label{eb20}
  f*g(p) = {(d/2-1)!\over 2\pi^{d/2+1}(d-2)!}
  \int_0^\infty\!dr\int_0^\pi\!d\theta\,
  f(r)g(\sqrt{p^2-2|p|r\cos\theta+r^2})(r\sin\theta)^{d-2}r
\eeq
It is again evaluated with Gaussian integration methods. The second
advantage is that there are no problems with slowly decaying functions.
Sometimes there are large cancellations between different terms.
In this case it is essential to use nonadaptive integration methods
because they will not result in a loss of accuracy.

The explicit reduction of the various correlators to the basic
forms (\ref{eb10}) and (\ref{eb20}) is more or less
trivial. The selfconsistency equation has been solved by iteration.
The results are plotted in figure \ref{fig71} - \ref{fig76}.



\chapter{Figures}\label{Chf}

\stepcounter{section}
\addcontentsline{toc}{section}{\Alph{chapter}.\arabic{section}$\;$
   Panorama Function}

\stepcounter{section}
\addcontentsline{toc}{section}{\Alph{chapter}.\arabic{section}$\;$
   Constituent Quark Mass in Regular, Singular and Axial Gauge}

\stepcounter{section}
\addcontentsline{toc}{section}{\Alph{chapter}.\arabic{section}$\;$
   Pseudoscalar Triplet Correlator ($\pi$)}

\stepcounter{section}
\addcontentsline{toc}{section}{\Alph{chapter}.\arabic{section}$\;$
   Pseudoscalar Singlet Correlator ($\eta'$)}

\begin{figure}
  \epsfbox[93 0 480 550]{p55s.out}
  \caption[Pseudoscalar Singlet Correlator ($\eta'$)]{\label{fig72}
     \it Pseudoscalar singlet correlator normalized to the
     free massless quark correlator. There is a strong repulsion
     in this channel and no boundstate is formed.
     The theoretical curve is compared to a curve obtained from a
     pure continuum spectrum above $E_{\eta^\prime}$.
           }
\end{figure}

\stepcounter{section}
\addcontentsline{toc}{section}{\Alph{chapter}.\arabic{section}$\;$
   Scalar Triplet Correlator ($\delta$)}

\begin{figure}
  \epsfbox[93 0 480 550]{p11t.out}
  \caption[Scalar Triplet Correlator ($\delta$)]{\label{fig73}
     \it Scalar triplet correlator normalized to the
     free massless quark correlator. There is a strong repulsion
     in this channel and no boundstate is formed.
     The theoretical curve is compared to a curve obtained from a
     pure continuum spectrum above $E_\delta$.
           }
\end{figure}

\stepcounter{section}
\addcontentsline{toc}{section}{\Alph{chapter}.\arabic{section}$\;$
   Scalar Singlet Correlator ($\sigma$)}

\begin{figure}
  \epsfbox[93 0 480 550]{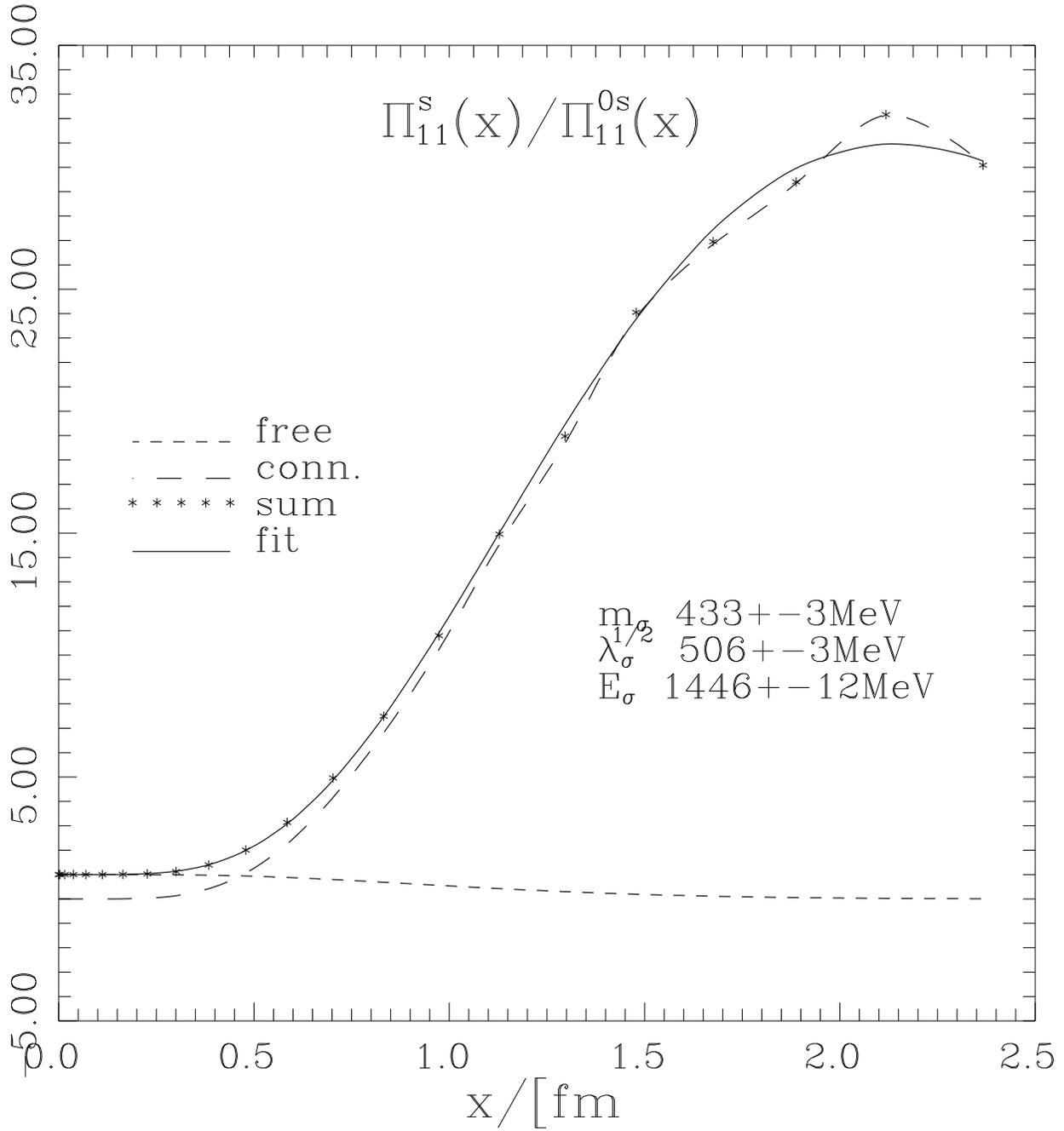}
  \caption[Scalar Singlet Correlator ($\sigma$)]{\label{fig74}
     \it Scalar singlet correlator normalized to the
     free massless quark correlator.
     The $\sigma$ mass $m_\sigma$ and coupling $\lambda_\sigma$
     and the threshold $E_\sigma$ are obtained from a
     spectral fit.
           }
\end{figure}

\stepcounter{section}
\addcontentsline{toc}{section}{\Alph{chapter}.\arabic{section}$\;$
   Axial Vector Correlator ($a_1,f_1$)}

\begin{figure}
  \epsfbox[93 0 480 550]{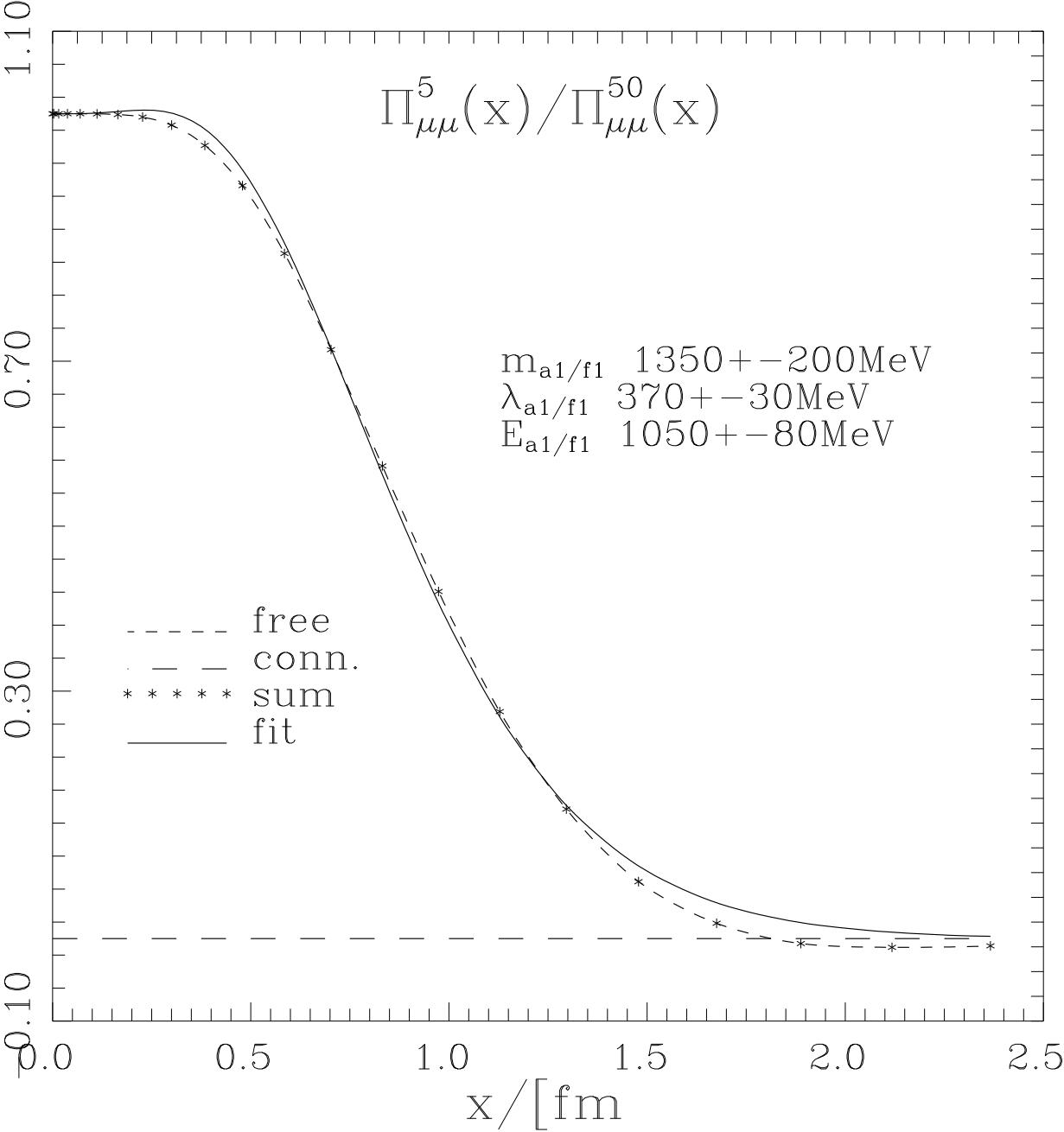}
  \caption[Axial Vector Correlator ($a_1,f_1$)]{\label{fig75}
     \it Axial vector correlator normalized to the
     free massless quark correlator. The triplet and singlet
     correlator are equal because the connected part has been neglected.
     The $a_1$ and $f_1$ mass, coupling and threshold
     are obtained from a spectral fit.
           }
\end{figure}

\stepcounter{section}
\addcontentsline{toc}{section}{\Alph{chapter}.\arabic{section}$\;$
   Vector Correlator ($\rho,\omega$)}

\begin{figure}
  \epsfbox[93 0 480 550]{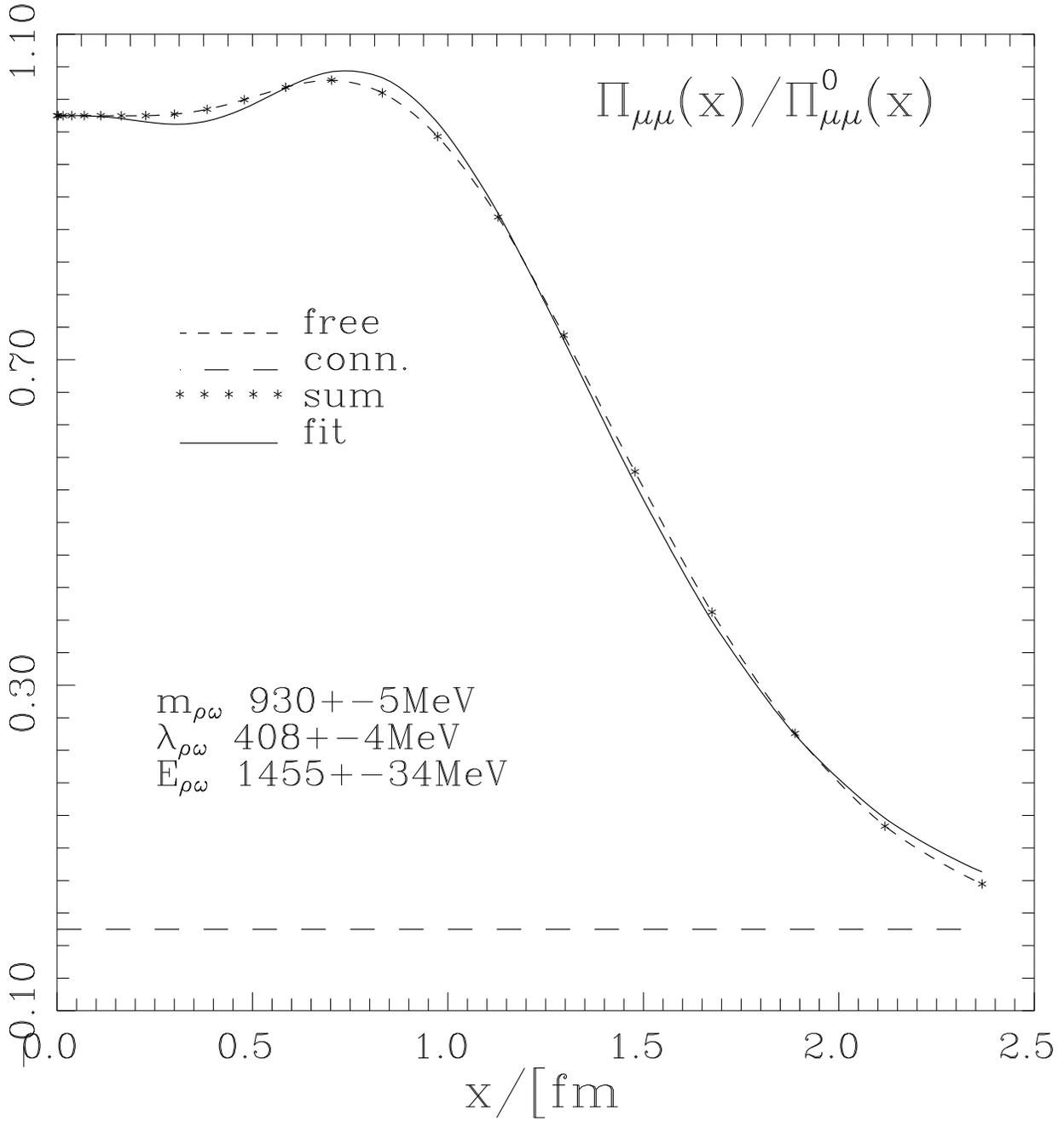}
  \caption[Vector Correlator ($\rho,\omega$)]{\label{fig76}
     \it Vector correlator normalized to the
     free massless quark correlator. The triplet and singlet
     correlator are equal because the connected part is zero.
     The $\rho$ and $\omega$ mass, coupling and threshold
     are obtained from a spectral fit.
           }
\end{figure}
\clearpage



\stepcounter{chapter}
\addcontentsline{toc}{chapter}{\Alph{chapter}$\;$ References}
\parskip=0ex plus 1ex minus 1ex

\newpage

\centerline{\Huge\bf Curriculum Vitae} \vspace*{15mm}

{\Large\bf Personal}
\begin{list}{\hfill}{\labelwidth=40mm\leftmargin=42mm}
\item[Name:\hfill]         Marcus Hutter
\item[Adress:\hfill]      Josephsburgstr. 59 \\
               D-81673 Munich
\item[Birth:\hfill]      14. April 1967 in Munich
\end{list}

{\Large\bf Schooling}
\begin{list}{\hfill}{\labelwidth=40mm\leftmargin=42mm}
\item[Sep.1973 - Jul.1977\hfill] Primary School in Munich and USA
\item[Sep.1977 - Mai.1987\hfill] High School in Munich and Markt-Schwaben \\
                     Majors in Math \& Physics
\item[20.Mai.1987\hfill]         Degree: Abitur
\end{list}

{\Large\bf Degree in Computer Sciences}
\begin{list}{\hfill}{\labelwidth=40mm\leftmargin=42mm}
\item[Nov.1987 - Mai.1992\hfill] Computer Science with minors in Math
                     at the Technical University  Munich.
                     Main area of study: Artificial Intelligence,
                     Neuronal Nets, Genetic Algorithms
\item[Master Thesis:\hfill] Implementation of a Classfier System
\item[26.Mai.1992\hfill]         Degree: Masters (Diplom)
\end{list}

{\Large\bf Degree in Physics}
\begin{list}{\hfill}{\labelwidth=40mm\leftmargin=42mm}
\item[Nov.1989 - Jul.1993\hfill] General Physics at the Technical University Munich
\item[Nov.1992 - Jan.1996\hfill] Ph.D. thesis in theoretical
                     particle physics. Supervisor Prof. H. Fritzsch.
                     Topic: Instantons in QCD
\end{list}


\end{appendix}
\end{document}